\pdfoutput=1
\documentclass[useAMS,usenatbib]{mn2e}
\usepackage{aas_macros}
\usepackage[a4paper,centering, totalwidth=520pt, totalheight=700pt]{geometry}
\usepackage{graphicx}
\usepackage{amssymb}
\usepackage{amsmath}
\usepackage{enumerate}

\title[Multi-scale initial conditions]{Multi-scale initial conditions for cosmological simulations}
\author[O. Hahn \& T. Abel]{Oliver Hahn$^{1}$\thanks{E-mail:\,ohahn@stanford.edu} and Tom Abel$^{1,2,3}$ \\
$^{1}$Kavli Institute for Particle Astrophysics and Cosmology, SLAC/Stanford University, 2575 Sand Hill Road, Menlo Park, CA 94025, USA\\
${^2}$Zentrum f\"ur Astronomie der Universit\"at Heidelberg, Institut f\"ur Theoretische Astrophysik, Albert-Ueberle-Str. 2, 69120 Heidelberg, Germany\\
${^3}$Heidelberg Institut f\"ur Theoretische Studien, Schloss-Wolfsbrunnenweg 35, 69118 Heidelberg, Germany}
\begin{document}

\date{MNRAS in press}
\pagerange{\pageref{firstpage}--\pageref{lastpage}} \pubyear{2011}
\maketitle

\label{firstpage}

\begin{abstract}
  We discuss a new algorithm to generate multi-scale initial 
  conditions with multiple levels of refinements for cosmological
  ``zoom-in'' simulations. The method uses an adaptive convolution of
  Gaussian white noise with a real space transfer function kernel
  together with an adaptive multi-grid Poisson solver to generate
  displacements and velocities following first (1LPT) or second order
  Lagrangian perturbation theory (2LPT).  The new algorithm achieves
  RMS relative errors of order $10^{-4}$ for displacements and velocities in
  the refinement region and thus improves in terms of errors by about
  two orders of magnitude over previous approaches. In addition,
  errors are localized at coarse-fine boundaries and do not suffer
  from Fourier-space induced interference ringing.  An optional hybrid
  multi-grid and Fast Fourier Transform (FFT) based scheme is
  introduced which has identical Fourier space behaviour as
  traditional approaches. Using a suite of re-simulations of a galaxy cluster
  halo our real space based approach is found to reproduce correlation functions,
  density profiles, key halo properties and subhalo abundances with
  per cent level accuracy.  Finally, we generalize our approach for two-component
  baryon and dark-matter simulations and demonstrate that the power spectrum evolution
  is in excellent agreement with linear perturbation theory. For initial baryon density fields, it is
  suggested to use the local Lagrangian approximation in order to
  generate a density field for mesh based codes that is consistent
  with Lagrangian perturbation theory instead of the current practice
  of using the Eulerian linearly scaled densities.  
\end{abstract}



\section{Introduction}

Cold dark matter density perturbations drive all of cosmological
structure formation \citep[e.g.][]{1982ApJ...263L...1P,
  1985ApJ...292..371D} in the current paradigm. Their non-linear
growth is followed with numerical simulations over an enormous range
of scales. From scales of hundreds of Megaparsecs
\citep[e.g.][]{2005Natur.435..629S,2010MNRAS.401..705P} studying
cosmological large-scale structure; to several hundred kilo parsecs
modeling the formation and evolution of galaxies including aspects of
the interstellar medium, molecular cloud-formation and black-hole
accretion processes \cite[e.g.][]{2008ApJ...676...33D,
  2008ApJ...685...40W, 2009ApJ...695..292C,2010arXiv1007.2566B}; and
even the formation of the first proto-stars formed at scales of a few
solar radii \citep[e.g.][]{2002Sci...295...93A, 2008Sci...321..669Y,
  2009Sci...325..601T} are studied in cosmic context.

The fluctuations in the standard model would extend from solar system
size to giga parsec scales. No single calculation would be able to
capture all these scales. One may therefore strive to carry out
simulations in as large volumes as possible to study a range of
representative environments for the formation of particular
cosmological objects. At the same time, however, high resolution is
necessary to sample the underlying gravitational potential wells.  A
popular way to achieve large simulation boxes with as high as possible
resolution for individual objects 
is the so-called ``zoom-in'' technique \citep[following versions of
e.g.][]{1994MNRAS.270L..71K,1994MNRAS.267..401N}. In this technique, a
small region encapsulating the formation history of the object of
interest is studied with much higher resolution inside of a lower
resolution representation of its larger-scale cosmic environment
interacting with the object through long-range gravitational forces.
The aim is to represent both the variance of the fluctuations on
scales of the object (sample variance) due to the use of a large
volume which provides a fair sample of cosmic environments
\citep[cf. eg.][]{2009MNRAS.399.1773C,2010MNRAS.405..274H}, as well as
the smaller scale fluctuations affecting the structure of the object
directly.  

Typically, two approaches have been followed for initial conditions of 
``zoom-in simulations''. Either, initial conditions are generated on a 
 uniform grid at the full resolution \cite[which can amount to a huge need
 for computational resources for deeply nested grids, see e.g.][]{
 2008ApJS..178..179P,2009MNRAS.398L..21S} and subsequently degraded
 by either averaging over regions of $2^{3d}$ particles, where $d$ is the de-refinement 
 factor outside the high resolution region, or by Fourier resampling. There are several disadvantages 
 with this approach: First, the computational requirements are immense compared 
 to the solution sought; and second, the simple averaged velocity fields are neither 
 continuous nor differentiable across coarse-fine boundaries, leading to 
 spurious shocks in baryonic simulations. Alternatively, e.g. the {\sc Grafic-2} 
 code \citep{2001ApJS..137....1B} is used which employs discrete Fourier transforms 
 to add small scales perturbations to coarser perturbations. The {\sc Grafic-2} approach 
 however requires the use of an anti-aliasing filter which damps out small-scale power
 and shows oscillatory errors at the few per-cent level that result from the combination 
 of discrete Fourier transforms of different resolution. Furthermore, it does not 
 incorporate the sampling of the real-space transfer function advocated by 
 \cite{1997ApJ...490L.127P} and \cite {2005ApJ...634..728S}. This latter aspect is a 
 feature of all codes that generate initial conditions by an inverse Fourier transform of a 
 sampled power spectrum on a three-dimensional lattice and thus also plagues most 
 applications of the previous approach. \cite {2005ApJ...634..728S} has shown
 conclusively that this conventional method can lead to a significant error in the 
 real-space statistical properties, in particular for small cosmological box sizes.

This paper extends on previous work by proposing a new algorithm to
generate multi-scale nested initial conditions
\citep[as][]{2001ApJS..137....1B}, but using 
a real-space convolution kernel to represent the Fourier space transfer function
of density perturbations
\citep[cf.][]{1996ApJ...460...59S,1997ApJ...490L.127P,2005ApJ...634..728S}
in a multi-scale convolution algorithm combined with an adaptive
multi-grid algorithm to determine initial velocities, particle
positions and densities compatible with Lagrangian cosmological
perturbation theory up to second order. More specifically, the approach uses 
mass conservation constraints to determine the convolution kernel from an accurate
real space representation of the transfer function kernel on a hierarchy of 
levels, which are used in FFT convolutions to obtain the density field from a hierarchical
white noise field, and a two-way interface multigrid Poisson solver that achieves smooth
gradients across coarse-fine boundaries. The resulting errors are
confined to the boundaries and at the level of a few $10^{-4}$ in
units of the standard deviation of the respective fields for the
interior of the refined region.

We furthermore show that using a hybrid Poisson solver (using
multi-grid for the hierarchy of grids and FFT for the finest grid),
our method produces particle distributions that closely match the
input power spectrum also on the smallest scales and make our 
approach attractive also for uni-grid simulations.  All the methods
discussed in this paper are implemented in the initial conditions code
{\sc Music} (MUlti-Scale Initial Conditions). The code also includes
an implementation of a linearized Boltzmann solver based on {\sc Linger}
\citep{1995ApJ...455....7M} that allows the calculation of cosmological
density and velocity perturbations for dark matter and baryons.

The paper is organised as follows. In Section \ref{sec:densities}, we discuss a 
multi-scale convolution method that produces a nested density field that is 
consistent with both the input correlation function and the input power spectrum, 
and which is then used as the source field for Lagrangian perturbation theory. 
Computing displacement and velocity fields in Lagrangian perturbation theory 
(LPT) requires a numerical solution to Poisson's equation. In Section 
\ref{sec:dispvel}, after briefly summarizing LPT, we describe the adaptive 
multi-grid algorithm that we use to obtain displacement and velocity fields. We 
also introduce a hybrid Fourier-space/multi-grid based Poisson solver which 
has better Fourier-space properties on small scales compared to the pure
multi-grid algorithm. In Section \ref{sec:error_analysis}, we discuss the 
performance of our method compared to initial conditions generated purely in 
Fourier-space both in unigrid and nested grid set-ups. We generalize our 
approach to initial conditions for two-component dark-matter and baryon 
simulations in Section \ref{sec:baryon_ics} and propose the usage of the local 
Lagrangian approximation to generate inital baryon density fields for grid 
codes before we summarize and conclude our results in Section \ref{sec:summary}.


\section{Generating density perturbations}
\label{sec:densities}
In this section, we summarize methods to generate Gaussian random fields that 
follow a prescribed power spectrum and act as source terms for density and velocity 
perturbations in Lagrangian perturbation theory. We give particular attention to  a 
convolution kernel based approach which has favourable properties in a multi-scale 
nested grid set-up. 

\subsection{Computing the seed density field}
Consider an over-density field $\delta({\bf r})$ that is completely described by its power 
spectrum $P(k) \equiv < \widetilde{\delta}({\bf k})\, \widetilde{\delta}^\ast({\bf k})>$, 
where the tilde denotes the transformed function of a Fourier transform pair. The amplitude 
of density fluctuations is usually expressed in terms of the transfer function $\mathcal{T}(k)$, which is defined 
such that
\begin{equation}
P(k) = \alpha\,k^{n_s} \, \mathcal{T}^2(k),
\end{equation} 
where $n_s$ is the (constant) power spectrum spectral index after inflation, and $\alpha$ is a 
normalization constant. The first step in setting up initial conditions for cosmological
simulations thus consists in generating a white noise sample of random values $\mu({\bf r})$ 
(typically sampled from a Gaussian distribution with zero mean and unit variance) 
and to require that their amplitudes follow a specific power spectrum $P(k)$. This can 
be achieved by multiplying the Fourier transformed white noise field $\widetilde{\mu}(\bf k)$ 
with the square root of the power spectrum, i.e. for all ${\bf k}$ representable on a grid of 
given resolution set
 \begin{equation}
 \widetilde{\delta}({\bf k}) = \sqrt{P(\left\| {\bf k} \right\|)}\,\widetilde{\mu}({\bf k}) = 
 	\alpha\left\|{\bf k}\right\| ^{n_s/2}\,\mathcal{T}(\left\| {\bf k} \right\|)\,\widetilde{\mu}({\bf k}).
 \label{eq:delta_amplitude_k}
 \end{equation}
 The real space over-density field $\delta({\bf r})$ is then obtained by inverse Fourier 
 transformation, and we call this procedure ``$k$-space sampling''.
 
 It is interesting to note that a product in Fourier space simply corresponds to a convolution 
 in real space \citep[cf. also][]{1985ApJS...57..241E,1996ApJ...460...59S}, 
 i.e. equation (\ref{eq:delta_amplitude_k}) is equivalent to
 \begin{equation}
 \delta({\bf r}) = T(\left\| {\bf r}\right\|)\,\star\,\mu({\bf r}),
  \label{eq:delta_amplitude_r}
 \end{equation}
 where $T(r)$ is the real-space counterpart of $\widetilde{T}(k)\equiv \alpha k^{n_s/2}\,\mathcal{T}(k)$, 
 and ``$\star$'' denotes a convolution.
 
 It is thus mathematically equivalent whether equation
 (\ref{eq:delta_amplitude_k}) is evaluated in Fourier space, followed
 by an inverse transform, or whether equation
 (\ref{eq:delta_amplitude_r}) is evaluated using an inverse transform
 of $\widetilde{T}(k)$ followed by the convolution. Most cosmological initial
 conditions generators follow the first approach
 \citep[e.g.][]{2001ApJS..137....1B}, while
 e.g. \cite{1996ApJ...460...59S}, \cite{1997ApJ...490L.127P} and
 \cite{2005ApJ...634..728S} use the second or variations thereof. The
 discrete realizations of the density fields derived with the two
 approaches will have significant differences. In particular,
 employing eq. (\ref{eq:delta_amplitude_k}) forces periodicity of the
 real space transfer function on box scales that leads to an
 underestimation of the two-point correlation function (see Section
 \ref{sec:tf_real}, and e.g. \citealt{1997ApJ...490L.127P} and
 \citealt{2005ApJ...634..728S} for a detailed discussion).  This is
 particularly relevant for small cosmological boxes ($L\lesssim
 100\,h^{-1}{\rm Mpc}$), where the periodic contribution is
 non-negligible.
 
 The convolution kernel $T(r)$ can be thought of as a ``propagator'' of 
 a single Gaussian white noise fluctuation $\mu_\mathbf{q}(\mathbf{x})\equiv\delta_D(\mathbf{x}-\mathbf{q})$
 at location $\mathbf{q}$, where $\delta_D$ is the Dirac $\delta$-function. The time evolution
 of each single such perturbation is given by the time evolution of $T(r)$. Instead of a variation of the 
 traditional method of sampling Fourier modes according to a given power spectrum, this real-space 
 picture has thus an intuitive physical motivation: The convolution operation imprints the 
 density perturbations due to the white noise $\mu(\mathbf{x})$ onto space.
 Apart from the better behaviour of the two-point correlation function, we follow the 
 convolution approach since it allows for an easier treatment of nested grids that 
 are separated by sharp boundaries in real space. To achieve this, a method will 
 be developed that correctly overlaps the ``propagator'' with such 
 coarse-fine boundaries. We give an 
 account of how we numerically compute eq. (\ref{eq:delta_amplitude_r}) in the next section 
 and of the multi-scale convolution
 approach for nested grids in Section \ref{sec:nested_convolution}.

\subsection{The real-space transfer function}
\label{sec:tf_real}
\begin{figure}
\begin{center}
\includegraphics[width=0.4\textwidth]{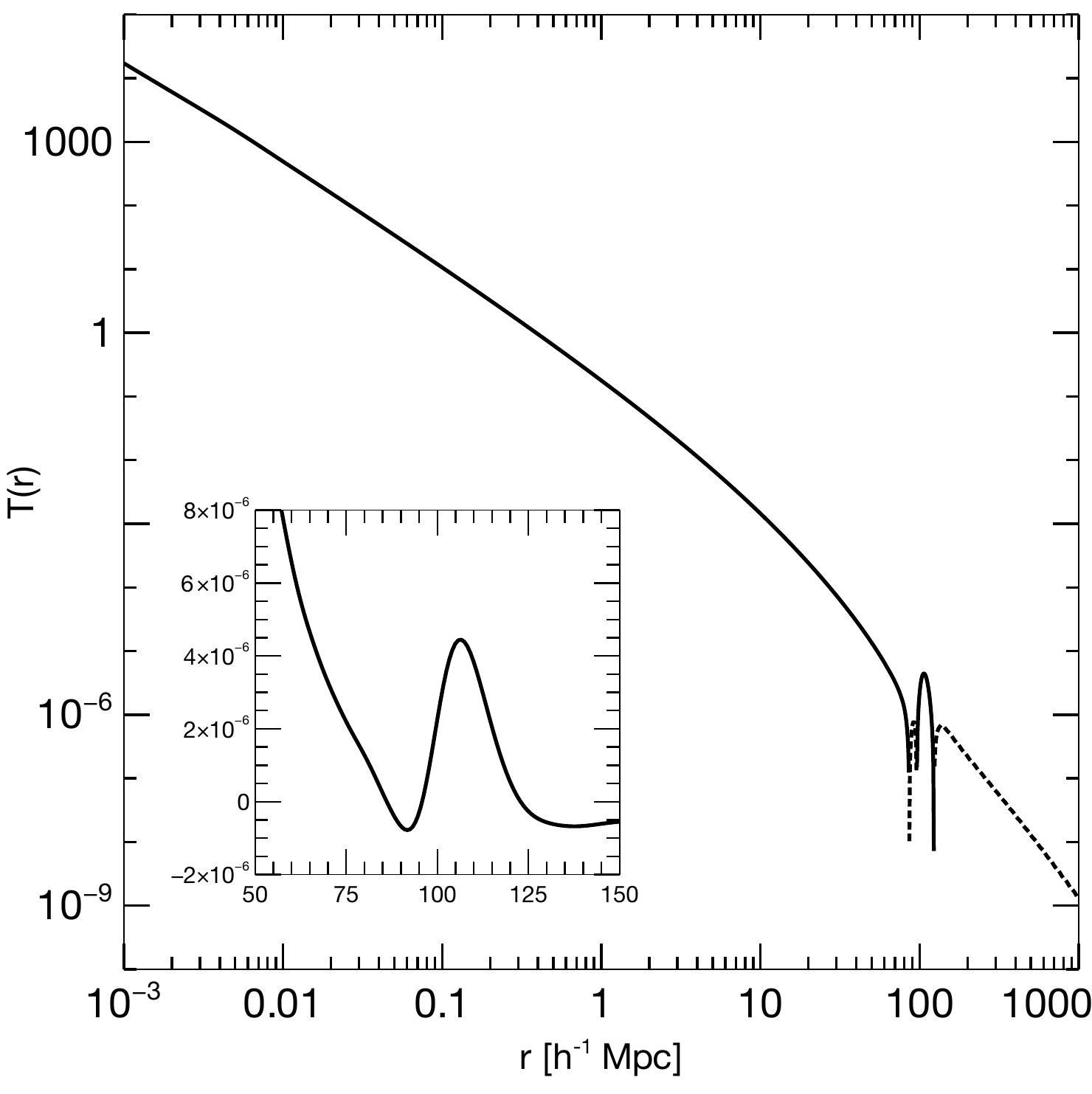}
\end{center}
\caption{\label{fig:transfer_real}The real space transfer function $T(r)$ at $z=50$ for 
total matter density perturbations obtained using the {\sc FFTlog} method from a 
tabulated k-space transfer function $T(k)$. The transfer function is positive where the 
line is solid and negative where it is dotted. The inset shows the baryon oscillation 
peak in linear scale where the change of sign occurs.}
\end{figure}
Since the transfer function $\widetilde{T}(k)$ is typically computed numerically using a 
spectral Boltzmann solver -- e.g. the Boltzmann solver included with our code, 
{\sc Linger} \citep{1995ApJ...455....7M}, {\sc CMBfast} \citep{1996ApJ...469..437S} 
or {\sc Camb}\footnote{{\sc Camb}, written by A.~Lewis and A.~Challinor, can be 
obtained from http://camb.info}, or taken from a fitting formula
(e.g. \citealt{1998ApJ...496..605E}) -- its Fourier transform is not available a priori. Thus, in the 
approach we follow, an accurate real-space transfer function $T_{\mathcal{R}}(r)$ is first
calculated which is subsequently applied as a convolution kernel to the 
Gaussian white noise field using FFT convolution. Assuming a spherically
symmetric transfer function $\widetilde{T}(k)$, one has
\begin{eqnarray}
T_{\mathcal{R}}(r) & = & \frac{1}{(2\pi)^3}\int_\mathbb{R}\widetilde{T}(k)\,\exp(i\,{\bf x}\cdot{\bf k})\,{\rm d}^3k\\
& = & \frac{4\pi}{(2\pi)^3}\int_0^\infty \widetilde{T}(k)\frac{\sin(kr)}{kr}\,k^2\,{\rm d}k.
\end{eqnarray}
This real space transfer function, $T_{\mathcal{R}}(r)$, can be
computed using the  {\sc FFTlog} algorithm
\citep{1978JCoPh..29...35T,2000MNRAS.312..257H} from an input k-space
transfer function $T(k)$.  {\sc FFTlog} 
uses the fact that the Fourier transform of a log-log function can be written as a Hankel 
transform which can be carried out by FFT. The resulting transform is highly accurate over many orders of magnitude. 
We kindly refer the reader to \cite{2000MNRAS.312..257H} for details. The $r=0$ component needs to be computed separately.
It is obtained by a three dimensional numerical integration over the range of Fourier modes included in the simulation.
This $r=0$ component sets the white noise level of the density perturbations.

Figure \ref{fig:transfer_real}, shows the result of the transform from
tabulated k-space transfer functions at $z=50$ for
matter density perturbations generated with a 1D linear Boltzmann solver
\citep[cf.][]{1995ApJ...455....7M}. 

Finally, the convolution in eq. (\ref{eq:delta_amplitude_r}) has to be computed numerically. To this end,
we compute a discretisation  $T(\mathbf{x}_{ijk}) =  \mathcal{T}_{\mathcal{R}}(\|| \mathbf{x}_{ijk} \||)$, 
with $\mathbf{x}_{ijk}=(ih, jh, kh)$, $(i,j,k)\in\left[ -N/2+1, N/2 \right]^3 $ over the entire simulation volume 
of length $L=hN$. Next, a white noise random field $\mu(\mathbf{x}_{ijk})$ is created on the same discretisation.
Then both $T(\mathbf{x}_{ijk})$ and $\mu(\mathbf{x}_{ijk})$ are transformed by FFT, multiplied element-by-element with one 
another and the result is inverse transformed to yield $\delta(\mathbf{x}_{ijk})$.
Note that when sampling on a finite grid, the real-space sampled transfer function is 
no longer spherically symmetric when transformed by FFT to three dimensional Fourier space.
However, equivalently, the traditional $k$-space sampled transfer function is not spherically symmetric
in real space.

\begin{figure}
\begin{center}
\includegraphics[width=0.4\textwidth]{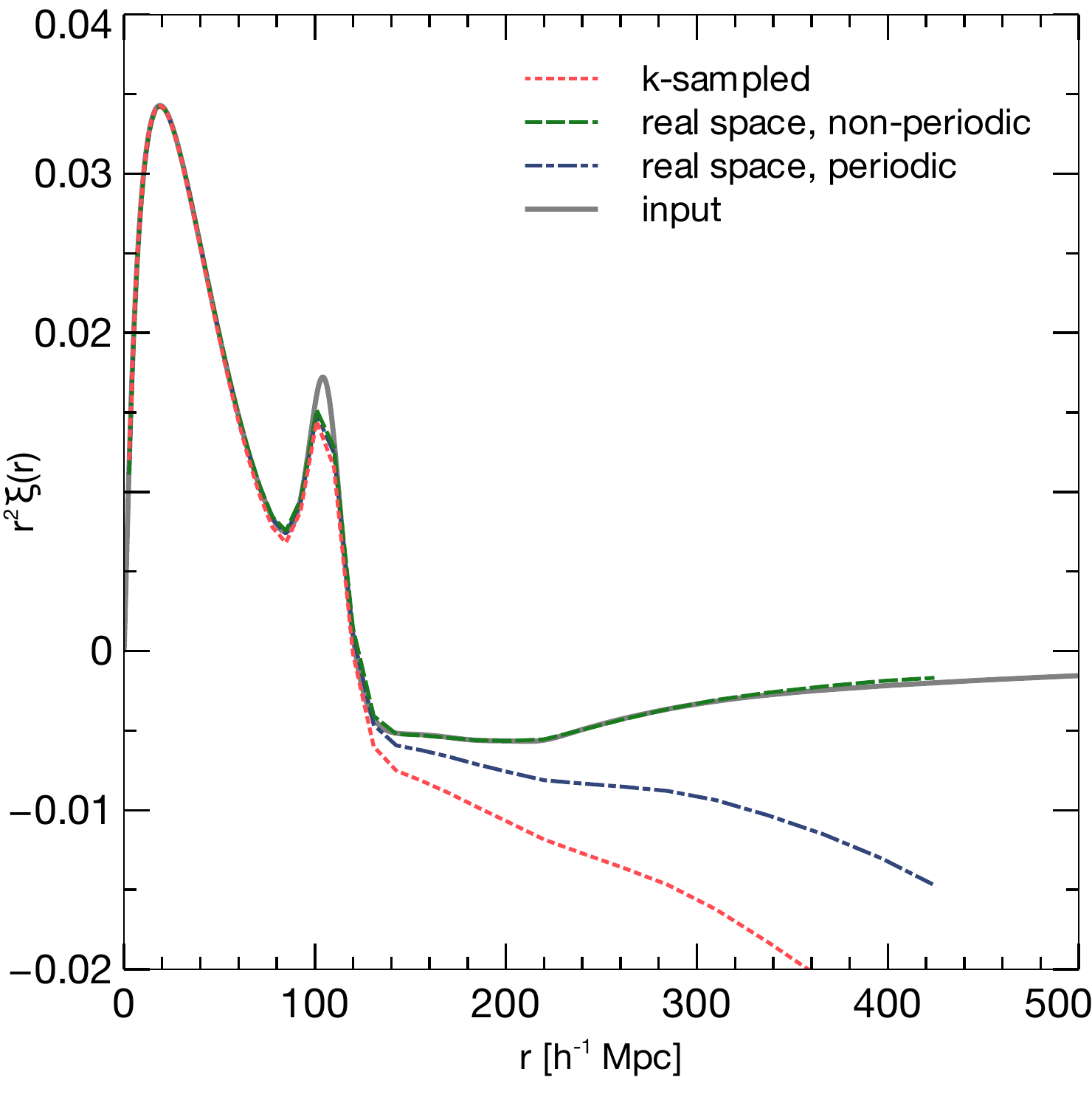}
\end{center}
\caption{\label{fig:corrfun_ic}Auto-correlation function of the transfer function (equivalent to the two-point correlation function) for $k$-space
sampled (dotted red) and real space transfer functions that are non-periodic (dashed green) and periodic on the box length (dot-dash blue).
The gray line indicates the two-point correlation function determined from the input power spectrum at $z=45$. All auto-correlation functions
have been computed for a $500\,h^{-1}{\rm Mpc}$ box with $256^3$ resolution. }
\end{figure}

The analysis performed in this paper uses a real-space transfer
function periodically replicated on the box length. This abandons the
favourable properties of a truncated correlation kernel discussed in
detail by \cite{2005ApJ...634..728S}. However, this choice allows
comparison with the classical approach that samples the k-space
transfer function. For actual applications it is recommended that the truncated transfer functions
should be used (which is a simple parameter in our code). 

In Figure \ref{fig:corrfun_ic}, we show the auto-correlation function of the
transfer function kernels (equivalent to the two-point correlation
function). As demonstrated by \cite{2005ApJ...634..728S}, the
non-periodic real space kernel perfectly reproduces the input
two-point correlation function. The
periodic correlator, for which we have not subtracted the mean in this case,
underestimates the correlation function at large
radii and the $k$-space sampled kernel shows an even stronger drop at
large radii. Note that the big drop in the $k$-space sampled
correlation function is somewhat spurious. It is partly from  the
periodic component, as the periodic real space kernel shows a similar
suppression. However, it is mainly due to an additional integral
constraint. For the $k$-space sampled kernel, $P(k=0)=0$, so that
\begin{equation}
0=\int_\mathcal{D} \xi(\mathbf{x})\,{\rm d}^3x\simeq 4\pi\int \xi(r)\,r^2\,{\rm d}r,
\end{equation}
where $\mathcal{D}$ is the simulation volume and the latter equality holds in 
the case of a spherically symmetric correlation function and amounts to the usual integral
constraint on the two-point correlation function. This integral constraint however holds
for a finite $\mathcal{D}$ in the traditional $k$-space sampling approach (while $\mathcal{D}\simeq\mathbb{R}^3$
in our case) and the correlation function is thus offset by an
additive constant which is equal to the integral over the correlation function outside the simulation volume. This additive
constant leads to the additional deviation seen between the $k$-space sampled correlation function and the periodic 
real-space equivalent which is simply amplified by the multiplication with $r^2$. Any box with zero mean overdensity will therefore
fulfill the integral constraint over the box instead of over an infinite volume leading to a similar discrepancy. Boxes with a
non-zero mean overdensity would circumvent this problem and
the difference in mean overdensity can be incorporated in a change to different effective cosmological parameters. We will
not investigate further the possibility of simulations with non-vanishing mean overdensity but kindly refer the reader to \cite{2005ApJ...634..728S}
for a detailed discussion of this possibility.


\subsection{Generating a nested initial density field} 
In order to generate an accurate multi-scale representation of an initial density field, we will
now describe our algorithm to perform the convolution in eq. (\ref{eq:delta_amplitude_r}) on nested
grids. Particular care is taken to achieve mass conservation, as violations would affect
the entire domain when the density field is used to source the displacement and velocity fields 
(cf. Section \ref{sec:dispvel}). As in the {\sc Grafic2} approach\citep{2001ApJS..137....1B},
this amounts to linear constraints between the levels. However, unlike {\sc Grafic2}, these
constraints are not only top-down (i.e. the coarse level provides constraints for the fine),
but partially bottom-up to achieve a more conservative algorithm and high-resolution 
sampling of the real-space convolution kernel.

\begin{figure}
\begin{center}
\includegraphics[width=0.25\textwidth]{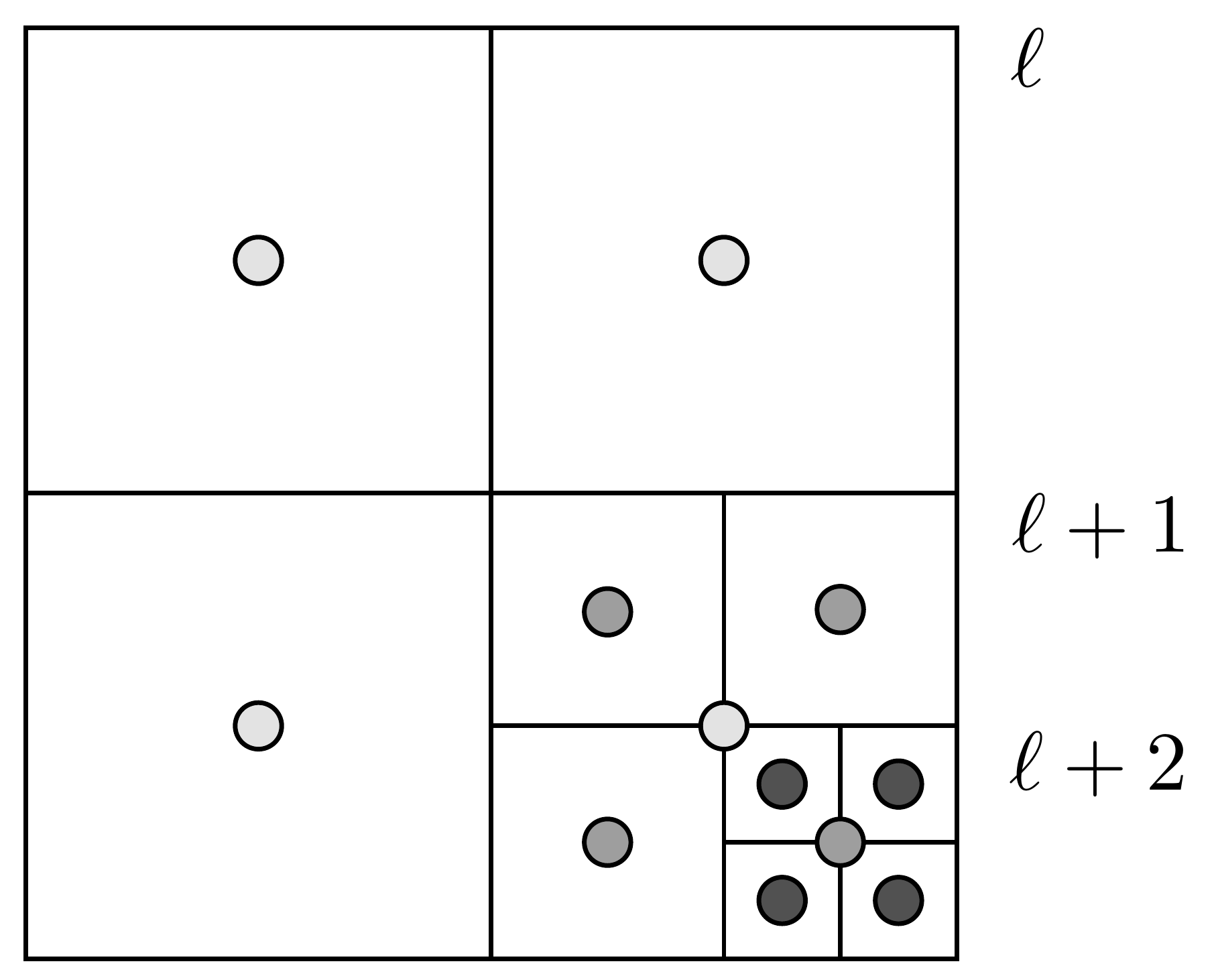}
\end{center}
\caption{\label{fig:grid_structure}The volume subdivision based nested grid structure used in 
the multi-scale algorithms. Shown are two additional refinement levels in which a parent cell 
can be refined into eight children cells. Note that centers of children do not coincide with the 
parent grid cell centres.}
\end{figure}

\label{sec:nested_convolution}
\subsubsection{White noise refinement}
In order to generate initial conditions for a nested subdomain on the volume subdivision grid structure
used in our approach (Figure \ref{fig:grid_structure}), the white noise for the subgrid has to be consistent 
with that of the coarse grid. Two separate approaches can be followed that are locally mass-conserving (in
the sense that the children cells average to the parent cell locally rather than that only the total mass in the subgrid 
domain is conserved):

The first approach uses the Hofmann-Ribak algorithm \citep{1991ApJ...380L...5H}. First an unconstrained
white noise field $\omega^{\ell+1}$ is generated for level $\ell+1$ which has 8 times the variance of the
noise field for level $\ell$ in the case of a refinement factor of 2. Next, for each group of eight children cells on the fine grid,
the mean is replaced by the respective value of the noise field on the coarser level $\ell$ \citep[cf. also][]{1997ApJ...490L.127P,2001ApJS..137....1B}. 

Since this approach does not retain the Fourier modes present on the coarse grid, another approach 
can be devised that achieves this: (1) again, an unconstrained white noise
field $\omega^{\ell+1}$ is generated for level $\ell+1$ at 8 times the coarse level variance.
(2) In order to preserve Fourier modes that are present in the coarse grid $\ell$, an FFT of
both the fine grid and that part of the coarse grid that overlaps the fine grid is performed. Then all modes
$\mathbf{k}\leq\mathbf{k}_{{\rm Ny},\ell}$ up to the Nyquist wave number of the coarse grid are replaced
with the respective Fourier coefficients from the coarse grid, followed by an inverse transform. 
(3) In order to ensure conservation of mass, 
the Hoffman-Ribak algorithm is applied in reverse, i.e. the coarse grid white noise values are
replaced by the average over the eight children of a coarse cell inside the refinement region. 
Since the Fourier interpolation conserves the mass of the entire subgrid, as does the 
averaging over children, the global mass in the simulation domain is also conserved.

Further levels can be computed by repeating first step (2) for all levels at increasing resolution, followed by the correction
step that replaces all coarse cells at all coarser levels with the average over the eight fine cells, starting at the finest level.
Thus, the Hoffman-Ribak constraint is fulfilled while at the same time Fourier modes are preserved between
grids.

In the following, we will describe how the convolution with the real-space transfer function, 
eq. (\ref{eq:delta_amplitude_r}), is performed when a refinement region is present.

\subsubsection{Generating convolution kernels} 
In Figure \ref{fig:grid_scheme}, we show schematically the set-up for one additional refinement level. The top grid 
domain $\Omega$ consists of the domain $\Omega_2$ covered by a refinement grid and the non-refined part 
$\Omega\backslash\Omega_2$. The refinement region $\Omega_2$ at twice the resolution is given
by $\Omega^\prime$. $\Omega_{2,p}$ represents the padding around $\Omega_2$ to twice the length
per dimension that is needed to perform isolated Fourier convolutions.

The next step in our method is to generate convolution kernels $T(r)$ for all required levels. In contrast to the {\sc Grafic2}-approach, 
we construct these kernels purely in real space, starting from the finest level. For the finest level $\ell$, this consists
of a simple evaluation of the real-space transfer function $\mathcal{T}_{\mathcal{R}}(r)$ on $\Omega^\prime\cup\Omega^\prime_{p}$,
i.e. $T^{\ell}(\mathbf{x}_{ijk})=\mathcal{T}_{\mathcal{R}}(\|| \mathbf{x}_{ijk} \||)$ for all $\mathbf{x}_{ijk}\in\Omega^\prime\cup\Omega^\prime_{p}$
 with the origin $\mathbf{x}=0$ placed at the centre. 

As for the white noise, care has to be taken to maintain local conservation of mass. Thus, rather than sampling at coarser resolution, for the next
coarser level $\ell-1$, the already discretized part of the convolution kernel for the fine-level grid needs to be restricted to the coarser grid
in a conservative way.
This can be achieved by averaging the contributions due to all eight children cells when convolved with the kernel, i.e. we compute
\begin{equation}
T^{\ell-1}(x,y,z) = \frac{1}{8}\sum_{i,j,k\in\{-\frac{1}{2},+\frac{1}{2}\}}T^{\ell}(x+i,y+j,z+k),
\label{eq:transfer_average}
\end{equation}
for those grid points $(x,y,z)$ that correspond to the refined part of the grid. This provides the convolution
kernel for a volume of the size of the refined region including the padding, i.e. for $\Omega_2\cup\Omega_{2,p}$. 
The remaining cells outside that region, i.e. on $\Omega_1\backslash(\Omega_2\cup\Omega_{2,p})$ are again sampled 
directly from the real-space transfer function: $T^{\ell-1}(\mathbf{x}_{ijk})=\mathcal{T}_{\mathcal{R}}(r)$ for
all $\mathbf{x}_{ijk}\in\Omega_1\backslash(\Omega_2\cup\Omega_{2,p})$. This last step is in violation with strict
mass conservation since eq. (\ref{eq:transfer_average}) does not hold outside the refined region. Since most of the
mass of the convolution kernel is close to the centre, this approximation produces however negligible errors and avoids
the substantial computational overhead of an exact evaluation of $\mathcal{T}_\mathcal{R}(r)$ 
over the entire domain at the finest resolution. The procedure of averaging and newly sampling uncovered volume
 is repeated until the coarsest level is reached.

\subsubsection{Noise convolution: The first refinement level}

The density field on the top grid level $\ell$ is determined
as in the unigrid case by computing $\delta^\ell = T^\ell\star\mu^\ell$ on $\Omega$ with periodic boundary conditions automatically
satisfied by the FFT. For the refined region, several contributions will be co-added:

\begin{figure}
\begin{center}
\includegraphics[width=0.45\textwidth]{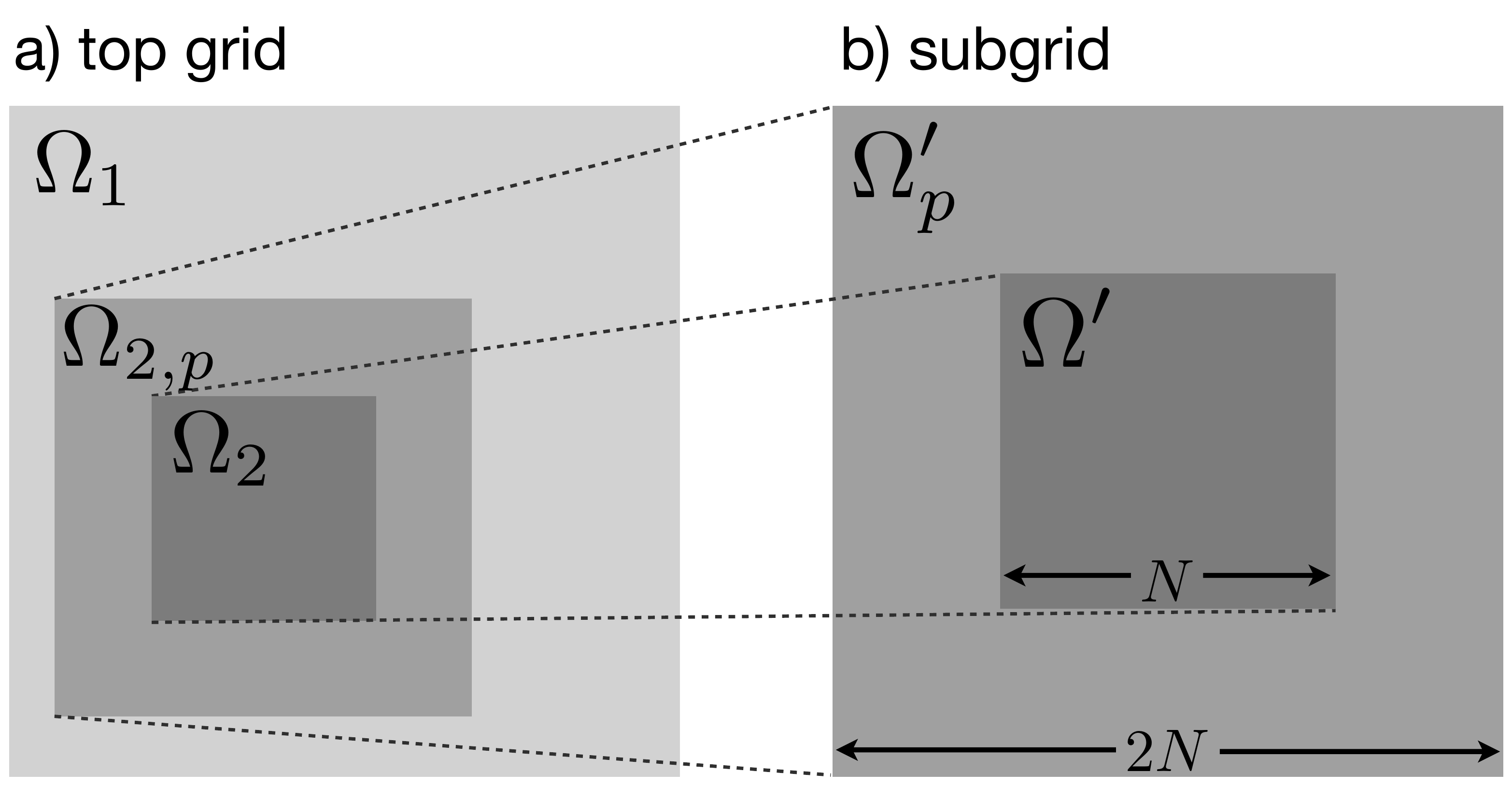}
\end{center}
\caption{\label{fig:grid_scheme}Schematic representation of a set-up with one refinement level. The left panel a) shows
the top grid $\Omega$ which consists of a region $\Omega_2$ covered by a refinement grid and the non-refined
part $\Omega\backslash\Omega_{2}$. The right panel b) shows the sub-grid domain $\Omega^\prime$ which is equivalent to $\Omega_2$
in the left panel but has twice the resolution. To this sub-grid a padding region $\Omega^\prime_p$ will be added when
performing the FFT convolution with isolated boundaries and is denoted by $\Omega_{2,p}$ on the top grid.}
\end{figure}

\begin{enumerate}[1.]
\item The coarse grid contribution $\delta^{\ell+1}_{\rm coarse}$ to the refinement region is computed by zeroing $\mu^\ell$ 
on $\Omega_2$ to obtain $\mu_{\Omega\backslash\Omega_2}^{\ell}$, computing the convolution $\delta_1^{\ell}=T^\ell\star\mu_{\Omega\backslash\Omega_2}^{\ell}$
and interpolating $\delta_1^{\ell}$ from $\Omega_2$ to $\delta^{\ell+1}_{\rm coarse}$. We use tri-cubic interpolation for
this purpose.
\item On level $\ell+1$ isolated boundary conditions apply, so $\Omega^\prime$ has to be padded to twice its size $2N$ so that an 
FFT-based convolution is still possible. In order to find the density $\delta^{\ell+1}_{\rm self}$ due to the sub-grid alone, 
$\mu^{\ell+1}$ on $\Omega_p^\prime$ is set to zero to obtain $\mu^{\ell+1}_{\Omega^\prime}$, followed by computing the convolution
$ \delta^{\ell+1}_{\rm self} = T^{\ell+1}\star\mu^{\ell+1}_{\Omega^\prime}$.
\item In order to reduce errors at the boundary, a correction term $\delta^{\ell+1}_{\rm bnd}$ is added that accounts for the 
fluctuations just outside $\Omega^\prime$. To this end, the coarse grid value of the white noise field is subtracted from each
of the 8 children cells, equivalent to ``unapplying'' the Hoffman-Ribak method. We then zero the result of this operation on $\Omega^\prime$,
so that it is non-zero only on $\Omega^\prime_p$ and obtain $\hat{\mu}^{\ell+1}_{\Omega^\prime_p}$. Since boundary conditions are
isolated, it would be necessary to truncate the transfer function in order to have a non-periodic unconstrained white noise field. 
We however found that a truncation introduces larger errors than assuming periodicity (on scales larger than the subgrid).
Finally, the FFT-convolution $\delta^{\ell+1}_{\rm bnd}=T^{\ell+1}\star\hat{\mu}^{\ell+1}_{\Omega^\prime_p}$ is evaluated.
\item Finally, the result of the previous step is restricted also to the coarse grid, i.e. onto $\Omega_{2,p}$, in order to include
information about fluctuations on smaller scales. 
\end{enumerate}
Finally, all three contributions are added to find the refined density field
\begin{equation}
\delta^{\ell+1} = \delta^{\ell+1}_{\rm self} + \delta^{\ell+1}_{\rm coarse} + \delta^{\ell+1}_{\rm bnd}
\end{equation}
on $\Omega^\prime$. In order to further reduce errors due to the boundary, we allow for optional additional padding 
of the subregion with a few grid cells and cut out the desired region at the end.

\subsubsection{Noise convolution: Further refinement levels}
For further refinement levels, the same steps 1-4 from above are repeated at level $\ell+i$, $i\ge2$, with the only
difference being that the coarse grid contribution $\delta^{\ell+i}_{\rm coarse}$ is computed using isolated boundary
conditions on $\ell+i-1$, i.e. with a zero padded random field $\mu^{\ell+i-1}$. Furthermore, all coarse contributions
have to be interpolated down to this level. Let $\mathcal{P}$ be the interpolation operator, then
\begin{equation}
\delta^{\ell+i}=\delta^{\ell+i}_{\rm self} + \delta^{\ell+i}_{\rm bnd} + \sum_{j=0,\dots,i-1}\mathcal{P}^{i-j}\left[\delta^{\ell+j-1}_{\rm coarse}\right],
\end{equation}
where $\mathcal{P}^q\left[\cdot\right]$ indicates a successive interpolation between $q$ levels. We use a conservative tri-cubic
interpolation for this purpose, i.e. we first perform a normal tri-cubic interpolation and then correct the eight children
cells with an additive constant to ensure that their average equals the coarse cell value.


\section{Initial particle positions and velocity fields}
\label{sec:dispvel}
In this Section, we briefly summarize the application of first and second order Lagrangian perturbation theory to obtain
the initial displacement and velocity fields which are based on solutions of Poisson's equation. For details, we kindly refer the reader
to the wide field of existing literature on Lagrangian perturbation theory \citep[e.g.][]{1994A&A...288..349B,1995A&A...296..575B,
1998MNRAS.299.1097S,2002PhR...367....1B}. We then summarize the
multi-grid algorithm which solves Poisson's equation numerically before we discuss its extension to the adaptive multi-grid
algorithm which provides solutions to Poisson's equation on nested grids. Finally, we examine several methods to obtain
velocity and displacement fields that have the same behaviour at large wave numbers as those obtained with the traditional
$k$-space sampling.

\subsection{Lagrangian perturbation theory}

\subsubsection{First order perturbations}
\label{sec:firstlpt}
Lagrangian perturbation theory describes the evolution of density perturbations in the rest-frame of the fluid. The 
position ${\bf x}$ of a fluid element at time $t$ with respect to its initial position ${\bf q}$, and the respective fluid
velocity, can then be written as
\begin{equation}
{\bf x}(t) = {\bf q} + {\bf L}({\bf q},t),\label{eq:LagrangeCoord}\quad
\dot{\bf x}(t) = \frac{{\rm d}}{{\rm d}t} {\bf L}({\bf q},t) 
\end{equation}
where $\bf{L}({\bf q})$ we call the ``displacement field'' which is derived using perturbation theory. 

It can be easily shown that at first order in the perturbations \citep[cf.][]{1970A&A.....5...84Z}, the displacement field ${\bf L}$
can be written as the gradient of a potential $\Phi$ which is proportional to the gravitational potential $\phi$,
\begin{equation} 
{\bf L}({\bf q}) = - \frac{2}{3 H_0^2 a^2 D_+(t)}\,\boldsymbol{\nabla}_{\bf q} \phi({\bf q},t) \equiv D_+^{-1}(t)\,\boldsymbol{\nabla}_{\bf q} \Phi({\bf q},t),
\label{eq:zeldovich}
\end{equation}
where $H_0$ is the Hubble constant, $a$ is the expansion factor at time $t$, $D_+(t)$ is the
growth factor of linear density perturbations (i.e. the time evolution of the growing mode) and $\phi$
is the gravitational potential obeying Poisson's equation
\begin{equation}
\Delta_{\bf q} \phi({\bf q},t) = \frac{3}{2}H_0^2 a^2 \delta({\bf q},t).
\end{equation}
Since the velocities are given by the gradient of a potential, velocities are irrotational,
i.e. $\boldsymbol{\nabla}\times \dot{\bf x}(t) = 0$, in this approximation.

Note that the Gaussian over-density field $\delta$ is the source field of the displacements. It is not
the density field that is measured after displacing the fluid elements, which is non-Gaussian. We
give a derivation of the latter in Section \ref{sec:ICdens}. The Gaussian field $\delta$ should 
not be used to impose an initial density field for the baryonic component.

In order to obtain the displacement vectors from the initial over-density field $\delta$, Poisson's equation
has to be solved numerically. The most common approaches use an FFT based Poisson solver
\citep[e.g.][]{2001ApJS..137....1B}, while we chose a multi-grid based Poisson solver as it can be
easily extended to an adaptive multi-grid solver which is able to handle nested adaptive grids
in a very natural way.

\subsubsection{Second order perturbations}
\label{sec:seclpt}
Several studies have shown \cite[see e.g.][]{1994ApJ...436..517M,1998MNRAS.299.1097S,2006MNRAS.373..369C,2007JCAP...12..014T} that first order Lagrangian perturbation 
theory (cf. Section \ref{sec:firstlpt}) might not be
accurate enough for current simulations as it leads to e.g. significantly underestimated higher order moments
of the density probability distribution functions at early times. In particular, \cite{1994ApJ...436..517M} have
shown that 2LPT matches the skewness of the density field for top hat collapse, while 3LPT would
in addition also match the kurtosis, and so on.
At second order, the displacement field contains not only contributions from the 
gravitational potential, but also from a second-order potential that is sourced by the off-trace
components of the deformation tensor, i.e.
\begin{equation}
{\bf L}({\bf q},t) = D_+(t)\,\boldsymbol{\nabla}_{\bf q} \Phi({\bf q},t) + D_2(t)\,\boldsymbol{\nabla}_{\bf q} \Psi({\bf q},t), 
\end{equation}
where $\Psi$ obeys the Poisson equation $\Delta_{\bf q} \Psi({\bf q},t) = \tau({\bf q},t)$, with 
\begin{equation}
\tau({\bf q},t) = -\frac{1}{2}\sum_{i,j}\left[ \left(\partial_{q_i}\partial_{q_j}\Phi\right)^2-\left(\partial_{q_i}\partial_{q_i}\Phi\right) \, \left(\partial_{q_j}\partial_{q_j}\Phi\right)\right],
\end{equation}
and $D_+(t)$ is the growth factor of linear perturbations, and $D_2(t)\simeq\frac{3}{7}D^2_+(t)$.
Adding second order perturbation theory is thus algorithmically identical to repeating the steps for
first order displacements: After computing the source-field
$\tau$ using finite differences, Poisson's equation can be solved numerically using the (adaptive) multi-grid
scheme. A similar adaptive approach to generate initial conditions at second order has been made by
\cite{2010MNRAS.403.1859J} who use a tree-PM method to evaluate the second order contribution.
Note that while our method allows for 2LPT, we only use 1LPT in this paper to aid the comparison with the typical
practice in the field.


\subsection{Multi-grid solution of Poisson's equation}
\label{sec:multigrid}
\begin{table*}
\begin{center}
\begin{tabular}{|r|c|c|}
Order $n$ & Laplacian ${\rm L}$ & Gradient ${\rm \bf G}$\\
\hline
\hline
exact: & $\partial^2_x$ & $\partial_x$ \\
$2$: & $\left[\begin{array}{ccc}1&-2&1\end{array}\right]$& $\frac{1}{2}\left[ \begin{array}{ccc}-1& 0 & 1\end{array}\right]$\\
$4$: & $\frac{1}{12}\left[\begin{array}{ccccc}-1&16&-30&16&-1\end{array}\right]$& $\frac{1}{12}\left[\begin{array}{ccccc}1&-8&0&8&-1\end{array}\right]$\\
$6$: & $\frac{1}{180}\left[\begin{array}{ccccccc}2&-27&270&-490&270&-27&2\end{array}\right]$&$\frac{1}{60}\left[\begin{array}{ccccccc}-1&9&-45&0&45&-9&1\end{array}\right]$ \\
\hline
exact: & $-k^2$ & $-i\,k$ \\
$2$: & $-2\left[-\cos(k)+1\right]$ & $-i\,\sin(k)$\\
$4$: & $-\frac{1}{6}\left[\cos(2k)-16\,\cos(k)+15\right]$ & $-\frac{i}{6}\left[-\sin(2k)+8\sin(k)\right]$\\
$6$: & $-\frac{1}{90}\left[-2\cos(3k)+27\,\cos(2k)-270\,\cos(k)+245\right]$ & $-\frac{i}{30}\left[\sin(3k)-9\sin(2k)+45\sin(k)\right]$\\
\hline
\end{tabular}
\end{center}
\caption{\label{tab:fd_approx}Finite difference stencils in one dimension for the Laplacian ${\rm L}$ and gradient operators ${\rm \bf G}$ up to 6th order (top rows) and their respective 
Fourier transforms $\widetilde{\rm L}$ and $\widetilde{\rm \bf G}$ (bottom rows).}
\end{table*}%
Both first and second order Lagrangian perturbation theory for velocity and displacement fields require the
numerical solution of Poisson's equation followed by calculating gradients. This can be achieved by use of
the multi-grid algorithm \citep[e.g.][]{RPFedorenko_1961a,ABrandt_1973a}. In order to solve Poisson's equation
\begin{equation}
\Delta \phi({\bf x}) = f({\bf x})\quad\textrm{on domain $\Omega$},
\label{eq:poisson_2}
\end{equation}
with periodic boundary conditions in our case, we cover $\Omega$ with a hierarchy of grids $M^0,M^1,\dots,M^m$ of 
respective grid spacing $h^0,h^1,\dots,h^m$, where $h^\ell/h^{\ell+1}=2$, i.e. a refinement factor of 2
between multi-grid levels (cf. also Figure \ref{fig:grid_structure}). 

Define ${\rm I}_\ell^{\ell-1}$ as the restriction and ${\rm I}_{\ell}^{\ell+1}$ as the injection operator.
We use the Full Approximation Scheme \citep[FAS --][]{ABrandt_1977b}, see also \cite{trottenberg2001}, to solve the discrete form of 
equation (\ref{eq:poisson_2}) on grid $\ell$ given by
\begin{equation}
{\rm L}^\ell u^\ell(x) = f^\ell(x)\quad\textrm{for $x\in M^\ell$},
\end{equation}
where ${\rm L}$ is a finite difference approximation to the Laplacian (as in Table \ref{tab:fd_approx}) and $u^\ell$ is an approximation to $\phi$ 
on grid $M^\ell$. The residual $r^\ell(x)$ can then be written as $r^\ell = f^\ell - {\rm L}^\ell u^\ell$. 
FAS then uses these residuals to correct the solution on level $\ell-1$, so that (note that operators do not commute)
\begin{equation}
{\rm L}^{\ell-1}u^{\ell-1} = {\rm I}_\ell^{\ell-1}r^\ell + {\rm L}^{\ell-1}{\rm I}_\ell^{\ell-1}u^\ell,
\end{equation}
is obtained. This is equivalent to solving
\begin{equation}
{\rm L}^{\ell-1}u^{\ell-1} = {\rm I}_{\ell}^{\ell-1}f^\ell + \tau_\ell^{\ell-1}
\end{equation}
on level $\ell-1$, where
\begin{equation}
\tau_\ell^{\ell-1} \equiv {\rm L}^{\ell-1}{\rm I}_\ell^{\ell-1}u^\ell-{\rm I}_\ell^{\ell-1}{\rm L}^\ell u^\ell. \label{eq:mg_source_term}
\end{equation}
The grid at level $\ell$ thus provides additional source terms $\tau$ to the equation at level $\ell-1$ in addition to
the restricted source $f^{\ell-1}\equiv{\rm I}_\ell^{\ell-1}f^\ell$ accounting for the non-commutative nature of the
operators -- eq. (\ref{eq:mg_source_term}) is just the commutator $\left[{\rm L},\,{\rm I}_\ell^{\ell-1}\right]$ of the Laplacian and the restriction operator.  

Within each level, corrections are propagated using a smoothing (or diffusion) scheme $S\left(u^\ell,\,f^\ell\right)$, 
typically a Gauss-Seidel sweep.

The complete FAS multi-grid scheme then consists of the following algorithm to solve ${\rm L}^\ell u^\ell=f^\ell$,
starting with $\ell=m$:
\begin{enumerate}[1.]
\item If $\ell=0$, set $u^0\equiv0$.
\item Apply $\nu_1$ smoothing steps, $u^\ell_i\equiv S\left(u^\ell_{i-1},\,f^\ell\right)$, $i=1\dots\nu_1$
\item Calculate the residual, $r^\ell \equiv f^\ell - {\rm L}^\ell u^\ell$
\item Restrict the residual, $r^{\ell-1} \equiv {\rm I}_\ell^{\ell-1} r^\ell$
\item Restrict the smoothed solution, $u^{\ell-1}\equiv {\rm I}_\ell^{\ell-1}u^\ell_{\nu_1}$
\item Apply the $\tau$-correction, $f^{\ell-1}\equiv r^{\ell-1}+{\rm L}^{\ell-1}u^{\ell-1}$
\item Apply FAS scheme recursively to solve ${\rm L}^{\ell -1}u^{\ell-1}=f^{\ell-1}$
\item Correct the solution $u^{\ell} \equiv u^\ell_{\nu_1}+{\rm I}^\ell_{\ell-1}\left(u^{\ell-1}-{\rm I}^{\ell-1}_\ell u^\ell_{\nu_1}\right)$
\item Apply $\nu_2$ smoothing steps, $u^\ell_i\equiv S\left(u^\ell_{i-1},\,f^\ell\right)$, $i=1\dots\nu_2$
\end{enumerate}
The first step ensures that the mean of the potential vanishes and that the algorithm converges in the case
of periodic boundary conditions. For non-periodic boundary conditions, a direct solution would need to be
computed. Throughout each cycle we enforce periodic boundary conditions, whenever $u$ is changed.
The scheme is to be repeated until the norm of the residual, computed after step 9, falls
below some desired threshold. Since we only call FAS once in step 7, our approach uses only V-cycles.

We found excellent convergence, i.e. a reduction of the residual by at least one order of magnitude
per iteration, for all finite difference approximations of the Laplacian that we tested (up to
sixth order, cf. Table \ref{tab:fd_approx}) using
the red-black Gauss-Seidel method as the smoothing operation $S\left(u^\ell,\,f^\ell\right)$ with $\nu_1=\nu_2=2$
sweeps, simple averaging
over the 8 child cells of one coarse cell as the restriction operation ${\rm I}^{\ell-1}_\ell$ and 
straight injection of the coarse cell value into the 8 child cells as the prolongation operation
${\rm I}_{\ell-1}^\ell$.

The FAS scheme has been adopted as it operates on the solution itself on each level, while
the standard multigrid operates on residuals. This allows us to incorporate conservative
boundary conditions at coarse-fine boundaries more easily (cf. Section \ref{sec:bound_interp}).


\begin{table}
\begin{center}
\begin{tabular}{|r|c|}
Order $n$ & Flux operator ${\rm F}$\\
\hline
\hline
$1$: & $\left[\begin{array}{cc}-1&1\end{array}\right]$\\
$3$: & $\frac{1}{12}\left[\begin{array}{cccc}-1&15&-15&1\end{array}\right]$\\
$5$: & $\frac{1}{180}\left[\begin{array}{cccccc}2&-25&245&-245&25&-2\end{array}\right]$\\
\hline
\end{tabular}
\end{center}
\caption{\label{tab:fd_fluxes}Finite difference flux operators for the Laplacian. Convolved with $\left[\,-1 \,\,\, 1\,\right]$, these become the respective Laplacians
of order $(n+1)$.}
\end{table}%

\subsection{Adaptive multi-grid}
\label{sec:bound_interp}
We will now describe the extension of the FAS multi-grid algorithm described above to additional nested adaptive grids, 
$M^{\prime\,m+1},\dots M^{\prime\,m+k}$,
covering non-coextensive subdomains $\Omega^{\prime\,i}$, $i=1,\dots,k$ with $\Omega^{\prime\,i+1}\subset\Omega^{\prime\,i}$.
In our case, the $M^\prime$ are simply rectangular grids. Two modifications have to be made:
First, restriction ${\rm I}^{\ell-1}_\ell$ and prolongation ${\rm I}_{\ell-1}^\ell$ only operate on overlapping regions  $\Omega^{\prime\,i}\cap\Omega^{\prime\,i+1}$
of the domains. The remainder $\Omega^{\prime\,i}\backslash\Omega^{\prime\,i+1}$
is treated as if it resided at the finest level. Second, Poisson's equation on additional sub-grids is solved with 
the coarse grid solution $u^{\ell}$ acting as a boundary condition for the finer level Poisson equation ${\rm L}^{\ell+1}u^{\ell+1}=f^{\ell+1}$.
The boundary condition is realised by adding ghost zones to the sub-grids $M^\prime$ to which boundary values are
interpolated. 

For these coarse-fine boundaries, we first use polynomial interpolation using only coarse grid information parallel to the
fine grid surface. Using these intermediate values together with values inside the fine grid, another polynomial
interpolation step is performed normal to the fine grid surface \citep[similar to][]{martin1996}. The order of the interpolating polynomial is chosen 
identical to the order of the finite difference scheme for the Laplacian. This serves as a ``guess'' for the solution on the boundary
and is thus a weak Dirichlet boundary condition. ``Weak'' as we will not actually use the boundary cell solution as
the boundary condition.

In a second step, we apply additional constraints to the interpolated ghost zone values so that the coarse flux matches the fine flux across
the coarse-fine boundary. This procedure can be easily understood by rewriting Poisson's equation  in the following way:
\begin{equation}
\Delta\phi(\mathbf{x}) = \boldsymbol{\nabla}\cdot\boldsymbol{\nabla}\,\phi(\mathbf{x})  = f(\mathbf{x})
\end{equation}
can be integrated over one grid cell (in one dimension for notational simplicity) to yield
\begin{equation}
\int_{x_i-h/2}^{x_i+h/2}\partial_x\,{F}_\phi(x')\,{\rm d}x'  =  m_i,
\end{equation}
where $m_i=\int\,f(x_i)\,{\rm d}x'$ is the mass contained in cell $i$, and ${F}_\phi\equiv\partial_x\phi$ is the potential flux.
Applying the divergence theorem, we find that
\begin{equation}
{F}_\phi(x_i+h/2)-F_\phi(x_i-h/2) = m_i,
\end{equation}
i.e. the mass in cell $i$ generates a flux through the cell surfaces given by $F_\phi$. In order for the adaptive multi-grid scheme to 
be conservative, the flux across common surfaces has to be identical (since the mass contained in a coarse cell is taken to be
identical to the mass inside its eight children). This is trivially fulfilled for inner cells but care has to be 
taken at outer coarse-fine boundaries.

The flux operators ${\rm F}$ are matched to the order of the Laplacian and given in Table \ref{tab:fd_fluxes}.
They are gradient operators of order $n-1$ on the cell boundaries. When convolved with the first order gradient, $\left[\begin{array}{cc}-1&1\end{array}\right]$
(which corresponds to a determination of the net flux), the
respective Laplacian operator of order $n$ is recovered. Using the flux operator, the 4 fluxes due to the fine 
grid can be evaluated and then subtracted from the coarse flux through the same surface element. This flux difference is subtracted 
from the ghost zone values contributing to the fine grid flux so that the sum of the fine fluxes now matches the coarse flux. A flux-correction is necessary
since otherwise the sub-grid induces artificial source terms through the ghost zone interpolation and the multi-grid algorithm 
does not converge (in the sense that the residual does not vanish on all levels). Note that this flux matching can be thought of as constraints on the interpolation scheme. We thus obtain mixed boundary conditions:  a von~Neumann boundary condition 
due to the flux matching, which constrains only one degree of freedom for the boundary cells at a single coarse-fine boundary, 
and a subordinate Dirichlet
boundary condition, which constrains all degrees of freedom, but is only used to evaluate the fluxes on the fine grid 
\citep[cf. also][]{Miniati:2007:BSA:1297418.1297547}. This results
in a two-way interface between coarse and fine cells. These boundary conditions ensure that gradients
across coarse-fine boundaries are smooth. Note that this is not the case for the one-way interface multigrid solvers currently
employed in most cosmological simulation codes due to the need for adaptive time-stepping.
Using this conservative procedure we maintain multi-grid convergence 
(i.e. the residual reduces by at least an order of magnitude per iteration) also with an arbitrarily deep hierarchy of
adaptive sub-grids.

Note that our approach to use an adaptive Poisson solver is in similar spirit as \cite{1996ApJ...460...59S} who used a 
tree to compute displacements and velocities, or \cite{2010MNRAS.403.1859J} who also used a tree to compute the
second order term for 2LPT. The advantage of the multi-grid method is however that it has
a well controlled residual to the equation to be solved so that errors are easily controllable by setting the
convergence criterium in terms of the residual norm rather than by tuning opening angles for the tree.
The performance of a tree-based Poisson solver is particularly problematic for density fields with low constrast.
Using a tree has the further disadvantage that periodic boundary conditions have to be incorporated in
a hybrid way (e.g. by using FFT for the top grid). This leads to an additional source of errors arising from the
long-range/short-range split. 

\subsection{Fourier analysis of the finite difference operators}
Operating in real space requires the use of a finite difference approximation to the Laplacian ``${\rm L}$'' and gradient operators ``${\rm\bf G}$''. In
Table \ref{tab:fd_approx}, we give the standard stencil representations for the one dimensional versions of these operators. The 
three-dimensional versions can be obtained by subsequent convolution of the one-dimensional operators along all
three Cartesian coordinate axes. In the bottom half, the Fourier transforms of the operators are given together with the 
exact Fourier transform of the continuous operators. Since a regularly spaced mesh is a Dirac comb, due to symmetry reasons, 
the Fourier transform of these operators takes the form of a cosine series, in the case of the Laplacian, and of a sine series, in the case 
of the Gradient. These series are approximations to the respective continuous and non-periodic transforms of the operators. 
The sine series representing the finite difference gradients vanish at the Nyquist wave number leading to a suppression of
small scale modes. Furthermore at low order, the finite difference approximation leads to an attenuation also at slightly large scales. 
We will investigate the influence of this attenuation on cosmological initial conditions in what follows. We analyze non-adaptive 
unigrid initial conditions in this section, in order to differentiate these effects from those due to adaptive initial conditions, which 
will be addressed in Section \ref{sec:nested_errors}.

\subsubsection{Damping of small-scale perturbations}
\label{sec:smalldamp}
In the Zel'dovich approximation, the displacement and velocity fields are proportional to the gradient of the potential (cf. eq. \ref{eq:zeldovich}). 
In this section, we investigate how the order of the finite difference approximation for the Laplacian and the gradient affects perturbations 
close to the Nyquist wave number $k_{\rm Ny}=\pi / h$, where $h$ is the grid spacing which we set equal to unity in this section
for convenience.

\begin{figure}
\begin{center}
\includegraphics[width=0.4\textwidth]{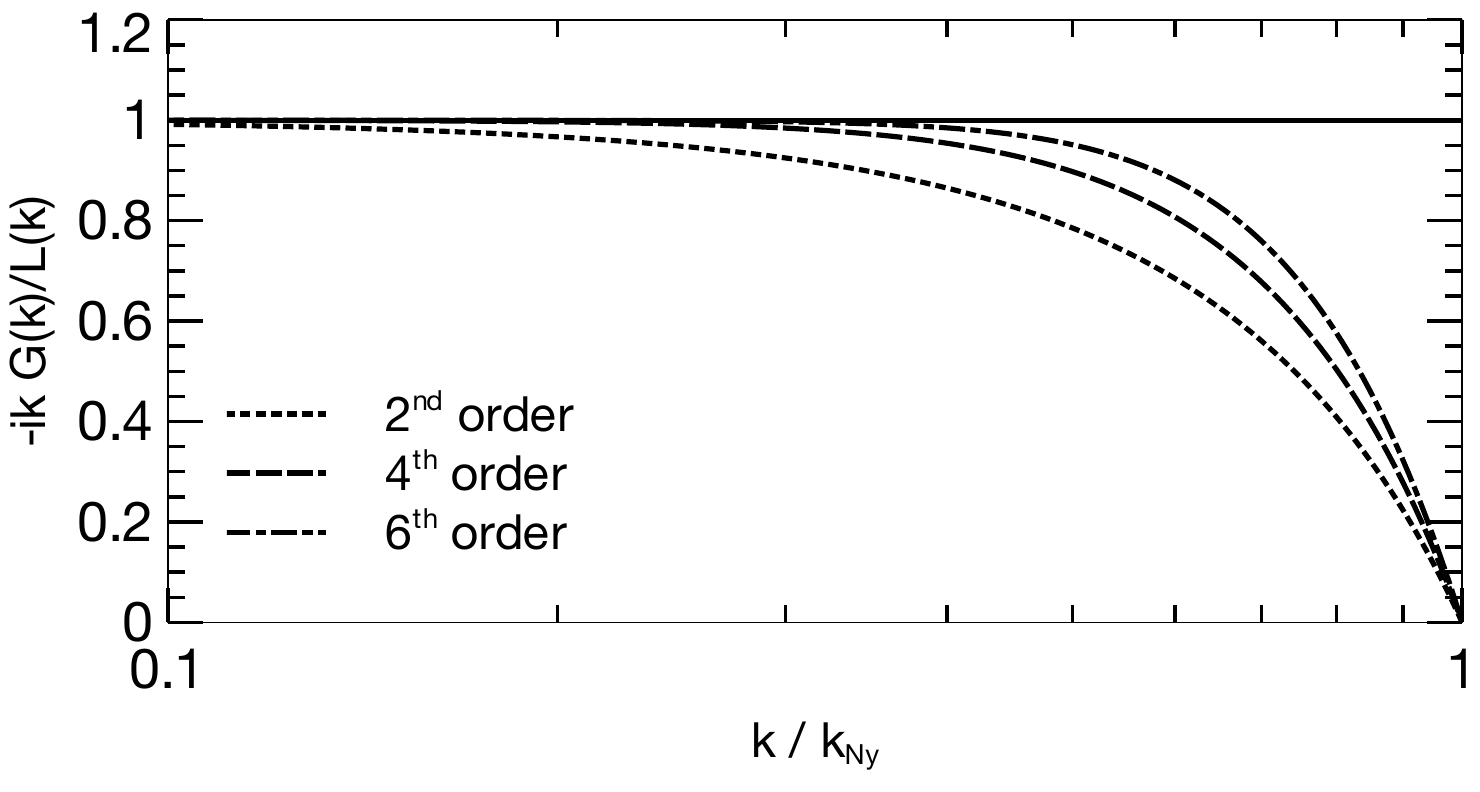}
\end{center}
\caption{\label{fig:fd_effects}Wavenumber dependent attenuation of perturbation amplitudes due to the finite difference approximations of the Laplacian
and gradient operators. Shown is the relative attenuation with respect to the exact solution for 2nd (dotted), 4th (dashed) and 6th 
order (dash-dotted) approximations in one dimension. The wavenumber is in units of the Nyquist wave number $k_{\rm Ny}$. }
\end{figure}

We define $\mathbf{v}$ as the gradient of the potential arising from a source field $f$. The potential is solved using the multi-grid scheme outlined above, and the gradient
operator is applied subsequently. The exact solution $\mathbf{v}$ in k-space is given by $\widetilde{\mathbf{v}}$, where the tilde represents the Fourier transform.
Expressed by the Fourier transforms of the operators, this is simply:
\begin{equation}
\widetilde{\mathbf{v}} = \frac{\widetilde{\rm \bf G}}{\widetilde{\rm L}} \widetilde{f} = \frac{i\mathbf{k}}{k^2}\widetilde{f},
\label{eq:vel_exact}
\end{equation}
where the last equality holds for the exact solution. We define the attenuation as the ratio between the one-dimensional finite difference solution -- using
approximations of a given order for ${\rm L}$ and ${\rm \bf G}$ and taking their Fourier transform as in Table \ref{tab:fd_approx} --  and the exact 
solution $\widetilde{\mathbf{v}}$ in k-space.

In Figure \ref{fig:fd_effects}, we show this attenuation as a function of wavenumber $k$ that is to be expected from the finite difference approximations
to the operators up to sixth order. All finite difference gradients have zero amplitude at $k_{\rm Ny}$ so that fluctuations at this scale can not
be represented in principle. A 2nd order approximation leads to significant attenuation of
$\sim78$ per cent at $k_{\rm Ny}/2$. However, for 6th order, attenuation is at the level of a few per cent at $k_{\rm Ny}/2$.
We thus expect a suppression of the power spectrum at large $k$ depending on the order of the finite difference operators employed. 
Note that $k_{\rm Ny}/2$ corresponds to scales of two grid cells and that the attenuation enters squared into the power spectrum.

\subsection{Recovering small-scale power}
Both the use of a transfer function evaluated in real space and finite difference methods lead to a loss of power at scales
close to the Nyquist wavenumber. We outline methods to solve this problem in what follows. Most of these methods
will lead to spectral leakage and the associated spurious oscillations (cf. \ref{sec:aliasing}) which poses no problem 
in the case of dark matter particle initial conditions but should be avoided for baryons. 

\subsubsection{Finite Volume Correction}
\label{sec:finite_vol_deconv}
As demonstrated in Section \ref{sec:bound_interp}, the multi-grid method is a finite volume approach. The solution is 
determined by computing flux balances across grid faces. This implies that the source field $f$ is simply a cell average.
Hence, implicitly, every discrete value $T_{\rm D}({\bf x})$ is a piecewise
constant approximation to $T_{\mathcal{R}}({\bf x})$ which is incorrect in the particle case. 
The piecewise constant averaging is given in terms of a kernel convolution
\begin{equation}
\delta_\textrm{grid}({\bf x}_i) = W_s({\bf x}) \star \delta_\textrm{cont.}({\bf x}).
\end{equation}

In order to restore the 
small-scale fluctuation amplitudes, it however suffices to perform a deconvolution with the cell assignment function
-- equivalent to nearest grid point (NGP) assignment, see e.g. \cite{1981csup.book.....H},
\begin{equation}
W_s({\bf x}) = \prod_{i=1\dots3} H\left(x_i+\frac{h}{2}\right)\,\left[1-H\left(x_i-\frac{h}{2}\right)\right],
\end{equation}
where $H$ is the Heaviside step-function and $h$ is the grid spacing. Since $H$ has an algebraic form for its Fourier transform
\begin{equation}
\widetilde{W}_s({\bf k}) = \prod_{i=1\dots3} \frac{2}{k_i}\sin\left(\frac{h}{2}k_i\right),
\end{equation}
we can perform this deconvolution in the Fourier domain and thus recover some of the sub-grid power -- see also
\cite{2005ApJ...620..559J} who use a similar procedure for power spectrum estimation. Note that -- by virtue of
restoring sub-grid power -- the deconvolution leads to spectral leakage. The associated ringing will however be completely 
filtered out by subsequent finite difference operations.

\subsubsection{A hybrid Poisson solver}
The attenuation of subgrid power due to the finite difference operators themselves (which however comes at the benefit of a non-oscillatory 
velocity field) may be considered undesirable as part of the velocity information is effectively destroyed by the finite difference approximation. Since the lack of small scale power is
only relevant on the finest grid, a simple solution to circumvent this problem can be devised. Since the Fourier transforms
of the finite difference operators are known, the fine grid solution can be simply replaced by the $k$-space exact solution. Setting the right-hand side 
of Poisson's equation $f(\bf{x})\equiv0$
on the boundary, we can recompute the self-gravity due to the fine grid and replace it with the one obtained with the exact $k$-space
method, i.e. using a grid zero-padded to twice its size, the correction is
\begin{equation}
\widetilde{v}_j^{\prime}(\mathbf{k}) = \left[i\frac{k_j}{k^2}-\frac{\widetilde{{\rm G}}^{(n)}_j}{\widetilde{{\rm L}}^{(n)}}\right]\,\widetilde{f}(\mathbf{k}),
\end{equation}
where $n$ is the order of the finite difference approximation employed and $j$ is the direction along which the gradient is taken.
The result is inverse transformed and added to the solution obtained with the finite difference method. The long-range part is
still provided by the adaptive multi-grid solution, which is correct on scales larger than two grid cells. Thus, by definition, small scale
power is recovered, while the long-range part remains unaffected. Note that for the hybrid solver, the finite volume correction
from Section \ref{sec:finite_vol_deconv} should not be applied since the $k$-space Poisson solver is not a finite volume method. 
Instead, the method outlined in the next Section should be employed.

\subsubsection{Averaging correction}
\label{sec:avg_deconv}
Even with the hybrid approach, a simple grid assignment of the real-space transfer function, determined as in Section \ref{sec:tf_real}, will lead
to damping at small scales since, unlike in the $k$-space transfer function case, no sub-grid information is present. We can however restore
this sub-grid power also for the hybrid case which itself corrects only the attenuation due to the finite volume approximation. 

We can compute each value $T_{\mathcal{R}}({\bf x})$ from an average over sub-grid scales of the highest level $\ell$ (imagined on the next higher refinement
level $\ell+1$ of the mesh). The value on level $\ell$ is then an average over 8 cells at twice the resolution, equivalent to a convolution with the sub-grid kernel
\begin{equation}
K_{\rm sg}({\bf x}) = \prod_{i=1\dots3} \frac{1}{2}\left[\delta_{\rm D}\left(x_i+\frac{h}{4}\right)+\delta_{\rm D}\left(x_i-\frac{h}{4}\right)\right],
\end{equation}
where $\delta_{\rm D}$ is the Dirac $\delta$-function and $h$ is the grid spacing. The discretized transfer function
$T_D(r_{ijk})$ is thus given by averaging over the true transfer function $T_{\mathcal{R}}$ at sub-grid scales
$T_D(r_{ijk}) = T_{\mathcal R}\star K_{\rm sg}$. The kernel $K_{\rm sg}$ can be explicitly
calculated in Fourier space to be
\begin{equation}
\widetilde{K}_{\rm sg}({\bf k}) = \prod_i\cos\left(\frac{h\,k_i}{4}\right).
\end{equation}
Power is thus reduced at large $k$ compared to higher resolution. 
Deconvolving with $\widetilde{K}_{\rm sg}$ will restore this power. The average over sub-grid cells is taken while computing the
real space transfer function kernel on the three dimensional grid.

\subsubsection{Effect of the corrections on the initial power spectrum}
\label{sec:corr_effects}
\begin{figure}
\begin{center}
\includegraphics[width=0.4\textwidth]{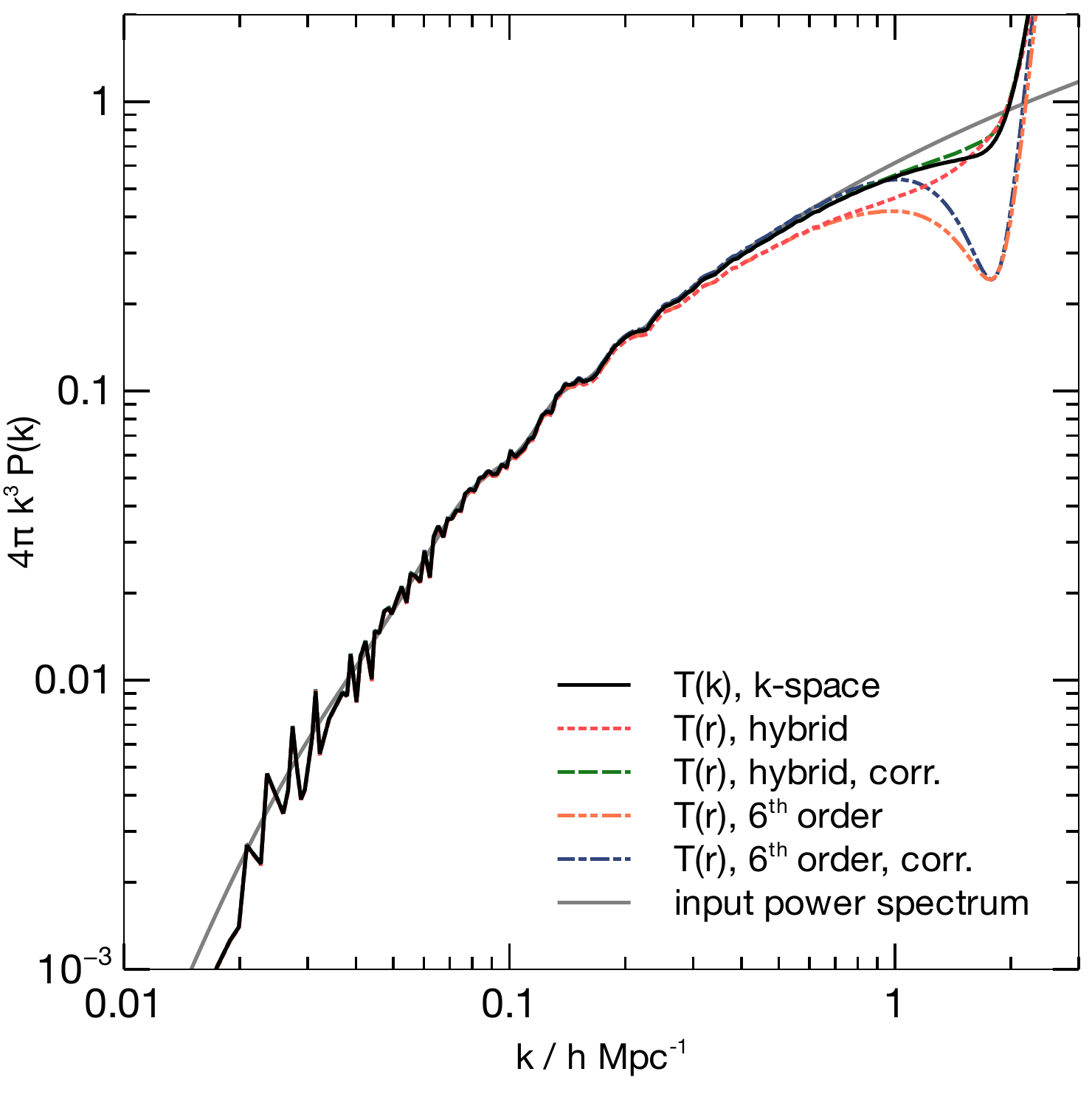}
\end{center}
\caption{\label{fig:power_ic}Influence of the real space transfer function, the finite difference approximations of the Laplacian
and gradient operators and the sub-grid corrections on the initial power spectrum. Shown is the FFT estimated power spectrum (using CIC particle interpolation
on a $1024^3$ grid for the $512^3$ particles in a $100\,h^{-1}{\rm Mpc}$ box)
for the classical exact k-space initial conditions (solid black), uncorrected real-space, 6th order (dash-dot-dot) and the corrected version thereof (dash-dot),
as well as the uncorrected hybrid (dotted) and the corrected hybrid (dashed). The solid gray line indicates the theoretical input spectrum. }
\end{figure}

In Figure \ref{fig:power_ic}, we show the influence of the finite difference order on the initial power spectrum compared to traditional initial conditions
using the k-space transfer function and a Fourier based Poisson solver. All initial conditions are for a $512^3$ particle discretisation of a $100\,h^{-1}{\rm Mpc}$
box. The power spectrum was computed using the FFT on a $1024^3$ cell density grid obtained via cloud-in-cell (CIC) interpolation from the particle positions. No deconvolution
with the grid assignment operator
was performed so that the CIC assignment leads to attenuation close to the Nyquist frequency of the mesh (outside of the plot) used to compute power spectrum 
in all cases \citep{2005ApJ...620..559J}. 
The uncorrected spectra using the real-space transfer function kernel show lack of power for wave numbers above $\sim\,0.1k_{\rm Ny}$ even when
using the hybrid Poisson solver which is equivalent to a pure FFT Poisson solver in the case of a periodic grid. This lack of power on such large scales
is undesirable as it will suppress the growth of well resolved haloes. 

Performing the finite volume correction in the case of the 6th order finite difference Poisson solver recovers all power below $\sim0.5\,k_{\rm Ny}$ 
but (naturally) retains the lack of power on even smaller scales.
In the case of the hybrid Poisson solver, the power is recovered on all scales down to $k_{\rm Ny}$. We observe a slight excess of power on the smallest scales.

Note that the motivation to apply these corrections is to match the properties of $k$-space sampled initial conditions as closely as possible. 
However, they should not be understood as mandatory and are treated as options in our code.
Applying the corrections for the respective Poisson solvers, the initial power spectra can be made to agree with the traditional exact $k$-space
sampled initial conditions with differences only on the smallest scales. We investigate the influence of these differences on $N$-body simulations
in Section \ref{sec:error_analysis}. For the remainder of the paper, the respective corrections are applied for each method without explicit reference.


\subsubsection{Spectral leakage and aliasing in FFT-Poisson solvers and Fourier space transfer functions}
\label{sec:aliasing}
\begin{figure}
\begin{center}
\includegraphics[width=0.47\textwidth]{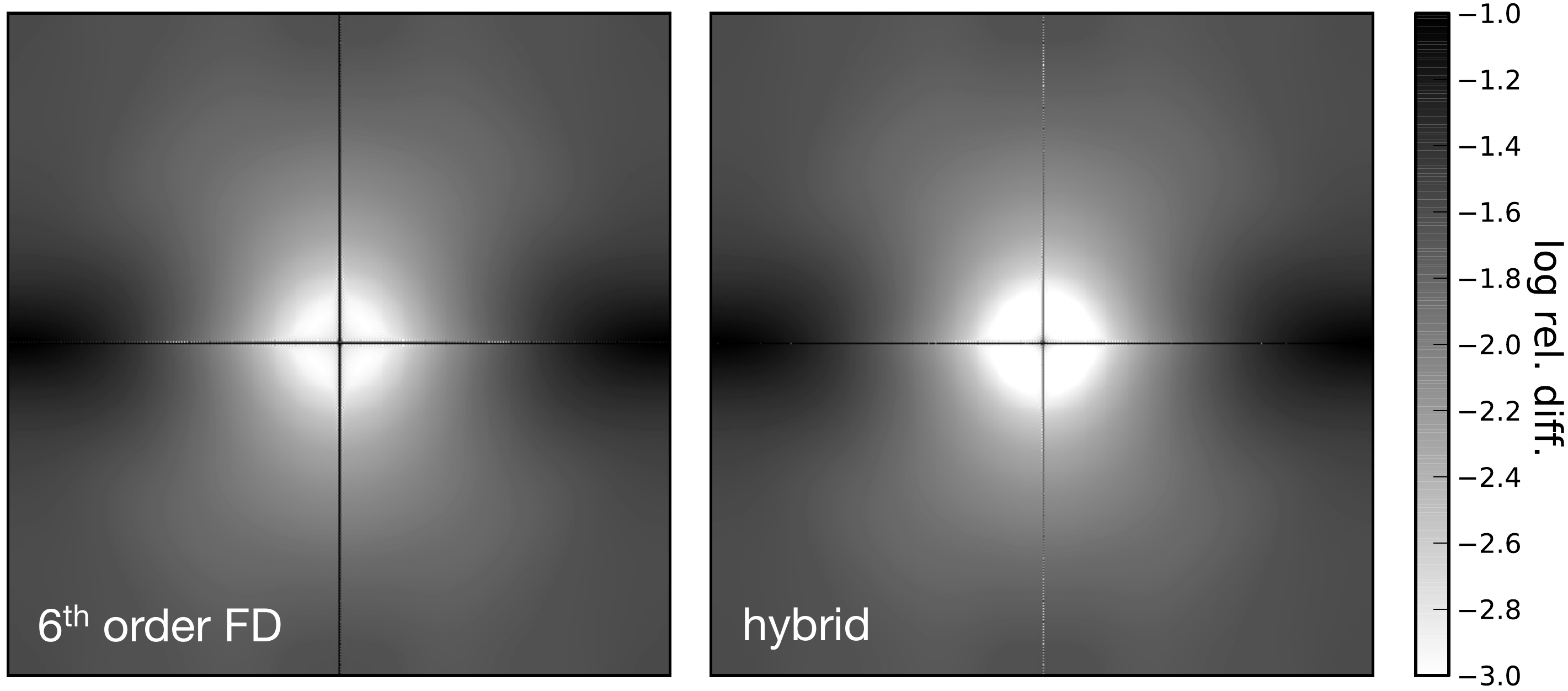}
\end{center}
\caption{\label{fig:error_vx_maps}Spatial distribution of the relative difference between the real space computed y-velocity field due to a single peak
and the velocity field computed using $k$-space sampling. The left panel shows the difference for the 6th order finite difference field and the right panel
for velocities determined with the hybrid solver. Use of the FFT poisson solver (in both the $k$-space sampled and the hybrid case) leads to pronounced 
oscillations along the y-axis through the peak. The use of the $k$-space transfer function leads to a loss of isotropy.
Figure \ref{fig:error_vx} shows the error along this axis in more detail.}
\end{figure}

\begin{figure}
\begin{center}
\includegraphics[width=0.4\textwidth]{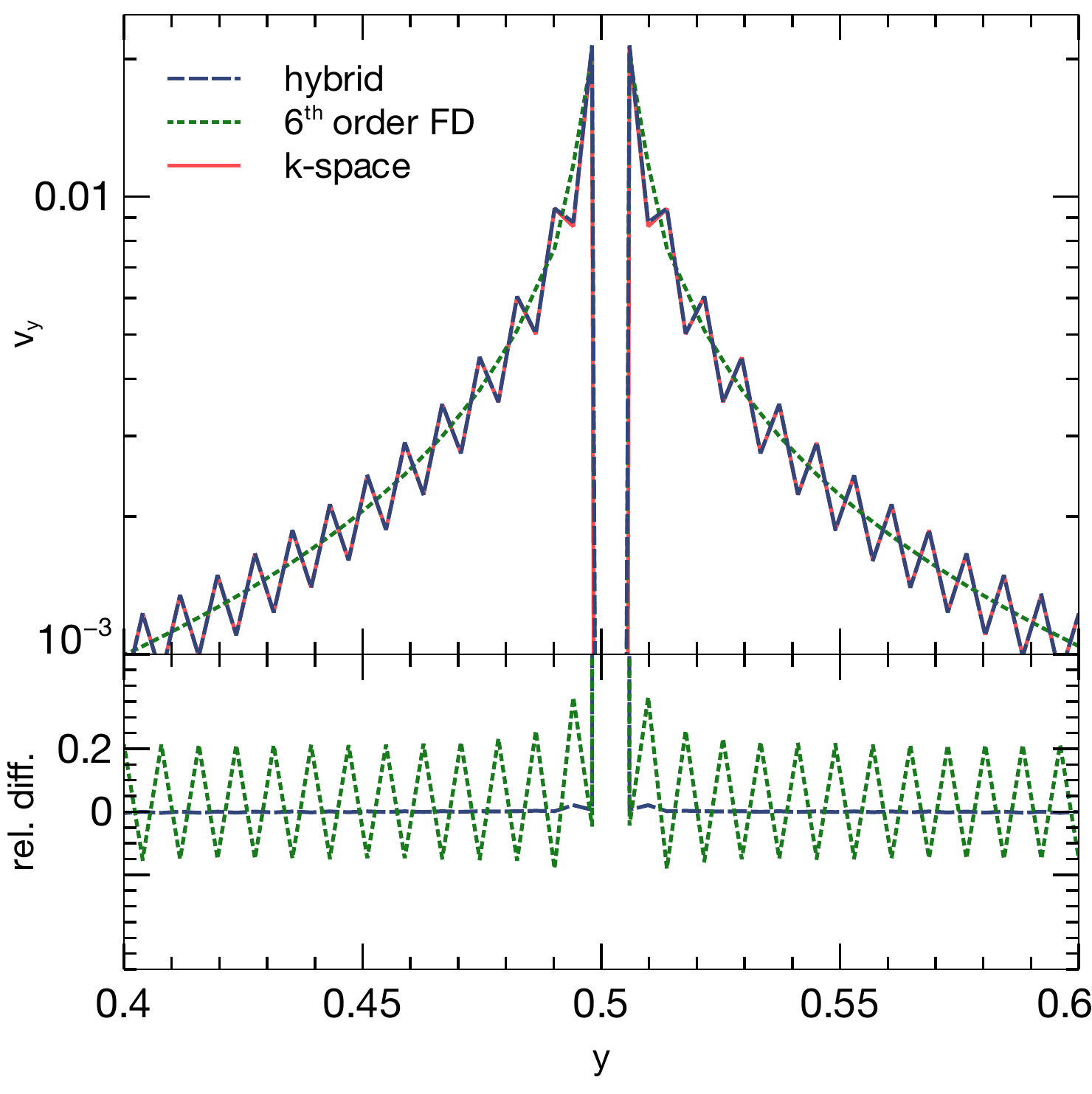}
\end{center}
\caption{\label{fig:error_vx}Slice through the central part of the velocity field induced by a single peak (cf. Figure \ref{fig:error_vx_maps}). The top panel
shows the y-velocity component using the three methods, the top panel shows the relative difference of the real-space methods with respect to the
$k$-space sampling. The left half of the top panel has positive sign, the right half is negative. }
\end{figure}

We investigate now the influence of using a real-space transfer function as well as a finite difference or hybrid Poisson solver compared to 
the standard method of using the $k$-space transfer function and computing velocities. 
Velocities computed with equation (\ref{eq:vel_exact}) are not periodic in $k$-space. Furthermore, the real space transform will be
periodic and suffer from aliasing due to its Fourier space discretisation.

In order to assess the associated effects, we compute the velocity perturbation associated with a single peak at the center of a $100\,h^{-1}{\rm Mpc}$ box
at $512^3$ resolution. The peak represents a Dirac $\delta$-function convolved with the transfer function. 
In Figure \ref{fig:error_vx_maps}, we show the difference between the velocity field obtained with the real space transfer function
using both 6th order multigrid and the hybrid Poisson solver, with respect to the velocity field obtained in the traditional way. We show 
two-dimensional slices in $(x,y)$ of the relative difference through the peak for the y-component of the velocity. Note that the velocity
field is zero on the horizontal axis through the centre.

We first note that the relative difference falls below $10^{-3}$ only in the central region. The 6th order multi-grid result has a slightly
smaller central region with such low difference to the $k$-space sampled results. Outside the central region, the relative difference rises to a maximum of $\sim2$
per cent (outside the plane where the velocity vanishes), which is independent of the Poisson solver employed and thus a feature of the real-space transfer function. Indeed, it is 
the
manifestation of the lack of spherical symmetry and aliasing (due to the finite number of samples) of the $k$-space transfer function when transformed back to real space on a finite mesh.

We also observe spectral leakage showing up as a vertical feature passing through the centre. Spectral leakage is due to the non-periodicity in Fourier-space and thus the truncation of a non-periodic function at a finite wavenumber. The truncation leads to an oscillation at the Nyquist wavenumber
in real space. In Figure \ref{fig:error_vx}, we show the central part
of the velocity fields along the vertical axis of Figure \ref{fig:error_vx_maps} revealing the oscillatory nature. Spectral leakage is present whenever a $k$-space exact Poisson 
solver is employed. Hence, the feature is not present in the multi-grid solution, which is periodic in Fourier space. 
Naturally, a similar feature is also present in the $k$-space transfer function itself, 
when transformed to real space on a three dimensional mesh \citep[see also][]{2001ApJS..137....1B}. 

Note that leakage in the velocity perturbations associated with a peak is likely to cause secondary correlations in the velocity field around rare high peaks
but is unavoidable if no window function is used with the FFT Poisson solvers to truncate the spectrum smoothly at the Nyquist frequency. The use of such a window function (as 
e.g.
the Hanning filter employed in {\sc Grafic-2}, \citealt{2001ApJS..137....1B}) however reduces power at high frequencies which might be undesirable
since it suppresses the small wavelength perturbations and thus the growth of haloes associated with them. We note that the finite difference approach
has similar filtering properties with a relatively sharp cut-off in $k$-space. Furthermore the finite difference operators are periodic in Fourier space.
We do not consider the use of other window functions but perform
our analysis of initial conditions generated with finite difference multi-grid and the hybrid Poisson solver in parallel throughout the remainder of this
paper. 


\begin{figure}
\begin{center}
\includegraphics[width=0.4\textwidth]{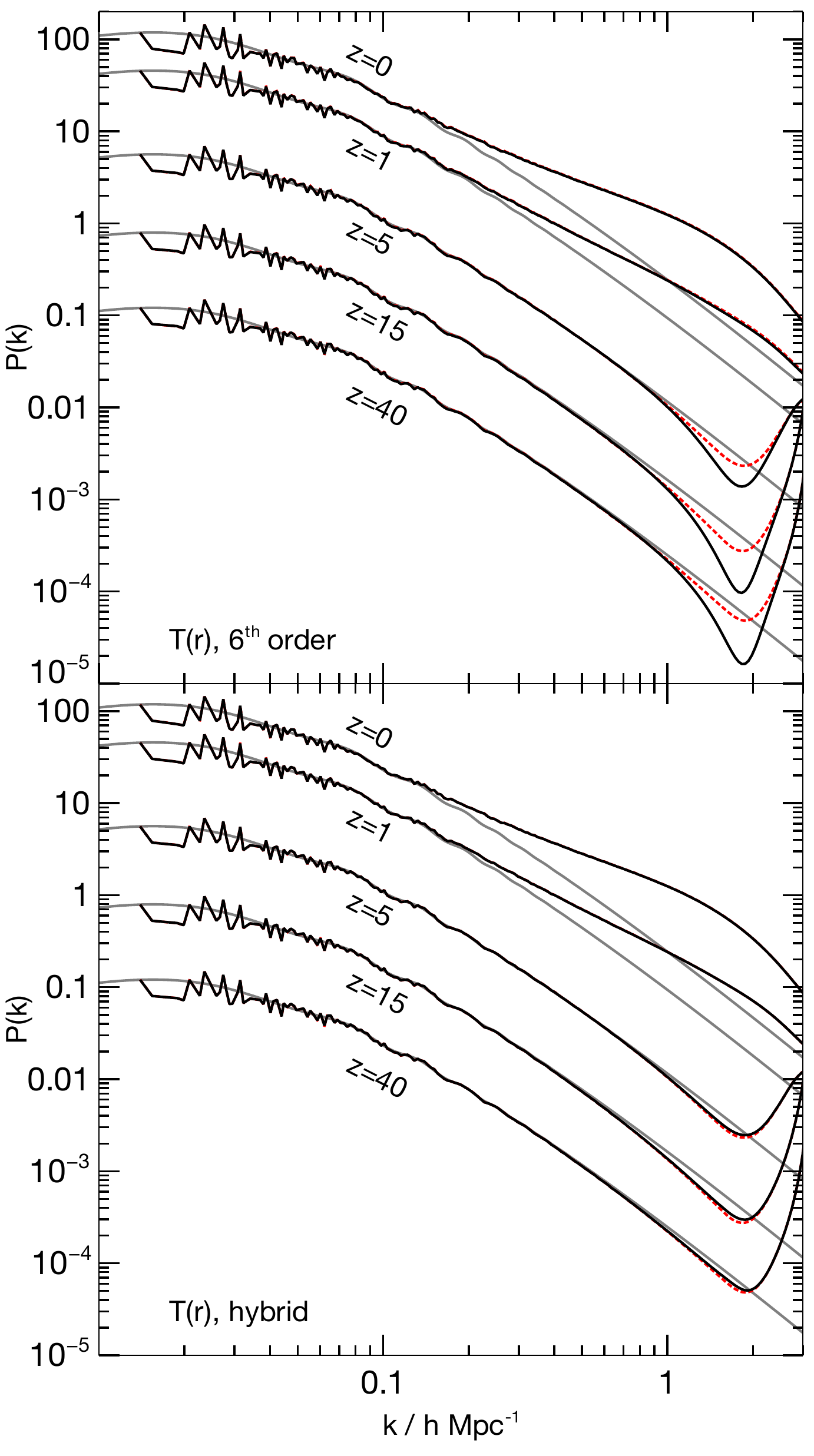}
\end{center}
\caption{\label{fig:power_evol}Redshift evolution of the power spectrum for initial conditions computed using a 6th order multi-grid Poisson solver (solid black line in top panel)
and using a hybrid Poisson solver (solid black line in bottom panel) compared to the traditional $k$-space sampling approach (dashed red lines) and the power spectrum
evolution predicted by linear perturbation theory for the input power spectrum (gray lines). Results are shown for a simulation of a $1\,h^{-1}{\rm Gpc}$ box with
$512^3$ particles. Power spectra have been computed after CIC mass assignment on a $1024^3$ grid.}
\end{figure}

\section{Error analysis}
\label{sec:error_analysis}
In this section, we first investigate the differences between the new methods described in Section \ref{sec:densities}
and \ref{sec:dispvel} and the traditional approach of operating exclusively in Fourier space on various observables
of cosmological $N$-body simulations evolved over cosmic time as well as of a single galaxy cluster in order to ensure
their proper performance in unigrid simulations. We then investigate the errors arising in nested ``zoom-in'' initial conditions 
generated with our method in both the initial conditions and in a re-simulation of a galaxy cluster.

\subsection{Statistical properties of the cosmological simulations}

In order to assess differences between initial conditions generated with the various methods, we ran {\sc Gadget-2} N-body simulation of a 
$1\,h^{-1}{\rm Gpc}$ and a $100\,h^{-1}{\rm Mpc}$ box with $512^3$ particles from redshift $z=45$ to $z=0$. First order initial conditions (1LPT) were generated using
the same random seeds. The power spectrum from which the initial conditions were generated in all cases has density parameters $\Omega_m=0.276$ and 
$\Omega_\Lambda=0.724$, a Hubble constant $H_0=70.3\,{\rm km}\,{\rm s}^{-1}\,{\rm Mpc}^{-1}$, a normalization
$\sigma_8=0.811$ as well as a spectral index of $n_s=0.961$ and was computed using the fitting formula of \cite{1998ApJ...496..605E}.
We use Plummer equivalent softening lengths $\epsilon$ of $0.09\,h^{-1}{\rm Mpc}$
for the $1\,h^{-1}{\rm Gpc}$ runs and $0.009\,h^{-1}{\rm Mpc}$ for the $100\,h^{-1}{\rm Mpc}$ runs, respectively. These box sizes were chosen
in order to probe both the highly non-linear regime, with the $100\,h^{-1}{\rm Mpc}$ box, and the mildly non-linear regime with the $1\,h^{-1}{\rm Gpc}$ box.

\subsubsection{Evolution of the power spectrum}

In Figure \ref{fig:power_evol}, we show the redshift evolution of the density power spectra for the $1\,h^{-1}{\rm Gpc}$ box. We compare the linear theory evolution 
with the evolving power spectrum from initial conditions generated with the traditional $k$-space sampling method, as well as generated with a real space 
transfer function convolution and both a 6th order finite difference and the hybrid Poisson solver. 

As previously discussed in Section \ref{sec:corr_effects}, the finite difference initial conditions show a significant suppression of power on the smallest scales.
We see however that the power is immediately restored once the relevant scales enter non-linear growth and power is transferred from larger
to smaller scales. The difference with respect to the $k$-space initial conditions
decreases over time and is no longer visible at $z\sim1$. No wavenumbers below $\sim0.5\,k_{\rm Ny}$ of the $512^3$ initial particle grid are affected at any time. 
Initial conditions generated with the hybrid scheme agree (almost) perfectly with the $k$-space version.
Again, the small initial difference also disappears completely once the relevant scales become non-linear.

Regarding large scales, we see no difference between the three methods, so that we focus our attention on a closer inspection of small scale behaviour in what follows.

\begin{figure}
\begin{center}
\includegraphics[width=0.47\textwidth]{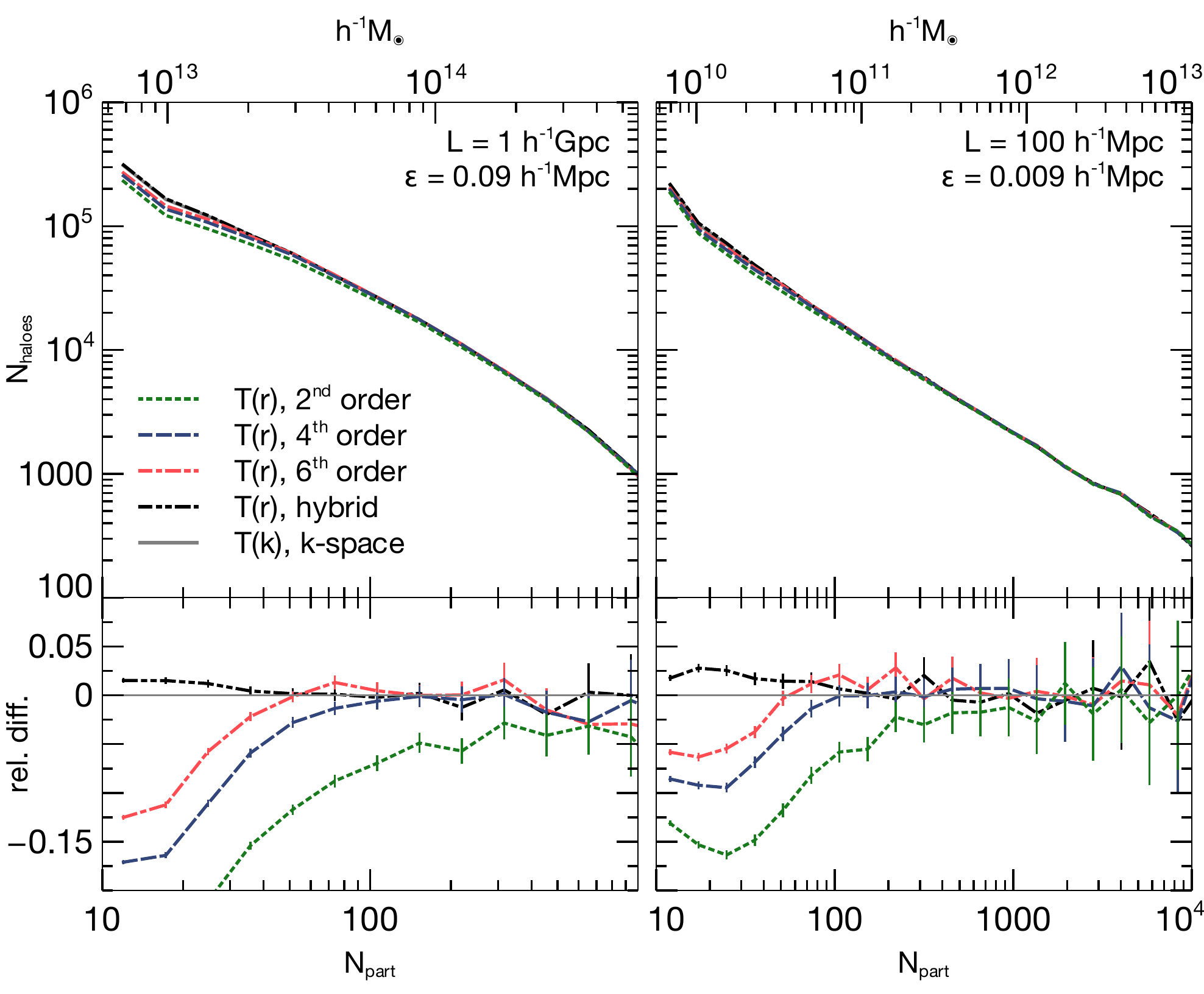}
\end{center}
\caption{\label{fig:fd_fofmf}The effect of the Poisson solver used to compute the initial conditions on the friends-of-friends halo abundance
 in N-body simulations at redshift $z=0$. Shown is the number of haloes as a function of their mass and the number of particles they consist of in the top panels.
 The bottom panels show the relative difference with respect to the ``exact k-space'' method. The left panels are for a $1\,h^{-1}{\rm Gpc}$ simulation box,
 the right panels for a $100\,h^{-1}{\rm Mpc}$ box. We show the results for a multi-grid Poisson solver using 2nd (dotted green), 4th (dashed blue), and
 6th (dash-dotted red) order finite difference approximations, as well as for the hybrid Poisson solver (dash-dot-dot black lines) and the $k$-space
 sampled initial conditions (solid gray). Errorbars in the lower panel indicate the Poisson errors on the halo counts.}
\end{figure}

\begin{figure}
\begin{center}
\includegraphics[width=0.47\textwidth]{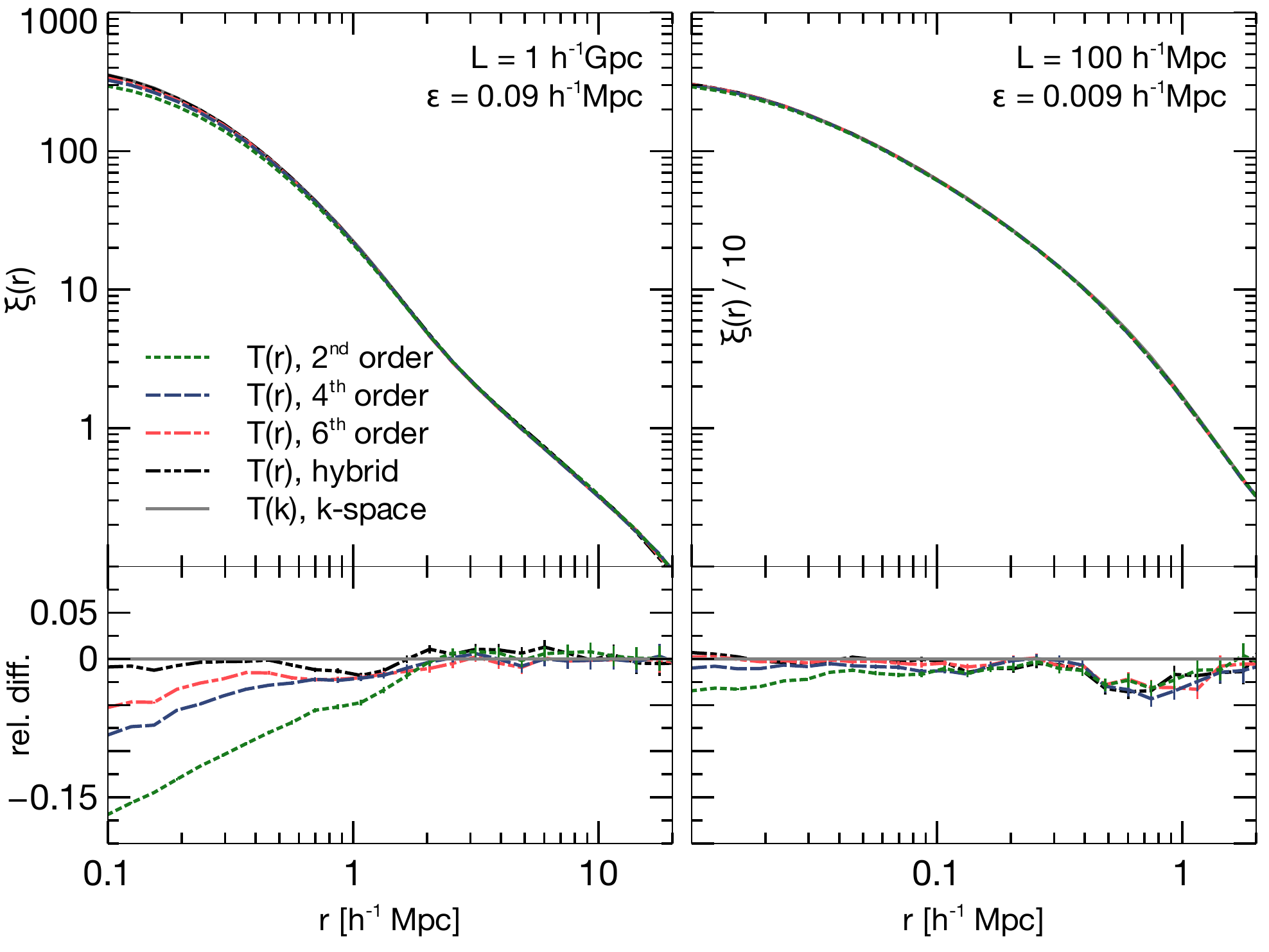}
\end{center}
\caption{\label{fig:fd_corrfun}Same as Figure \ref{fig:fd_fofmf} but showing the effect of the Poisson solvers used to compute 
the initial conditions on the small-scale two-point correlation function at redshift $z=0$. Errorbars in the lower panel
indicate the Poisson errors.}
\end{figure}

\subsubsection{The $z=0$ halo abundance}
\label{sec:test_halo_abundance}

We compare halo abundance at a given mass between the various initial conditions.
Haloes are identified using the Friends-of-Friends (FoF) algorithm \citep{1982ApJ...259..449P}, grouping particles which are separated by less than $0.2$ times the 
mean inter-particle distance.
Figure \ref{fig:fd_fofmf} shows the number of haloes $N_{\rm haloes}$ in the simulations as a function of both halo particle number $N_{\rm part}$ 
and halo mass. For both boxes and as expected, we see the strongest difference between the $k$-space sampled initial conditions and those with 2nd order accuracy of 
the finite difference operators. In fact, halo abundance is lower up to 1000 halo particles. The histograms agree above 
$\sim 100$ and above $\sim 40$ halo particles for 4th and 6th order, respectively, with that of the $k$-space method. The hybrid initial
conditions agree over all mass ranges within $\sim 2$ per cent. Since halo abundance deviates from
the Schechter function form also for the k-space initial conditions for haloes below 100 particles, it remains a question of future investigation whether
the overabundance of the smallest mass haloes is a spurious result of the $k$-space initial conditions.

\subsubsection{The $z=0$ small-scale correlation function}
\label{sec:test_corrfun}
\label{sec:err_xi}

In Figure \ref{fig:fd_corrfun}, we show the two-point correlation functions at $z=0$. 
The two-point correlation function $\xi(r)$ is defined as the 
excess probability (with respect to random) to find two particles in volume elements $dV_1$ and $dV_2$ separated by a co-moving distance $r$,
\begin{equation}
{\rm d}P_{12}(r) = \bar{n}^2\left[1 + \xi(r)\right] {\rm d}V_1\,{\rm d}V_2,
\end{equation}
where $\bar{n}$ indicates the mean particle number density. We computed the two-point correlation function using tree based sparse sampling
thus computing more close-pairs than distant pairs for a subset of 1 per cent of the particles. Thus, statistical errors increase on larger scales.

For the $1\,h^{-1}{\rm Gpc}$ box, our results indicate that $\xi$ is reduced by $\sim 17$ per cent at $r=\epsilon$ when using a 2nd order finite difference scheme,
while for 6th order the reduction is only around $\sim 5$ per cent. Both 4th and 6th order results deviate by less than $\sim 2$ per cent
at scales $r\gtrsim 6\epsilon$. For the highly non-linear regime probed by the $100\,h^{-1}{\rm Mpc}$ runs, we do not find any significant
differences. Gravitational interaction has wiped out any initial difference.
We conclude that differences are negligible in the highly non-linear regime and as soon as mildly non-linear scales are well resolved.

\subsection{Analysis of a $z=0$ galaxy cluster halo}
\label{sec:cluster_analysis}
In order to assess the influence of the real-space approach on the radial density profile of haloes, we investigate in more detail the properties of one cluster
halo forming from initial conditions generated in the various ways in the $100\,h^{-1}{\rm Mpc}$ box with $512^3$ particles. 

\begin{figure}
\begin{center}
\includegraphics[width=0.45\textwidth]{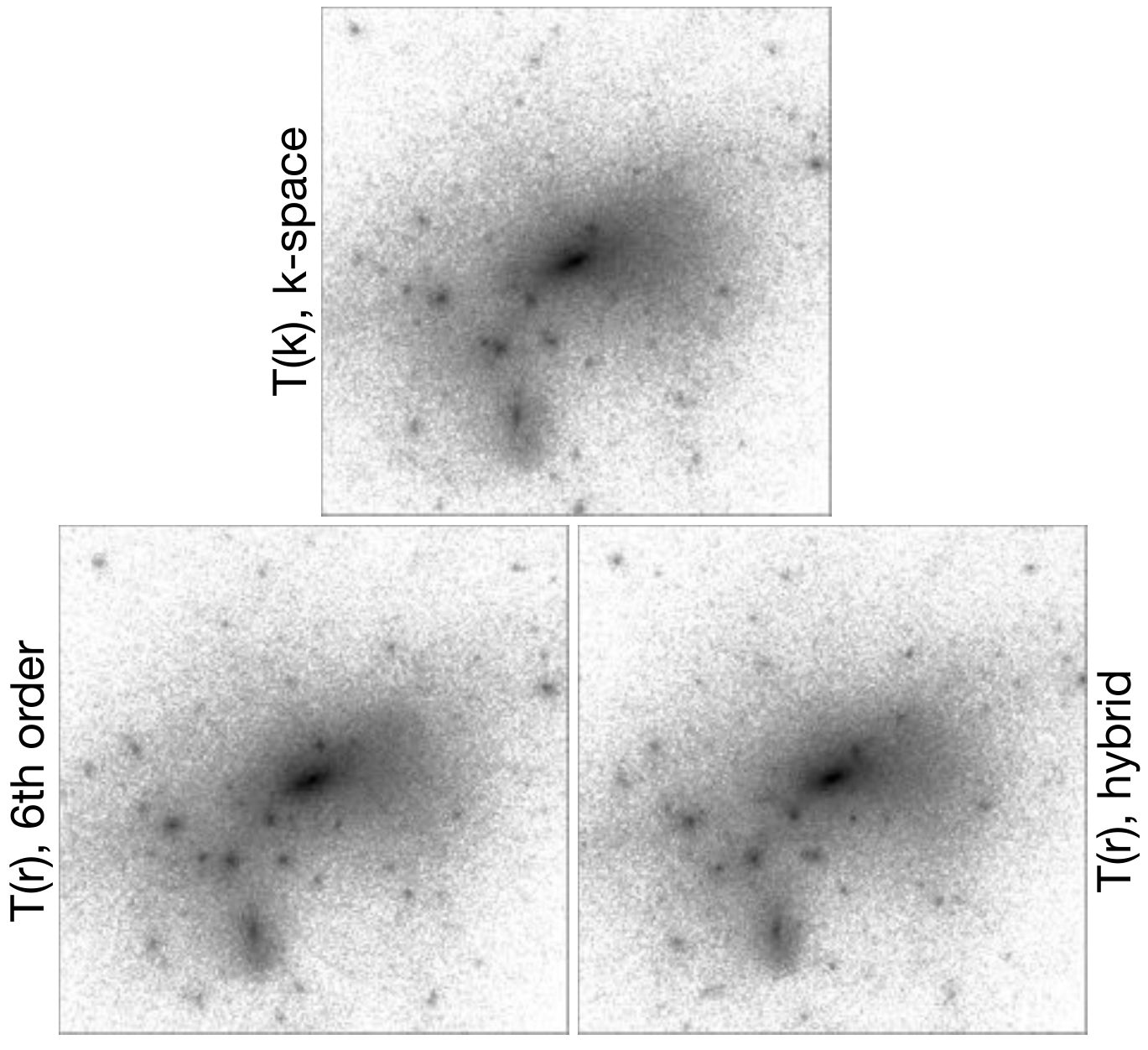}
\end{center}
\caption{\label{fig:halo_pics}Galaxy cluster of mass $2\times10^{14}\,h^{-1}{\rm M}_\odot$ at $z=0$ for the different methods to set up initial conditions. The
top panel shows the cluster formed from traditional $k$-space sampled initial conditions, while the bottom panels correspond to initial conditions using 
a convolution with the real space transfer function in conjunction with a 6th order finite difference Poisson solver (bottom left) and the hybrid Poisson solver (bottom right).
Each panel is the projection of a cube of $2\times2\times2\,h^{-3}{\rm Mpc}$ centered on the densest particle in the cluster.}
\end{figure}

In Figure \ref{fig:halo_pics}, images of the cluster at $z=0$ are shown for the three methods. Contrast in the images has been set in such a way as
to make substructure most visible, white areas are not actually devoid of particles. Visually, the only difference between the three panels
lies in the position of smaller substructure within the halo. All larger sub-haloes agree in their position. The overall shape of the halo is identical. 

\begin{figure}
\begin{center}
\includegraphics[width=0.3\textwidth]{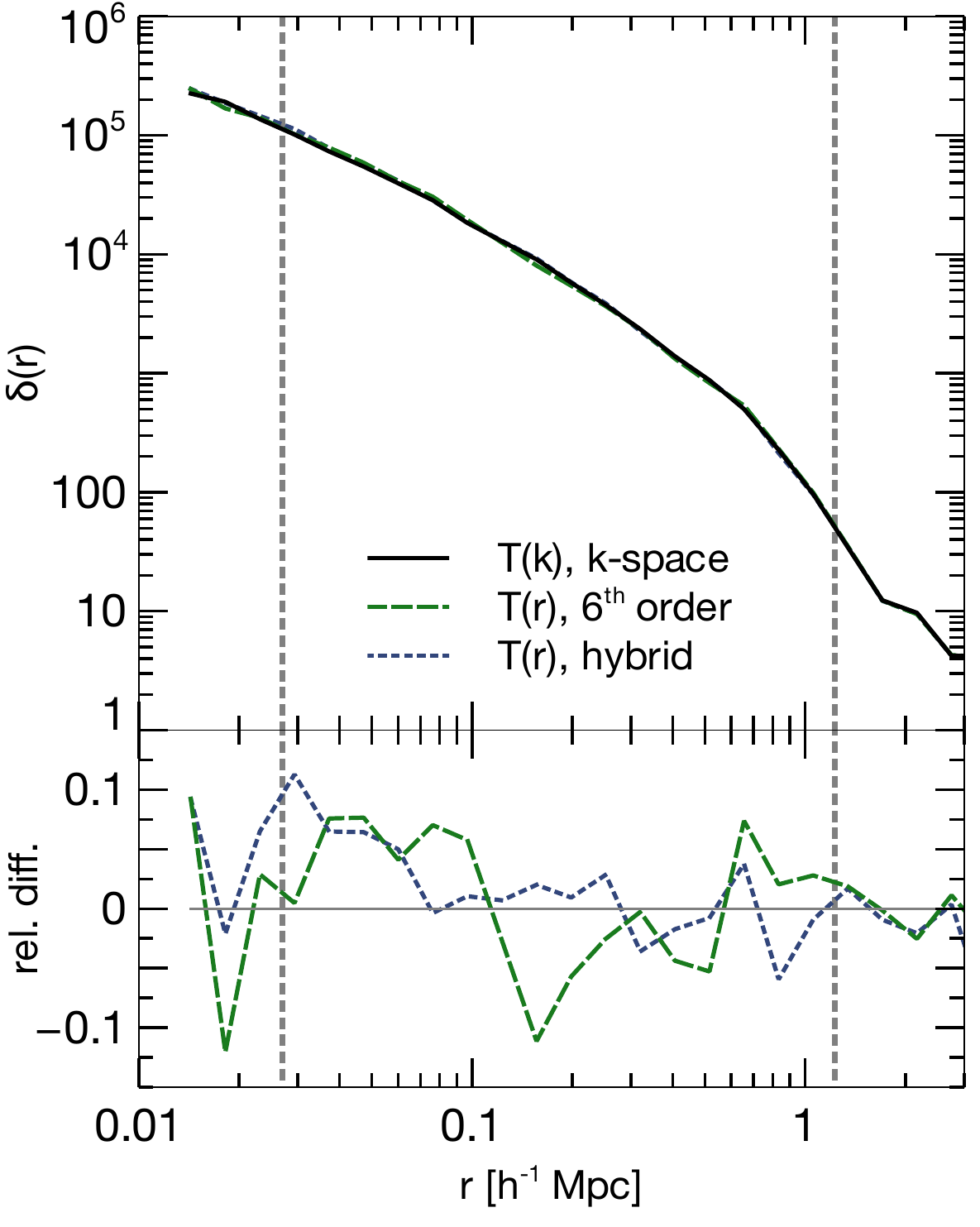}
\end{center}
\caption{\label{fig:halo_profiles}Radial density profiles for a $2\times10^{14}\,h^{-1}{\rm M}_\odot$ halo at $z=0$. The top panel shows the profiles
for initial conditions obtained with the 6th order multi-grid (dotted blue) and the hybrid Poisson solver (dashed green), as well as for the $k$-space sampled initial conditions
(solid black line). The bottom panel shows the relative difference between the profiles from real space and k-space initial conditions for the two methods.}
\end{figure}

In order to asses differences more quantitatively, we show in Figure \ref{fig:halo_profiles} the
radial density profile centered on the densest particle (determined by averaging over 32 neighbours and weighted by a spline kernel)
in each case. We find no systematic difference between either of the approaches. 
Differences are around 5 per cent and show no obvious bias. They are likely 
attributable to slight differences in the position of the substructure in the halo. These results further support the conclusion made in the previous section that
differences between the real and the k-space approach are completely negligible in the highly non-linear regime. 

The gross properties of the cluster halo are analyzed using the {\sc Amiga} halo finder (AHF) \citep{2009ApJS..182..608K}. In Table \ref{tab:res_haloprop_full},
we give the virial mass $M_{\rm vir}$ and radius $R_{\rm vir}$, maximum circular velocity $V_{\rm max}$, the radius $R_{\rm max}$ 
at which the circular velocity profile is maximal,  the dimensionless spin parameter $\lambda$ \citep{2001ApJ...555..240B}, the three-dimensional
velocity dispersion $\sigma_v$ as well as the minor-to-major axis ratio $c/a$, determined using the eigenvalues of the moment of inertia
tensor \citep[see e.g.][]{2007MNRAS.375..489H}. With the exception of $R_{\rm max}$, differences are around 1 per cent for all quantities 
investigated. It appears that $R_{\rm max}$ is very sensitive to small changes in the density profiles leading to a difference of a few per cent
in the finite difference case.

\begin{table}
\begin{center}
\begin{tabular}{|r|c|c|c|}
\hline
\hline
& $T(k)$, $k$-space & $T(r)$, finite. diff. & $T(r)$, hybrid \\
\hline
$M_{\rm vir}\,/\,h^{-1}{\rm M}_\odot$ & $2.140\times10^{14}$ & $2.141\times10^{14}$ & $2.130\times10^{14}$\\
$R_{\rm vir}\,/\,h^{-1}{\rm kpc}$ & $1231.2$ & $1231.4$ & $1229.4$\\
$V_{\rm max}\,/\,{\rm km}\,{\rm s}^{-1}$ & $941.5$ & $929.4$ & $940.6$ \\
$R_{\rm max}\,/\,h^{-1}{\rm kpc}$ & $680.6$ & $648.0$ &$669.5$ \\
$\lambda$ & $0.02510$ &$0.02497$ & $0.02503$\\
$\sigma_{\rm v}\,/\,{\rm km}\,{\rm s}^{-1}$ & $1000.1$ & $986.0$ & $1001.5$\\
$c/a$ & $0.7119$ & $0.7119$ & $0.7144$\\
\hline
\hline
\end{tabular}
\end{center}
\caption{\label{tab:res_haloprop_full}Properties of a cluster halo at $z=0$ for different Poisson solvers used to compute the initial conditions 
and real/$k$-space sampled density fields. }
\end{table}

\begin{figure}
\begin{center}
\includegraphics[width=0.35\textwidth]{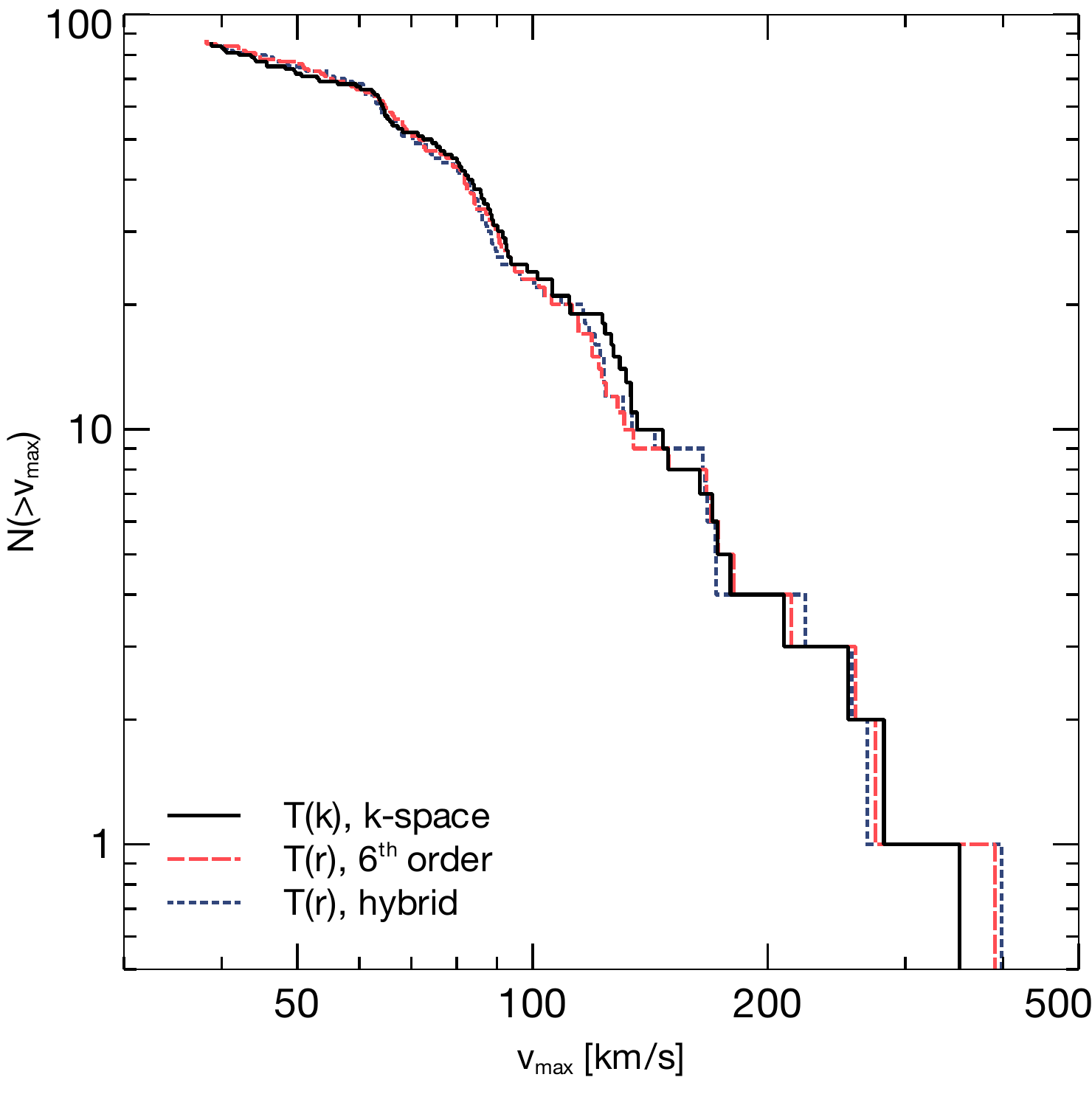}
\end{center}
\caption{\label{fig:subhmf_full}Cumulative distribution of maximum circular velocities $v_{\rm max}$ of sub-haloes of the cluster for the various methods.}
\end{figure}

Finally, using AHF, we investigate the sub-halo population of the cluster. Since the finite difference approach suppresses slightly the formation
of the smallest haloes (cf. \ref{sec:test_halo_abundance}), it is plausible that sub-haloes are affected as well, possibly even stronger. In Figure
\ref{fig:subhmf_full}, we show the cumulative distribution of maximum circular velocity $v_{\rm max}$ for all sub-haloes with at least 30 particles
contained within the virial radius of the
main cluster halo. We observe that the most massive substructure has a slightly lower $v_{\rm max}$ for the $k$-space initial conditions than for the
others. The following sub-haloes, at next lower $v_{\rm max}$ are identical. For the lower mass sub-haloes, the cumulative distributions agree, the
simulation starting from the multi-grid initial conditions
having slightly larger scatter than the one starting from hybrid initial conditions. In particular, we see no systematic deficiency of lowest 
mass haloes in the multi-grid case as might have been expected.

We conclude that differences in halo properties between the real-space and $k$-space conditions are at the per cent level for gross properties and
at the few per cent level for profiles and sub-halo mass functions. We do not observe any systematic bias.


\begin{figure*}
\begin{center}
\includegraphics[width=0.95\textwidth]{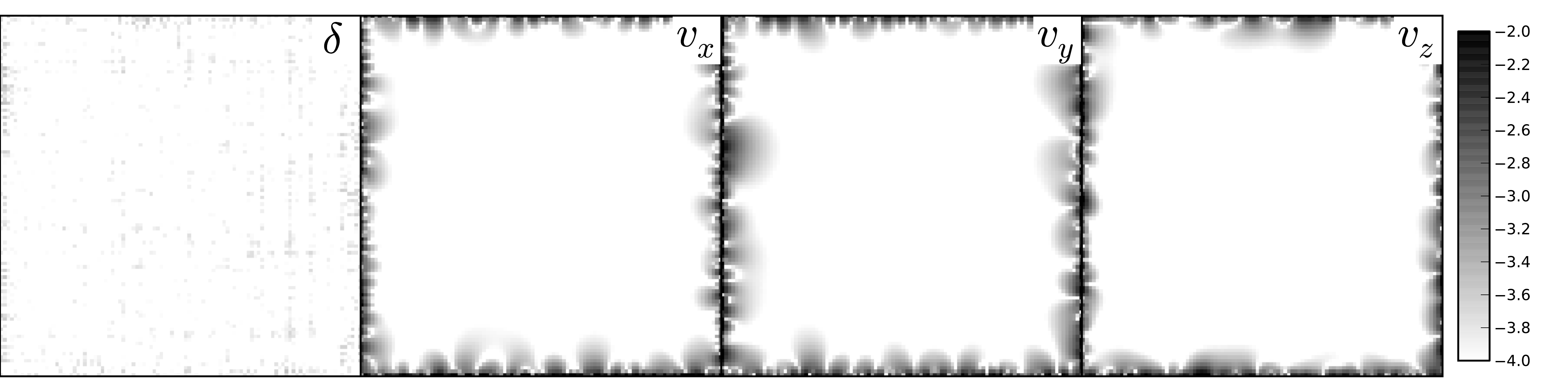}
\end{center}
\caption{\label{fig:err_1level_0pad}Error for the over-density (left-most panel) and the three components  
of the velocity ($v_x$, $v_y$, $v_z$ -- second to fourth panel, respectively) arising from the nested multi-scale convolution
approach for one additional refinement region. Shown is the difference between the respective quantity computed using nested convolution on
a $0.2\times0.2\times0.2$ $=102^3$ sub-volume at an effective resolution of $512^3$ embedded in a coarse grid of $256^3$ cells
 and the quantity in that region computed using the full grid of $512^3$
cells. We show the base 10 logarithm of the difference in units of the standard deviation of the respective field inside the refinement region on the $512^3$ grid. The
slices show the $x-y$ plane through the centre of the refinement region. }
\end{figure*}

\subsection{Errors in nested initial conditions: Initial redshift}
\label{sec:nested_errors}
In this section, we analyze the errors in nested initial conditions computed with our method compared to a full-resolution solution with no local refinement.
In order to compare our results with \cite{2001ApJS..137....1B}, we decided to implement an identical set-up, where we replace white noise in cells outside the refinement
region in the full-resolution set-up with the respective average value it has on the coarse grid in the ``zoom-in'' set-up. This ensures that identical white noise
information is used in all cases, i.e. all the information that is in principle representable in the refined hierarchy.
Note that naturally errors must arise at coarse-fine boundaries since the velocity and displacement fields must transition smoothly from fine to coarse resolution. 

\subsubsection{One refinement level}
First, initial conditions are generated on a uniform $512^3$ grid for a $100\,h^{-1}{\rm Mpc}$ box with white noise degraded to include identical information 
as in the ``zoom-in'' set-up to obtain the full-resolution answer. Next, another set of initial
conditions is generated for a region of 0.2 times the box length at an effective resolution of $512^3$ while the remaining volume is treated at a resolution of $256^3$.
The high-resolution region comprises $102^3$ grid cells. In Figure \ref{fig:err_1level_0pad}, we show the error for the density field and the three components of the velocity for a 
slice in the $x-y$~plane through the centre of the refinement region. Colors correspond to the logarithm of the difference between the multi-scale solution and the full-grid solution
in units of the standard deviation $\sigma_Q$ of the respective quantity $Q\in\left\{\delta,v_x,v_y,v_z\right\}$ on the full grid. 
In order to suppress the contribution from density convolution errors at the boundary, we padded the grids with additional 
$8$ grid cells at each boundary during the convolutions. This padding is however cut away before the potential is computed for the multi-scale hierarchy to ensure
that the final velocity field is continuous and differentiable across coarse-fine boundaries. All results are shown for 1LPT initial conditions, the respective errors 
in the 2LPT case are qualitatively identical. We used the 6th order multi-grid for the results shown in this and the next section. 

For the density, almost everywhere the difference is significantly below $10^{-4}\sigma_\delta$. We see that some errors at a few $\sim10^{-4}\sigma_\delta$ 
resulting from the interpolation of long-range components onto the finer grid. We find an RMS error of $\sim 7\times10^{-5}\sigma_\delta$.

For the velocity components, again, errors are significantly below $10^{-4}\sigma_{v_i}$ in the interior of the refinement region, apart from a few cells surrounding
the boundary. Errors at the boundary are at most $\sim10^{-2.5}\sigma_{v_i}$ and arise due to the smooth coarse-fine transition in the refined set-up which is
not present in the unigrid case. The RMS error is $\sim1.5\times10^{-3}\sigma_{v_i}$ for all three velocity components. Since the interior errors are below $10^{-4}\sigma$,
the RMS error over the entire refinement region is clearly dominated by only the errors at the boundary.

Errors for the hybrid Poisson solver are harder to compare since the long-range component from coarse grids is taken from the 
finite difference multi-grid solution so that it has no spectral leakage (see Section \ref{sec:aliasing}), while the same long-range
component in the unigrid situation will show the spectral leakage in the form of oscillations at the Nyquist frequency around the
smooth solution. This lack of spectral leakage in the long-range components completely dominates the error in the interior in a
direct comparison of the unigrid and the multi-scale solution at a level of $\sim 10^{-3}\sigma_{v_i}$, no other 
errors are however introduced. Since leakage is an artifact of the finite mesh used for the FFTs, these differences are
spurious and of no significance to our multi-scale approach to initial conditions.

\begin{figure}
\begin{center}
\includegraphics[width=0.25\textwidth]{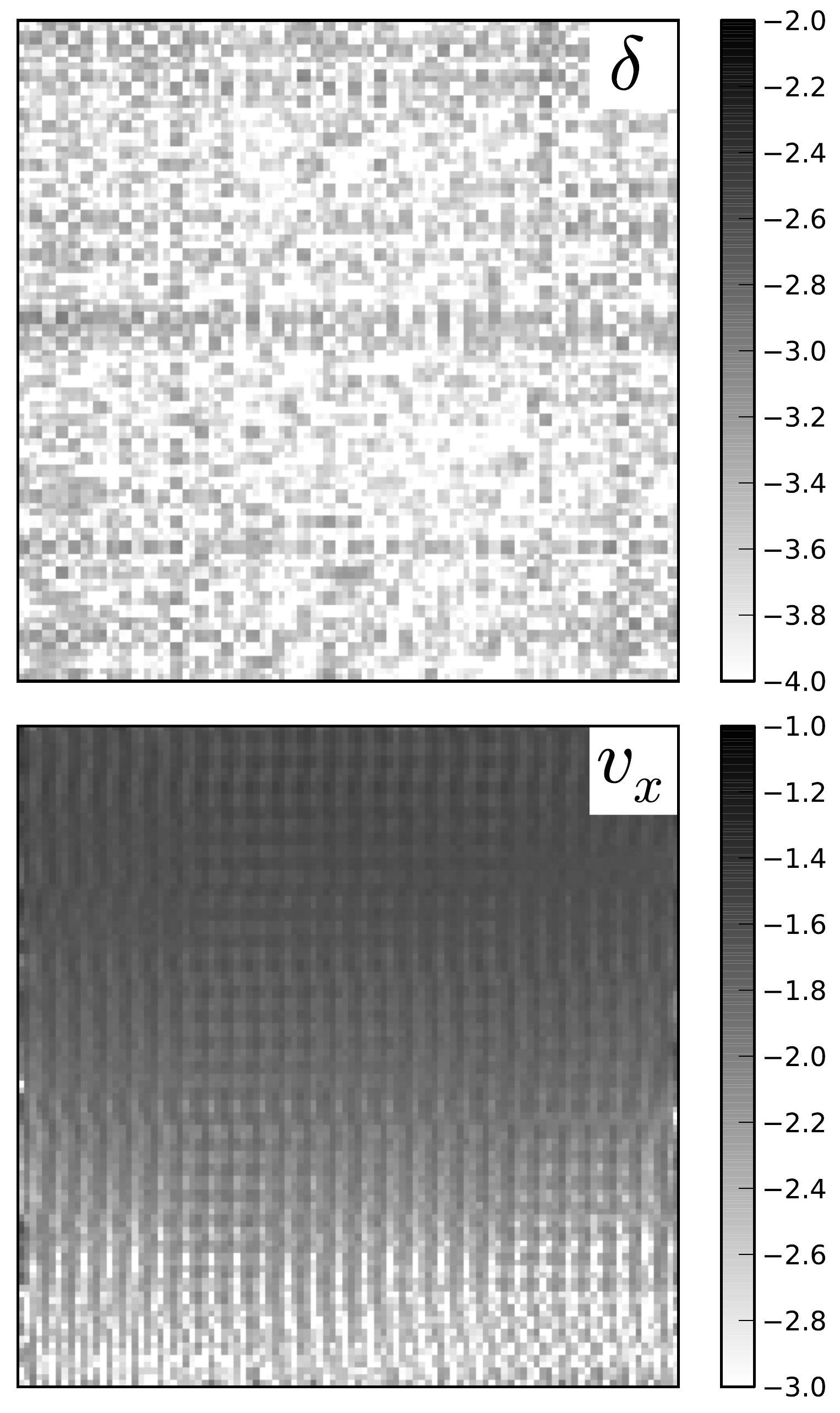}
\end{center}
\caption{\label{fig:err_grafic}Error for the density and the $v_x$ velocity component using {\sc Grafic2} for the same refinement set-up as in Figure \ref{fig:err_1level_0pad}.
Again, the base 10 logarithm of the difference in units of the standard deviation of the respective field inside the refinement region is shown.
Note that the colour scale for the $v_x$ error corresponds to a range one order in magnitude larger than in Figure \ref{fig:err_1level_0pad}.
The errors in the other velocity components are comparable to that for $v_x$.}
\end{figure}

In Figure \ref{fig:err_grafic}, we show, for comparison, the error in the density and $v_x$ velocity component obtained with {\sc Grafic2} 
for the identical refinement set-up and identical random numbers as in Figure \ref{fig:err_1level_0pad}. We have used version 2.101 of
the code and not removed or modified the standard antialiasing filter. Errors in density are about one, errors in velocity (and thus displacements) 
are about two orders of magnitude larger.
Furthermore, unlike with our approach, errors are oscillatory and spread over the entire refinement domain. In addition, there is a significant vertical
gradient at the level of several percent. We can only speculate about the sources of these errors. For the density field, our approach 
avoids the problem of aliasing and spectral leakage completely by operating purely in real-space. Furthermore, it produces by definition identical 
results for the $\delta^\ell_{\rm self}$ component in the refined and the unigrid case. In addition, we have 
ensured that mass is conserved between grids thus avoiding long-range errors in the velocity components. Finally,
the use of the adaptive multi-grid Poisson solver to compute velocities guarantees a solution of Poisson's equation
that is independent of the presence of a refined region if the mass field in the unigrid case is averaged with the
restriction operator and then re-injected with the respective injection operator (averaging and straight injection in our
case) in order to reflect the resolution in the refinement hierarchy.

As mentioned before, {\sc Grafic2} also uses an antialiasing filter which severely suppresses power on small scales,
so that the density and velocity fields generated are very smooth. We conclude that our new approach is an improvement by about two orders of magnitude 
in error reduction and the additional advantage that errors are completely confined to the boundaries. 

In summary, with our method, errors in the interior are below $10^{-4} \sigma$ with larger errors occurring at the boundary for the velocities due to the coarse-fine transition 
in the refinement hierarchy. It is particularly interesting to note that the velocity errors have no visible small scale components in the interior.

\begin{figure*}
\begin{center}
\includegraphics[width=0.95\textwidth]{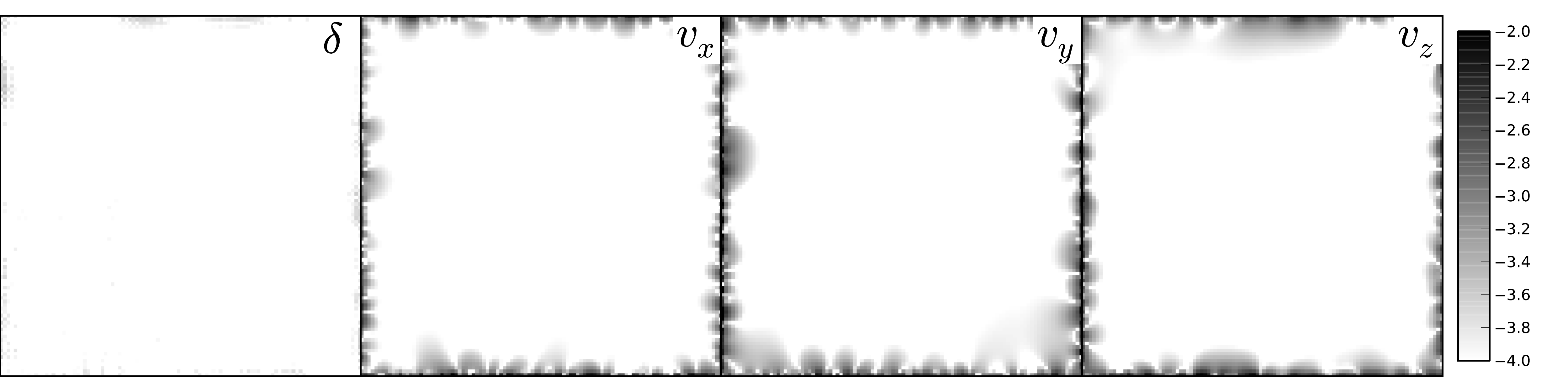}
\end{center}
\caption{\label{fig:err_2level_0pad}Same as Figure \ref{fig:err_1level_0pad} but with two refinement levels. The $102^3$ grid at $512^3$
effective resolution is embedded in an intermediate grid at $256^3$ effective resolution which is again embedded in a coarse grid
with $128^3$ cells. The intermediate grid pads the fine grid with 16 cells on each side. The errors are shown only for the finest grid.}
\end{figure*}

\subsubsection{Two refinement levels}

In this Section, we repeat the error analysis from above for a set-up of two nested 
refinement levels. The refinement region still has $102^3$ cells at an effective
 resolution of $512^3$ but is now embedded in an intermediate grid at $256^3$ 
 effective resolution which is in turn embedded in the coarse grid with $128^3$ 
 cells extending over the entire volume. The intermediate grid has an extent of 
 $16$ cells beyond the edges of the finest grid. As before, we pad each refinement 
 level with additional 8 boundary cells during the convolution step in order to avoid
 coarse-fine transition errors during the convolution. These boundary cells are 
 again cut off before the potential is computed.

In Figure \ref{fig:err_2level_0pad}, we show the differences in the high-resolution 
region between the full-grid solution and the nested multi-scale solution. Density 
and velocity errors in the high-resolution region are roughly identical to the one 
level case above and below $10^{-4}\sigma_Q$ in the interior. We find RMS errors 
of $\sim5\times10^{-5}\sigma_\delta$ for the overdensity field and 
$\sim1.8\times10^{-3}\sigma_{v_i}$ for the three velocity components.

To summarize, we see no significant impact on the velocity field due to the introduction 
of additional refinement levels. All errors are completely localized at the coarse-fine 
boundaries. Regarding memory consumption, using double precision variables (8 bytes), 
we observed a peak memory usage of $\sim2.7$~GBytes in unigrid mode (a $512^3$ 
double precision array requires 1~GByte of memory), $\sim670$~MBytes in the
case of the single refinement level set-up from above and $\sim270$~MBytes in the two-level case
from above, demonstrating the low memory footprint of our ``zoom-in'' approach
to generate initial conditions.


\subsection{Errors in nested initial conditions: Re-simulation of a galaxy cluster halo}
\label{sec:nested_errors_z0}

\begin{figure*}
\begin{center}
\includegraphics[width=0.85\textwidth]{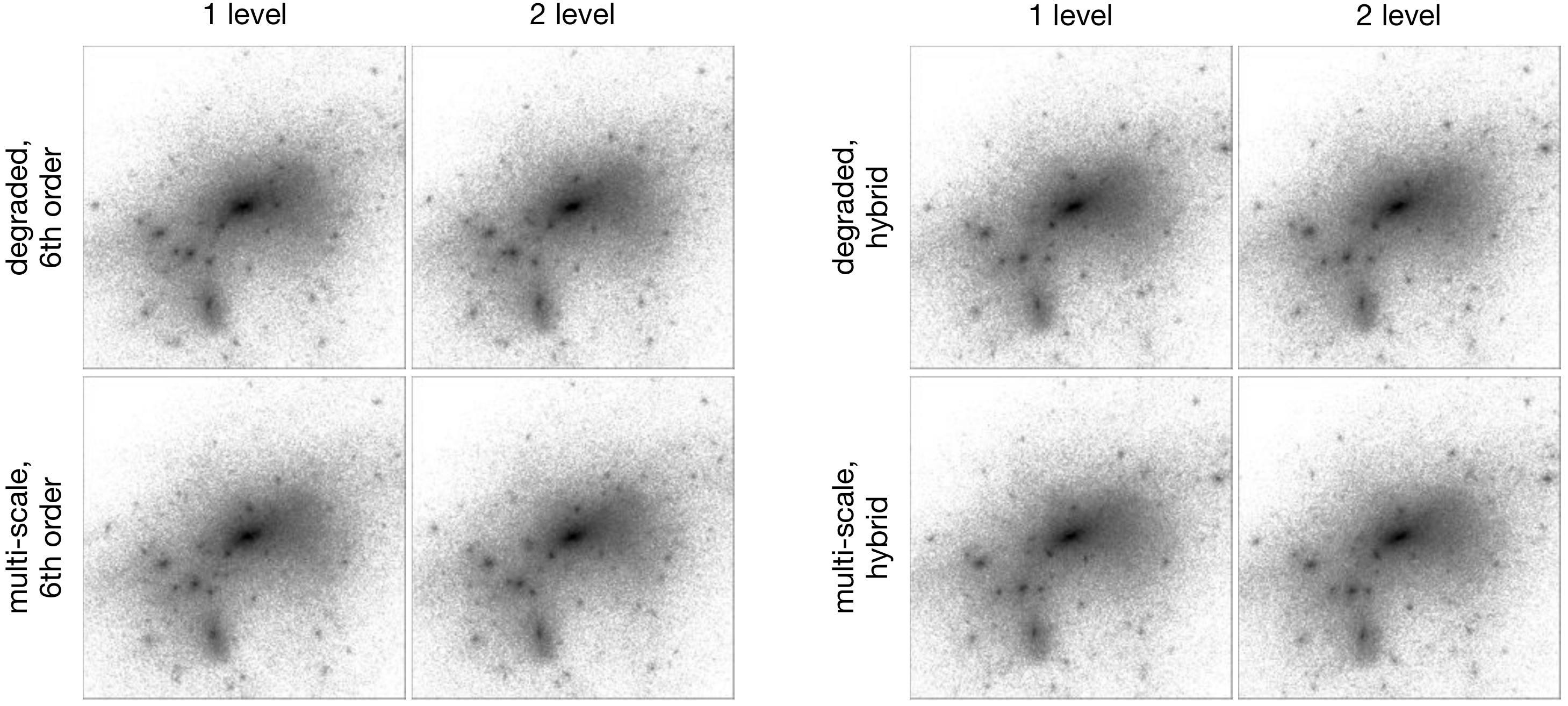}
\end{center}
\caption{\label{fig:cluster_err1}The resimulated cluster at $z=0$ with $512^3$ effective 
resolution. Panels in the upper row were obtained by degrading the initial conditions 
outside the refinement region, while initial conditions for the bottom row panels were 
generated using the multi-scale scheme. Left panels use the $6^{\rm th}$ order finite 
difference scheme, right panels the hybrid Poisson solver. In each of the two panels, 
the left column corresponds to a simulation set-up with $256^3$ effective resolution 
outside the high resolution region, while the right column shows result for a two-level 
zoom with a bounding region at $256^3$ resolution around the high resolution region 
and $128^3$ resolution in the remainder of the box. Each image represents a 
projected cube of $2\times2\times2\,h^{-3}{\rm Mpc}^3$, only high-resolution particles 
are shown. }
\end{figure*}

In order to assess the influence of multi-scale initial conditions on the formation of a re-simulated object, we generate multi-scale
initial conditions for one of the most massive clusters with mass $2\times10^{14}\,h^{-1}{\rm Mpc}$
in the $100\,h^{-1}{\rm Mpc}$ box discussed before. We deliberately set
the rectangular refinement region to include only the Lagrangian patch of the cluster halo. This will maximize the influence of
both the coarse sampling of the large-scale tidal field and boundary effects on the formation of the cluster halo and allow us
to estimate these effects. The refinement volume is a rectangular region of $23\times21\times21\,h^{-3}{\rm Mpc}^3$ and
was determined by following the FoF particles constituting the halo at $z=0$ back to the initial conditions and determining 
their bounding box rounded up/down to integer $h^{-1}\,{\rm Mpc}$. Note that a sphere at mean density containing the mass of the cluster would have a diameter
of $\sim 17.5\,h^{-1}{\rm Mpc}$. 

We perform {\sc Gadget}-2 simulations with one level of refinement where the effective resolution in 
the high resolution region is $512^3$ and $256^3$ in 
the remainder of the box, as well as with two levels of refinement, where the high resolution region is surrounded by a layer of 16 
particles thickness around each face at $256^3$ effective resolution that then drops to $128^3$ in the remainder of the box.
In addition, we perform each simulation two times for both the multigrid finite-difference and the hybrid approach: 
once with the initial density field determined using the multi-scale convolution
technique described in Section \ref{sec:nested_convolution}, and once with the density field determined at full resolution and then degraded (by
averaging) to the same resolution as in the multi-scale setup followed by solving Poisson's equation on the nested grid rather than the full grid. 
The latter produces completely negligible errors in the 
velocities and displacements inside the high resolution region. Differences from the full-grid initial velocities occur only in
2-3 cells at the boundary where the fields transition smoothly from fine to coarse resolution. All simulations are run
with {\sc Gadget}-2 and use a force softening length of $0.009\,h^{-1}{\rm Mpc}$ for the high-resolution particles and 
$0.09\,h^{-1}{\rm Mpc}$ for the other particles.

In Figure \ref{fig:cluster_err1}, we show the clusters at $z=0$ for all combinations of refinement set-up, Poisson solver and initial density
generation method. We observe that  the main halo as well as the most massive subhalos are consistent in position and size.
The positions of some of the smaller halos are shifted and some smaller haloes seem to have merged in one set-up while they
have not in some of the others. In general, visual differences are minimal. It is furthermore surprising that no systematic difference
between the 1-level and the 2-level set-up can be seen. This is most likely a result of the padding region in the 2-level simulations
at $256^3$ resolution which is identical to the 1-level simulations - the tidal influence of structures outside the padding region thus
appears negligible and highlights the importance of adding padding.

Comparing  Figure \ref{fig:cluster_err1} with Figure \ref{fig:halo_pics}, which shows the cluster in the unigrid simulations, we observe
no obvious systematic differences apart from smaller sub-halo positions being slightly shifted. The overall 
shape of the cluster halo agrees very well between the re-simulations and the full box simulations.

We ran the {\sc Amiga} halo finder (AHF) \citep{2009ApJS..182..608K} on
the resimulated clusters and quote the same key values as in Section \ref{sec:cluster_analysis} to quantify the gross properties
of the cluster halo in Table \ref{tab:res_haloprop}. As in the unigrid case, apart from $R_{\rm max}$, we find differences around 
1 per cent or below for all quantities investigated.

\begin{table*}
\begin{center}
\begin{tabular}{|l|c|c|c|c|c|c|c|}
\hline
& $M_{\rm vir}\,/\,h^{-1}{\rm M}_\odot$ & $R_{\rm vir}\,/\,h^{-1}{\rm kpc}$ & $V_{\rm max}\,/\,{\rm km}\,{\rm s}^{-1}$ & $R_{\rm max}\,/\,h^{-1}{\rm kpc}$ & $\lambda$ & $\sigma_{\rm v}
\,/\,{\rm km}\,{\rm s}^{-1}$ & $c/a$\\
\hline
\hline
\multicolumn{8}{|c|}{6th order finite difference}\\
\hline
1 level, degraded & $2.136\times10^{14}$ & $1230.4$ & $939.5$ & $646.5$ & $0.02543$ & $998.5$ & $0.7168$\\
1 level, multi-scale & $2.137\times10^{14}$ & $1230.6$ & $941.4$ & $667.7$ & $0.02545$ & $999.3$ & $0.7168$\\
2 level, degraded & $2.138\times10^{14}$ & $1230.9$ & $938.5$ & $652.3$ & $0.02552$ & $997.4$ & $0.7141$\\
2 level, multi-scale & $2.137\times10^{14}$ & $1230.6$ & $939.8$ & $612.8$ & $0.02550$ & $998.3$ & $0.7171$\\ 
\hline
\hline
\multicolumn{8}{|c|}{hybrid Poisson solver}\\
\hline
1 level, degraded & $2.135\times10^{14}$ & $1230.3$ & $942.4$ & $667.9$ & $0.02525$ & $1000.7$ & $0.7142$\\
1 level, multi-scale & $2.134\times10^{14}$ & $1230.1$ & $941.4$ & $649.8$ & $0.02530$ & $999.9$ & $0.7136$\\
2 level, degraded & $2.134\times10^{14}$ & $1230.0$ & $940.2$ & $665.1$ & $0.02532$ & $998.4$ & $0.7130$\\
2 level, multi-scale & $2.134\times10^{14}$ & $1230.0$ & $941.8$ & $669.6$ & $0.02534$ & $999.5$ & $0.7145$\\ 
\hline
\hline
\end{tabular}
\end{center}
\caption{\label{tab:res_haloprop}Properties of the cluster halo when resimulated with one or two levels of coarse resolution
particles outside the high-resolution region. The `degraded' runs started from initial conditions that were computed at the full resolution
of the high-resolution region, while the `multi-scale' runs use the adaptive zoom-in technique.}
\end{table*}%

\begin{figure}
\begin{center}
\includegraphics[width=0.47\textwidth]{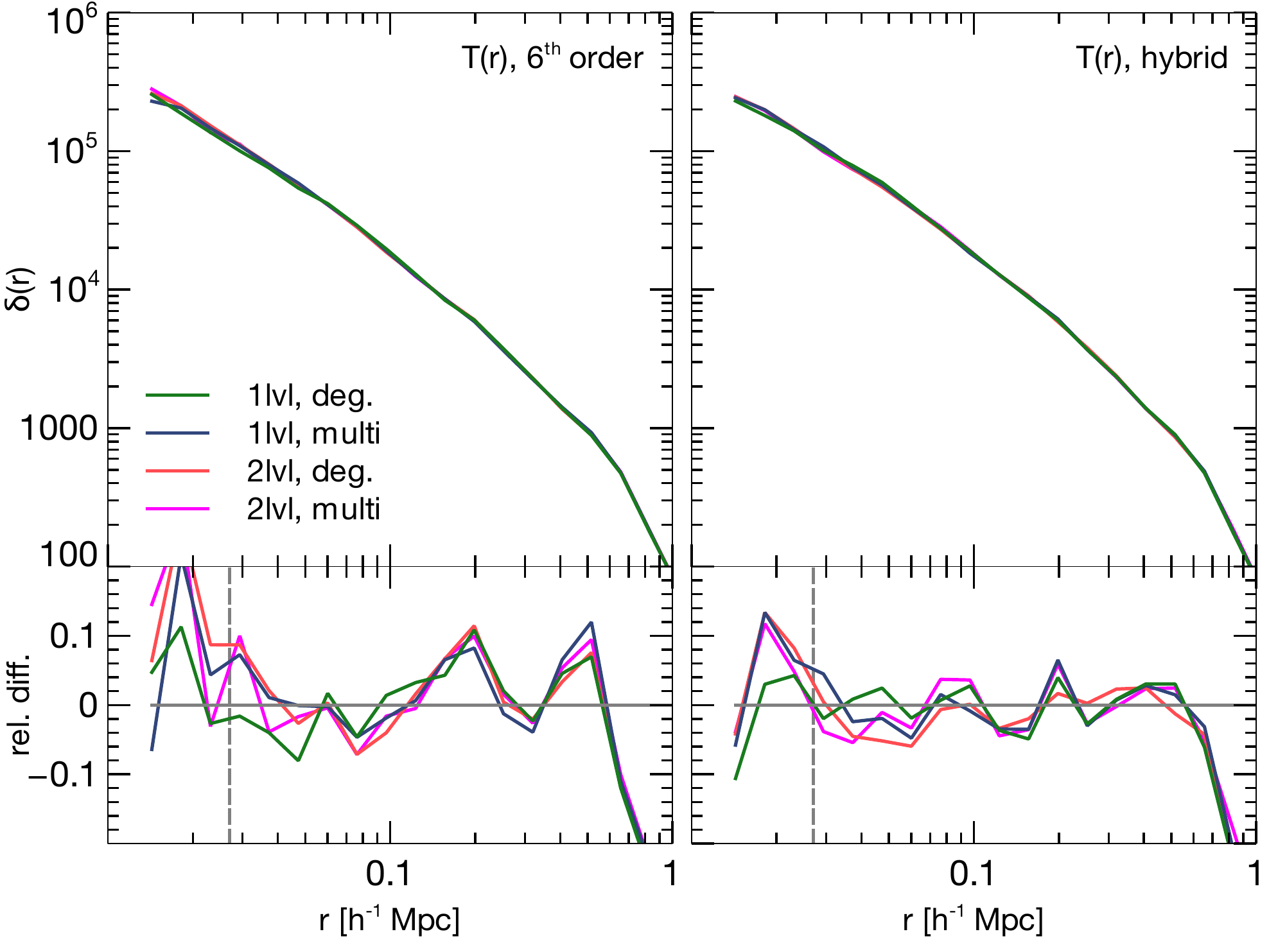}
\end{center}
\caption{\label{fig:cluster_profiles1}Overdensity profiles for the cluster at $z=0$ re-simulated with one and two refinement levels (top). 
The lower panel shows the relative difference with respect to the profile of the cluster in the unigrid simulation. Left panels show the 
result obtained from finite difference multi-grid initial conditions, the right panels correspond to the hybrid approach. The vertical light gray 
lines indicate the scale of three times the softening length.}
\end{figure}

\begin{figure}
\begin{center}
\includegraphics[width=0.47\textwidth]{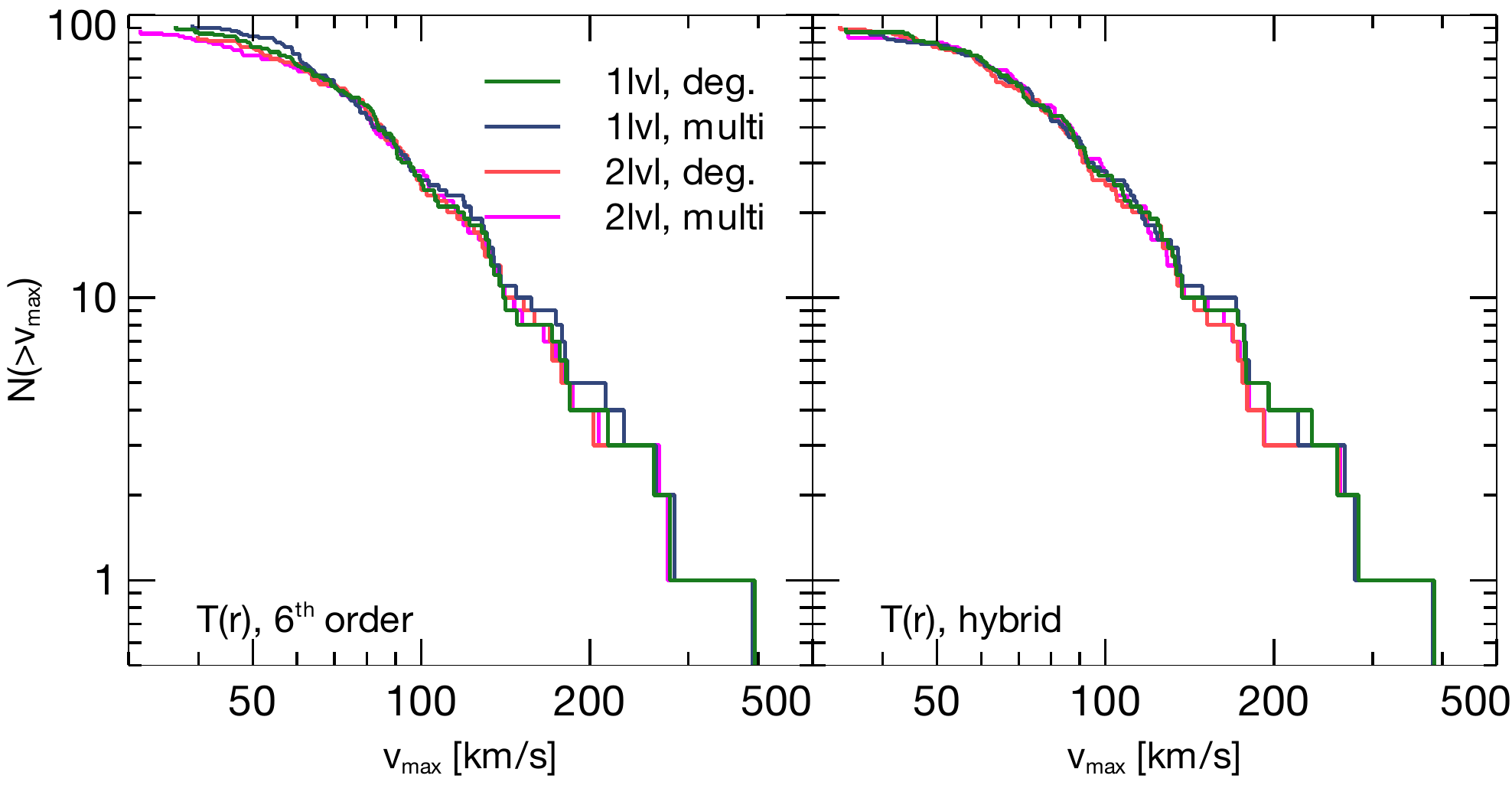}
\end{center}
\caption{\label{fig:cluster_subhmf1}Cumulative histogram of sub-halo maximum circular velocities $v_{\rm max}$ of the re-simulated cluster at $z=0$. The
left panel shows the result for finite difference multi-grid approach, the right panel the corresponding results for the hybrid approach.}
\end{figure}

In Figure \ref{fig:cluster_profiles1}, we show the radial over-density profiles for the various cases. The lower panel show the relative difference
with respect to the unigrid simulations. We observe no bias in either case. Scatter around the unigrid profile is larger for the finite
difference case than for the hybrid initial conditions. For both methods, the scatter is slightly larger for the 2-level than for
the 1-level set-up.
Note that all ``zoom-in'' density profiles fall below the unigrid profiles in the last bin. This is due to the too small size of our refinement
region as it is also present in the simulations generated by degrading the full density field. Despite the fact that it had been chosen 
too small, we find however not a single low resolution particle inside the virial radius, the first appearing at $\sim 1.3\,R_{\rm vir}$. 

In Figure \ref{fig:cluster_subhmf1}, we show the abundances of substructure as a function of maximum circular
velocity $v_{\rm max}$ within the virial
radius of the cluster for all of the re-simulations. Differences between the 1 and 2-level results are at the level of several per cent 
for the finite difference case. The hybrid initial conditions show 
better agreement between the four runs. It is hard and beyond the scope of this paper to investigate to what degree these
differences stem from simple changes of the sub-halo positions causing errors in determining their circular velocities
(or masses) in the sub-halo finder.

We observe that the errors due to our multi-scale method, particularly with the hybrid Poisson solver, agree very well with the 
degraded initial conditions in both the one and two refinement level set-up. In particular, the scatter between the results from multi-scale 
density fields and degraded density fields is never larger than the difference between degraded initial conditions at one or two refinement
levels itself. This leads us to conclude that the differences, apart from the introduction of an additional stochastic component, are
dominated by the late-time evolution and not the initial conditions. In particular, we find no evidence that our method introduces systematic
difference or bias in any of the investigated quantities. 

\begin{figure}
\begin{center}
\includegraphics[width=0.4\textwidth]{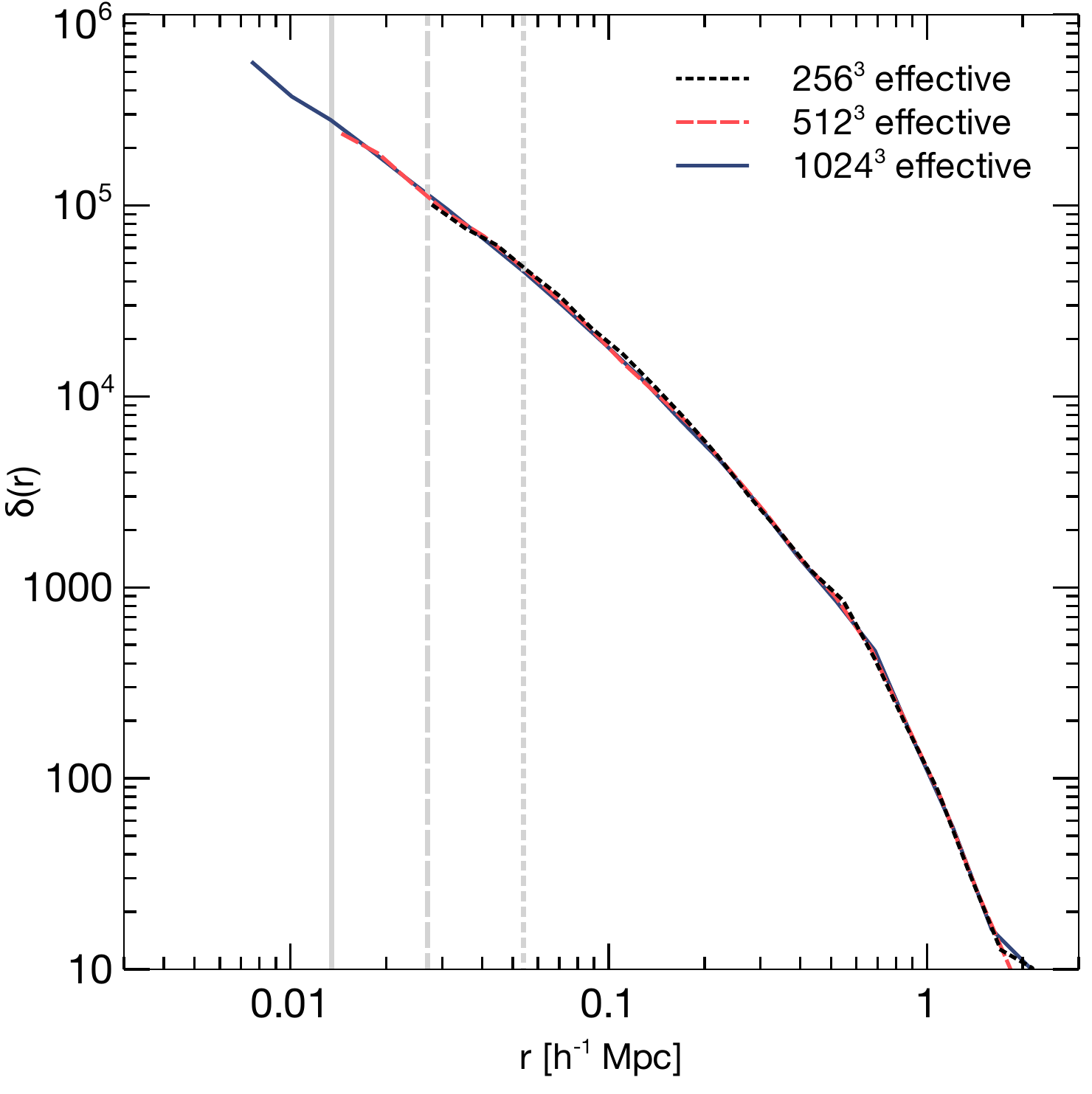}
\end{center}
\caption{\label{fig:cluster_prof_conv}Radial density profiles of the re-simulated cluster haloes at increasing resolution with a base grid of $128^3$ particles.
The profile for 1 refinement level, $256^3$ effective resolution is shown as a dotted black line; 2 levels, $512^3$ effective resolution as a dashed
red line; and 3 levels, $1024^3$ effective resolution, as a solid blue line. The vertical light gray lines in the corresponding line styles indicate
three times the softening length for each simulation.}
\end{figure}

Finally, we investigate the convergence of the density profile at even higher resolution for the hybrid Poisson solver case. In Figure \ref{fig:cluster_prof_conv}, we show the
radial density profiles for a series of re-simulations that all have a base resolution of $128^3$ particles and one to three additional 
refinement levels. The profiles trace each other almost perfectly.


\section{Baryon Initial Conditions}
\label{sec:baryon_ics}
In this section, we discuss the generation of initial conditions for a two-component -- dark matter and baryon -- fluid.
Linear perturbation theory predicts distinct amplitudes for baryon and CDM density fluctuations and also for their
respective velocity fluctuations. We demonstrate that our approach using real-space transfer functions can be
easily generalized for such a two-component fluid.

\subsection{Density and velocity transfer functions}

At high redshifts, at which initial conditions for cosmological simulations are typically generated, baryon density 
fluctuations are not yet exactly tracing the dark matter fluctuations \citep[e.g.][]{1998ApJ...501..442Y}. Furthermore, 
baryon fluctuations are exponentially suppressed beyond the photon diffusion scale \citep{1968ApJ...151..459S}.
These effects lead to a different shape of the baryon transfer function compared to the pure dark matter transfer function,
particularly on small scales below $\sim10^{-2}h^{-1}{\rm Mpc}$. 

\cite{2003MNRAS.344..481Y} demonstrated that the growth of 
density perturbations in the two-component fluid can only be correctly reproduced if besides the different initial amplitudes
of density perturbations also the difference in initial velocities between the two components are respected.
It is thus important that initial conditions for the two-component fluid reflect these important
differences between baryons and dark matter and are thus able to reproduce the correct growth of 
fluctuations in both components consistent with the
predictions from linear perturbation theories for those scales where perturbations are still small also with 
cosmological $N$-body + hydrodynamics simulation codes. We defer a more detailed comparison with
non-linear perturbation theory in two-component fluids \citep[e.g.][]{2010PhRvD..81b3524S} to later work.

\begin{figure}
\begin{center}
\includegraphics[width=0.4\textwidth]{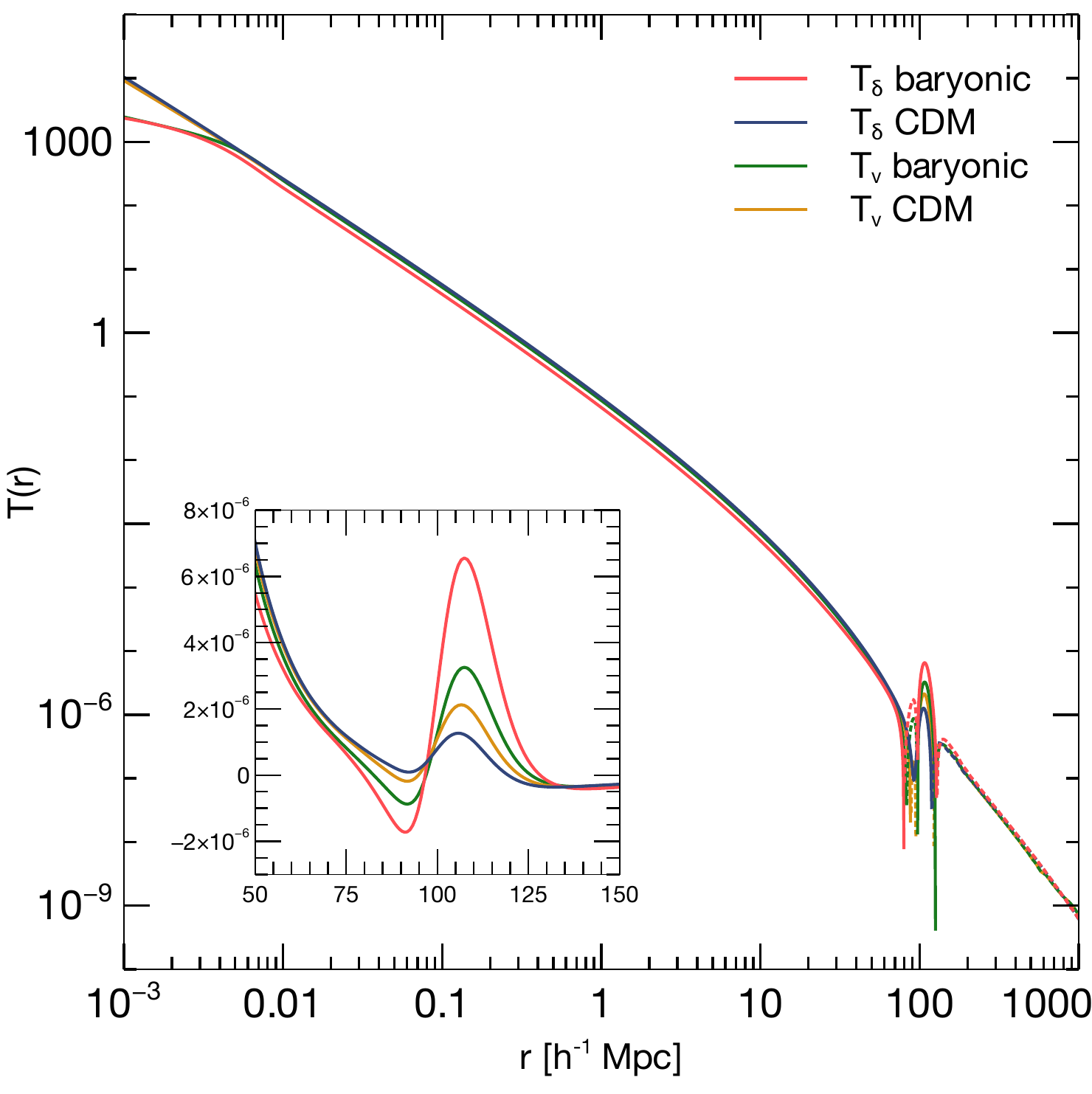}
\end{center}
\caption{\label{fig:transfer_all}The real space transfer function $T(r)$ at $z=100$ for baryon (red) and CDM (blue) density and
baryon (green) and CDM (orange) velocity perturbations. The inset shows the first baryon oscillation peak in linear scale. Note
the effect of  the Jeans scale for $r\lesssim10^{-2}\,h^{-1}{\rm Mpc}$ on the baryons.}
\end{figure}

For the two component fluid of baryons and dark matter, we determine both baryon and dark matter
velocities using transfer functions, equivalent to the density perturbation transfer functions, that we
construct in the following way. Using a Boltzmann solver, we solve the linearized equations of
relativistic perturbation theory to obtain the velocity perturbations of the CDM and the baryon
fluid. We adopt the notation of \cite{1995ApJ...455....7M} and 
kindly refer the reader to that publication for all details on the perturbation equations. We do not include
the baryon temperature fluctuations as suggested by \cite{2010arXiv1009.0945N} as it is mainly relevant to
perturbations with wavenumbers at and above the strongly Jeans damped regime. 
Using the results of linear perturbation theory, we define the two velocity transfer functions
\begin{eqnarray}
T_{v,c}(k) & \equiv & -\left(\frac{3}{2}\Omega_m H_0^2\right)^{-1}a\phi\\
T_{v,b}(k) & \equiv & T_{v,c}(k)-\frac{a}{\dot{a}}k^{-2}\,\theta_b,
\end{eqnarray}
where $\phi$ is the conformal Newtonian potential for spatial metric perturbations
and $\theta_b$ is the baryon velocity divergence.
These transfer functions are constructed in such a way that at linear order the relation
\begin{equation}
v_j(k) \propto \frac{ik_j}{k^2}T_v(k)\,k^{n_s/2}, \quad j=1,2,3
\end{equation}
holds for the amplitude of velocity perturbations, i.e. they simply replace the density transfer functions in our algorithm 
when computing initial velocities as in Section \ref{sec:firstlpt}. They furthermore allow the application of 2LPT 
(cf. Section \ref{sec:seclpt}), as they generate an ``effective'' source field for the velocity perturbations completely analogously to the 
relation between density and displacement.  The respective real space transfer functions for all four fields
are shown in Figure \ref{fig:transfer_all}.

\begin{figure*}
\begin{center}
\includegraphics[width=0.8\textwidth]{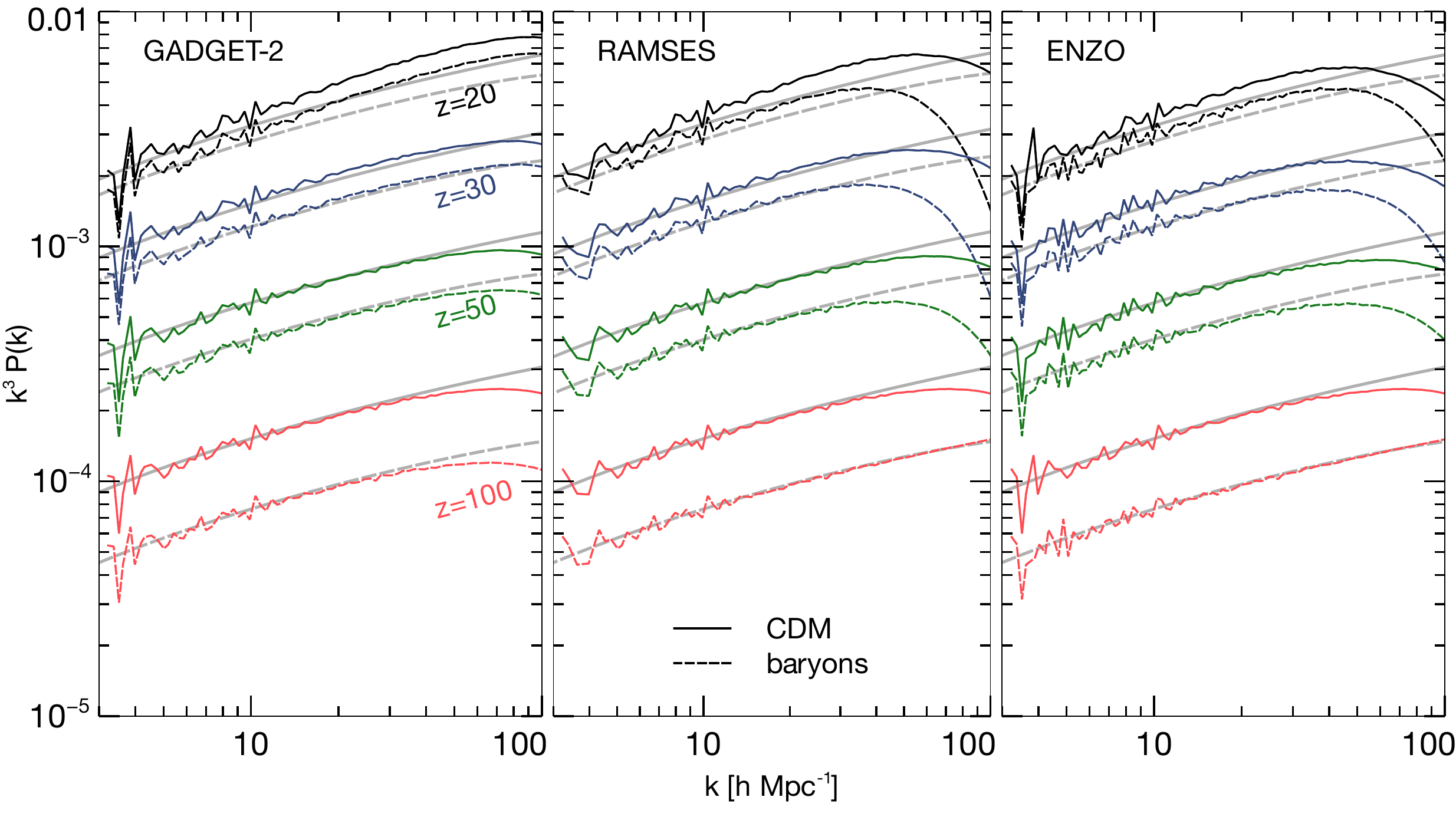}
\end{center}
\caption{\label{fig:evol_baryon}Evolution of the density power spectra of CDM (solid) and baryon (dashed) perturbations for different cosmological simulation
codes. The initial conditions have distinct amplitudes for baryon and CDM density and velocity perturbations. We show the power spectra
at redshifts $z=100$, $50$, $30$ and $20$ (bottom to top), the corresponding results from linear perturbation theory are shown as gray lines
in the background. Note that no adaptive mesh refinement was allowed for {\sc Ramses} and {\sc Enzo}.}
\end{figure*}

\subsection{Evolution of Baryon perturbations}
We now analyze the evolution of distinct baryon and CDM perturbations in two-component simulations using three commonly used cosmological
simulation codes: {\sc Gadget-2} \citep{2005MNRAS.364.1105S}, {\sc Ramses} \citep{2002A&A...385..337T} and {\sc Enzo} 
\citep{1997ASPC..123..363B,2001astro.ph.12089B,2004astro.ph..3044O}. In each case, the initial conditions are generated at $512^3$ resolution for a box
of comoving size $8\,h^{-1}{\rm Mpc}$ with the same cosmological parameters as before and $\Omega_{\rm baryon}=0.045$. The initial redshift
is set to $z=100$. For the grid codes ({\sc Enzo} and {\sc Ramses}), we do not allow for adaptive mesh refinement in order to avoid
multi-scale density fields when computing the matter power spectra. In the SPH case ({\sc Gadet-2}), 
we use a gravitational force softening of comoving $0.8\,h^{-1}{\rm kpc}$ for both baryons and CDM. Furthermore, SPH particles are placed on a
staggered grid with respect to the CDM particles. In all cases, we assume a polytropic
equation of state with adiabatic exponent $\gamma=5/3$, an initial temperature of $140 {\rm K}/\mu$, where $\mu$ is the mean molecular weight,
 and use neither cooling nor energy feedback. Note that {\sc Ramses}, in our set-up, uses
a piecewise linear method (PLM) for the hydrodynamics and mixed 2nd and 4th order gravity (2nd order in the Laplacian, 4th order in the gradients). 
Furthermore, we use the multidimensional monotonized central slope limiter \citep{1977JCoPh..23..263V}. {\sc Enzo}
uses a piecewise parabolic method (PPM) for the hydro and 2nd order gravity. We stop the simulations at $z=20$ and thus probe into the mildly
non-linear regime.

The power spectra of the two-component perturbations are shown in Figure \ref{fig:evol_baryon} at the initial and three subsequent redshifts, evolved
with the respective codes. These power spectra have been computed on a $512^3$ grid using FFT with CIC interpolation for all particles, and directly from
the baryon overdensity field for the grid codes. We do not correct for the loss of power due to the interpolation scheme and also do not smooth the 
SPH particles with their respective kernels. Thus, the baryon perturbations in the SPH case do not reflect the density perturbations seen by the
SPH scheme, resolution thus appears to be significantly higher than for the grid codes, so that our results should not be used as an argument in
favour or disfavour of either method.

We observe several differences between the codes. The SPH run with {\sc Gadget-2} shows excellent agreement with linear perturbation theory. The relative
amplitude between CDM and baryon perturbations at any given mode matches very well even in the mildly non-linear regime at $z=20$. It is not 
suprising that {\sc Gadget} performs very well in our test since hydrodynamical forces are negligible in the regime probed and we thus only observe the
performance of the tree-PM gravity solver.

In the case of {\sc Ramses} and {\sc Enzo}, CDM perturbations on the smallest scales grow slower than predicted by linear PT. This is however expected and a
direct comparison with the {\sc Gadget} results is unfair since we did not allow for refinement, and gravity forces are thus smoothed at
the grid scale. For {\sc Enzo}, the growth of CDM perturbations is more damped at small scales than in the {\sc Ramses} run. This is most likely the result of
the higher order of the gradient used by the {\sc Ramses} Poisson solver. 

For the baryon perturbations, small scales are growing slightly slower with {\sc Ramses} than with {\sc Enzo}. Again, this is to be expected since we use
{\sc Enzo} with PPM and {\sc Ramses} with PLM. Also the type of Riemann solver and slope limiter used in {\sc Ramses} has a slight effect, more 
diffusive limiters and solvers slightly decrease the power on the smallest scales. On large scales, all codes reproduce the correct growth of the 
perturbations, in very good agreement with linear PT.

To summarize, we observe that all three codes correctly evolve a baryon-CDM two-component fluid which starts with distinct density
and velocity perturbations for each component. Differences that we observe with respect to the results from linear perturbation theory
are attributable to the various numerical methods used in the simulation codes and not to the initial conditions.

\subsection{Local Lagrangian approximation for the baryon density field \label{sec:ICdens}}

The density field sourcing the displacements and velocities in Lagrangian perturbation theory 
-- which is a Gaussian random field -- is inconsistent with the density field of the displaced particles
-- which is non-Gaussian. Using the Gaussian perturbation field as the initial baryon density field
in grid based codes is thus inconsistent with the CDM perturbations.
In this section, we describe an approach to prescribe initial gas density for the mesh cells based
on the local Lagrangian approximation (LLA) that leads to consistency between the initial gas
and dark matter density fields \citep[see also][]{1991MNRAS.251..399B,1993ApJ...410..482P,1997MNRAS.284..425P}. Furthermore, 
it provides a natural way to prescribe the initial gas density on a grid also for 2LPT initial conditions.

The continuity equation is trivially fulfilled for a Lagrangian description of the fluid and simply reads ${\rm d}m={\rm const.}$
We can however express the continuity equation in terms of the evolving fluid coordinates ${\bf x}$. It then becomes
$\rho({\bf x},t)\,{\rm d}^3x = \bar{\rho}\,{\rm d}^3 q$,  
where $\bar{\rho}$ is the unperturbed mean density and $\rho({\bf x},t)$ is the 
density of the fluid element. Since the continuity equation is simply a volume integral, we can employ the
change of variables theorem to express the left-hand-side also in terms of the initial position ${\bf q}$,
\begin{equation}
\rho({\bf q},t)\,\left| \frac{\partial {\bf x}}{\partial {\bf q}}\right|\,{\rm d}^3 q = \bar{\rho}\,{\rm d}^3 q.
\end{equation}
Then, using eq. (\ref{eq:LagrangeCoord}), the formal evolution of the density at the initial position of the fluid elements can be simply written as
\begin{equation}
\rho({\bf q},t) = \frac{\bar{\rho}}{\det \left[ \delta_{ij} + \frac{\partial L_i}{\partial q_j}\right]}, \label{eq:gasdens}
\end{equation}
where $\delta_{ij}$ is the unity matrix. This implies that $\rho({\bf q},t)$ assumes a non-Gaussian
distribution in general. In fact, $\rho({\bf q},t)$ will become singular whenever an eigenvalue
of $\partial L_i/\partial q_j$ will become $-1$, corresponding to shell crossing along the respective
axis. Since we use a first or second order Lagrangian perturbation theory approximation for the
displacement field ${\bf L}({\bf q},t)$,
eq. (\ref{eq:gasdens}) is not exact and amounts to a ``local Lagrangian approximation''.
This means that, in general, mass will not be strictly conserved. We correct this by enforcing 
mass conservation by an a posteriori renormalization $\rho\rightarrow \rho + \tilde{\rho}$, where
$\tilde{\rho}$ is the measured deviation from mean density. Typically, for sufficiently high initial
redshift (as e.g. for the $z=100$ initial conditions used), the relative error is around
0.6 per cent and thus negligible (note that the error amounts to no more than $\sim 2$
per cent at much lower redshifts).

When computing ${\bf L}({\bf q},t)$ for baryons, $\Phi$ is sourced only by the 
baryon density perturbations, and so also $\Psi$ contains only baryonic contributions. 
Note that application of the LLA does not change the baryon power spectrum.

\begin{figure}
\begin{center}
\includegraphics[width=0.45\textwidth]{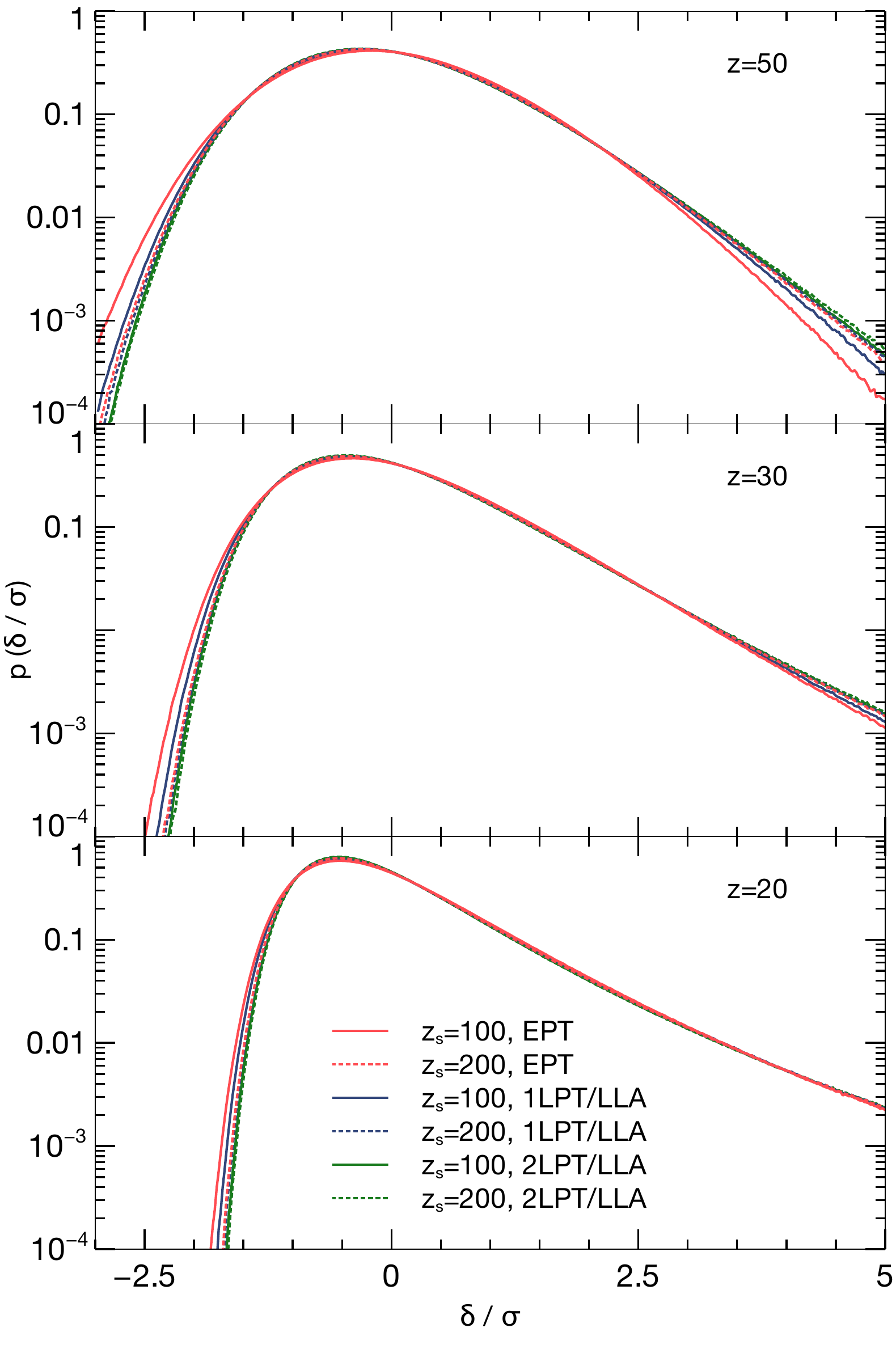}
\end{center}
\caption{\label{fig:pdf_evol}Evolution of the probability distribution functions (PDFs) of the baryon overdensity fields in units of the RMS
overdensity. For the $8\,h^{-1}{\rm Mpc}$ simulation box, the PDFs are given at three redshifts: $z=50$ (top panel), $z=30$ (middle panel)
and $z=20$ (bottom panel). Initial conditions were generated at two starting redshifts $z_s=100$ (solid lines) and $z_s=200$ (dotted lines)
with first order Lagrangian perturbation theory for dark matter and an initial Gaussian baryon density field following Eulerian perturbation
theory (EPT; red), with first order LPT for dark matter and baryons (using the LLA) (1LPT; blue) and with second order LPT for dark
matter and baryons (2LPT; green).}
\end{figure}

In Figure \ref{fig:pdf_evol}, we show the evolution of the probability distribution function (PDF) of the baryon overdensity field
in an {\sc Enzo} simulation using the same simulation parameters as above but for two different starting redshifts $z_s=100$ 
and $z_s=200$. For both starting redshifts, we generate initial conditions in three different ways:
\begin{description}
\item[EPT:] First order Lagrangian perturbation theory is used to initialize dark matter displacements and velocities as well
as baryon velocities. The initial baryon density field is taken from Eulerian perturbation theory (which is common practice in grid
based simulations). This baryon density field is initially Gaussian.
\item[1LPT/LLA:] Again 1LPT is used, but now the initial baryon density field is computed according to equation (\ref{eq:gasdens}),
so that it is consistent with 1LPT. This initial baryon density field is no longer Gaussian.
\item[2LPT/LLA:] Second order Lagrangian perturbation theory (cf. Section \ref{sec:seclpt}) 
is used for dark matter and baryons, using equation (\ref{eq:gasdens}) a consistent initial baryon density field
is generated.
\end{description}

We find that a similar transient behaviour as that between 1LPT and 2LPT for pure CDM simulations
\citep[cf. eg.][]{2006MNRAS.373..369C,2007JCAP...12..014T} can be observed. The skewness of the density
field is underestimated when starting at $z_s=100$ with the initially Gaussian field (EPT). This is somewhat
improved when starting at higher redshifts. When using an initial density field that is consistent with 1LPT, the
skewness is boosted so that the dependence of the PDF on the initial redshift of the simulation becomes smaller.
When using 2LPT and LLA, no significant dependence of the PDF on the starting redshift can be observed.
Note that we show the results for the baryon density field directly
taken from the simulations rather than a density field that was smoothed with an aperture filter. Filtering with a finite apperture 
however leads to almost identical results, the skewness is underestimated at high redshifts independent of scale.

In summary, Eulerian perturbation theory leads to an underestimation of the skewness of the baryon density PDF at
high redshifts. When using the LLA for the baryons, a PDF is imposed that is consistent with Lagrangian perturbation
theory. In particular, as has been shown already for dark matter simulations, the density PDF becomes much
less sensitive to the initial starting redshift when using 2LPT. Using the local Lagrangian approximation, consistent
initial baryon density fields can be construed.


\section{Summary \& Conclusions}
\label{sec:summary}
We have presented and implemented a method to generate real-space sampled initial conditions 
for nested grids, extending prior work by \cite{1997ApJ...490L.127P}, \cite{2001ApJS..137....1B}
and \cite{2005ApJ...634..728S}. These initial conditions are suitable for ``zoom-in'' cosmological simulations 
of structure formation with several levels of refinement that allow the study of single cosmological 
objects, such as e.g. groups and clusters of galaxies, single galaxies, or first stars, in a cosmological context
at very high resolution while maintaining a small computer memory footprint. 

Apart from the general advantage of real-space sampled initial conditions discussed by \cite{1997ApJ...490L.127P}
and \cite{2005ApJ...634..728S}, namely a correct reproduction of real-space statistical properties, such as
the two-point correlation function and mass variance in spheres, we demonstrate that our multi-scale convolution
approach for nested grids has very favourable error properties: (1) no interference
of different modes between different refinement levels, and (2) errors confined to coarse-fine boundaries. We find
RMS errors of velocity/displacement fields in the refined region at the order of $10^{-4}$ in units of the standard
deviation of the respective fields which is an improvement of about two orders of magnitude over previous
approaches. 

In order to determine particle displacement and velocity fields, at first and second order Lagrangian perturbation
theory, we followed two approaches. First, a pure multi-grid Poisson solver is used. We show that the finite volume
discretization it assumes leads to a suppression of perturbations at the smallest scales that can however be 
corrected by a deconvolution. Second, a hybrid Poisson solver is developed, which uses an adaptive multi-grid 
algorithm for inter-level gravity and an FFT-based Poisson solver for the finest grid. This hybrid approach leads to no suppression
of small scale perturbations. Analyzing statistical properties of unigrid simulations for which initial conditions have
been generated with the two different approaches, we find however that non-linear objects are not sensitive
to the lack of power at these very small scales and differences are only seen in the mildly non-linear regime
and at the lowest halo masses. We conclude that our real-space based approach is well suited also for
initial conditions for unigrid simulations.

In order to verify the accuracy of nested initial conditions generated with our method, we investigated the properties
of a galaxy cluster both in a unigrid simulation and a ``zoom-in'' simulation with one and two levels of refinement. We find
that the gross properties of the cluster, such as virial mass, radius, spin, shape and velocity dispersion, are recovered
at per cent level or better in the re-simulations. Density profiles show no bias with scatter at the level of a few per cent,
mainly attributable to slight changes in the positions of sub-structures in the re-simulations. We also find that the 
sub-halo mass function is recovered with an accuracy of a few per cent in all re-simulations, and observe that the 
hybrid Poisson solver performs slightly better than the pure multi-grid approach. We thus conclude that our algorithm
does not introduce bias or unreasonable scatter in observables that will be deduced from such a re-simulation 
 and will thus provide a reliable tool to study the internal structure of cosmological observables at
high resolution in large cosmological volumes. 

Finally, we study the inclusion of a baryonic component in our approach by generalising the use of real space transfer
functions also for distinct baryon and CDM density and velocity perturbations. We demonstrate that our approach
reproduces the evolution expected from linear perturbation theory correctly. Furthermore, we propose to set the initial
baryon density field based on a local Lagrangian approximation which is consistent with Lagrangian perturbation
theory of first or second order and which greatly reduces the dependence of the baryon density field on the starting
redshift of the simulation, thus reducing transient behaviour.


\section*{Acknowledgments}
The authors thank Hao-Yi Wu for help with testing the code and invaluable feedback, Romain Teyssier,
John Wise, Anatoly Klypin, Matthew Turk and Adrian Jenkins for comments and discussions that helped to improve both 
the paper and the code. Finally, the authors thank the referee, Simon Prunet, for very valuable suggestions to improve
the presentation of the paper. All simulations were run on the Orange cluster at KIPAC/SLAC. 
This work was partially supported by NASA ATFP grant NNX08AH26G,  NSF
AST-0808398 and NSF AST-0807075.
TA acknowledges financial support from the  Baden-W\"{u}rttemberg-Stiftung under grant P-LS-SPII/18.



\begin{thebibliography}{}

\bibitem[\protect\citeauthoryear{{Abel}, {Bryan} \& {Norman}}{{Abel}
  et~al.}{2002}]{2002Sci...295...93A}
{Abel} T.,  {Bryan} G.~L.,    {Norman} M.~L.,  2002, Science, 295, 93

\bibitem[\protect\citeauthoryear{{Bernardeau}, {Colombi}, {Gazta{\~n}aga} \&
  {Scoccimarro}}{{Bernardeau} et~al.}{2002}]{2002PhR...367....1B}
{Bernardeau} F.,  {Colombi} S.,  {Gazta{\~n}aga} E.,    {Scoccimarro} R.,
  2002, \physrep, 367, 1

\bibitem[\protect\citeauthoryear{{Bertschinger}}{{Bertschinger}}{2001}]{2001Ap%
JS..137....1B}
{Bertschinger} E.,  2001, \apjs, 137, 1

\bibitem[\protect\citeauthoryear{{Betancort-Rijo}}{{Betancort-Rijo}}{1991}]{19%
91MNRAS.251..399B}
{Betancort-Rijo} J.,  1991, \mnras, 251, 399

\bibitem[\protect\citeauthoryear{{Bouchet}, {Colombi}, {Hivon} \&
  {Juszkiewicz}}{{Bouchet} et~al.}{1995}]{1995A&A...296..575B}
{Bouchet} F.~R.,  {Colombi} S.,  {Hivon} E.,    {Juszkiewicz} R.,  1995, \aap,
  296, 575

\bibitem[\protect\citeauthoryear{{Bournaud}, {Elmegreen}, {Teyssier}, {Block}
  \& {Puerari}}{{Bournaud} et~al.}{2010}]{2010arXiv1007.2566B}
{Bournaud} F.,  {Elmegreen} B.~G.,  {Teyssier} R.,  {Block} D.~L.,    {Puerari}
  I.,  2010, \mnras, 409, 1088

\bibitem[\protect\citeauthoryear{Brandt}{Brandt}{1973}]{ABrandt_1973a}
Brandt A.,  1973, in Cabannes H.,  Teman R.,  eds, Proceedings of the Third
  International Conference on Numerical Methods in Fluid Mechanics Vol.~18 of
  Lecture Notes in Physics, Multi--level adaptive technique ({MLAT}) for fast
  numerical solution to boundary value problems.
Springer--Verlag, Berlin, pp 82--89

\bibitem[\protect\citeauthoryear{Brandt}{Brandt}{1977}]{ABrandt_1977b}
Brandt A.,  1977, Math. Comp., 31, 333

\bibitem[\protect\citeauthoryear{{Bryan}, {Abel} \& {Norman}}{{Bryan}
  et~al.}{2001}]{2001astro.ph.12089B}
{Bryan} G.~L.,  {Abel} T.,    {Norman} M.~L.,  2001, arXiv:astro-ph/0112089

\bibitem[\protect\citeauthoryear{{Bryan} \& {Norman}}{{Bryan} \&
  {Norman}}{1997}]{1997ASPC..123..363B}
{Bryan} G.~L.,  {Norman} M.~L.,  1997, in {D.~A.~Clarke \& M.~J.~West} ed.,
  Computational Astrophysics; 12th Kingston Meeting on Theoretical Astrophysics
  Vol.~123 of Astronomical Society of the Pacific Conference Series,
  {Simulating X-Ray Clusters with Adaptive Mesh Refinement}.
pp 363--+

\bibitem[\protect\citeauthoryear{{Buchert}, {Melott} \& {Weiss}}{{Buchert}
  et~al.}{1994}]{1994A&A...288..349B}
{Buchert} T.,  {Melott} A.~L.,    {Weiss} A.~G.,  1994, \aap, 288, 349

\bibitem[\protect\citeauthoryear{{Bullock}, {Dekel}, {Kolatt}, {Kravtsov},
  {Klypin}, {Porciani} \& {Primack}}{{Bullock}
  et~al.}{2001}]{2001ApJ...555..240B}
{Bullock} J.~S.,  {Dekel} A.,  {Kolatt} T.~S.,  {Kravtsov} A.~V.,  {Klypin}
  A.~A.,  {Porciani} C.,    {Primack} J.~R.,  2001, \apj, 555, 240

\bibitem[\protect\citeauthoryear{{Ceverino} \& {Klypin}}{{Ceverino} \&
  {Klypin}}{2009}]{2009ApJ...695..292C}
{Ceverino} D.,  {Klypin} A.,  2009, \apj, 695, 292

\bibitem[\protect\citeauthoryear{{Crain}, {Theuns}, {Dalla Vecchia}, {Eke},
  {Frenk}, {Jenkins}, {Kay}, {Peacock}, {Pearce}, {Schaye}, {Springel},
  {Thomas}, {White} \& {Wiersma}}{{Crain} et~al.}{2009}]{2009MNRAS.399.1773C}
{Crain} R.~A.,  {Theuns} T.,  {Dalla Vecchia} C.,  {Eke} V.~R.,  {Frenk} C.~S.,
   {Jenkins} A.,  {Kay} S.~T.,  {Peacock} J.~A.,  {Pearce} F.~R.,  {Schaye} J.,
   {Springel} V.,  {Thomas} P.~A.,  {White} S.~D.~M.,    {Wiersma} R.~P.~C.,
  2009, \mnras, 399, 1773

\bibitem[\protect\citeauthoryear{{Crocce}, {Pueblas} \& {Scoccimarro}}{{Crocce}
  et~al.}{2006}]{2006MNRAS.373..369C}
{Crocce} M.,  {Pueblas} S.,    {Scoccimarro} R.,  2006, \mnras, 373, 369

\bibitem[\protect\citeauthoryear{{Davis}, {Efstathiou}, {Frenk} \&
  {White}}{{Davis} et~al.}{1985}]{1985ApJ...292..371D}
{Davis} M.,  {Efstathiou} G.,  {Frenk} C.~S.,    {White} S.~D.~M.,  1985, \apj,
  292, 371

\bibitem[\protect\citeauthoryear{{Di Matteo}, {Colberg}, {Springel},
  {Hernquist} \& {Sijacki}}{{Di Matteo} et~al.}{2008}]{2008ApJ...676...33D}
{Di Matteo} T.,  {Colberg} J.,  {Springel} V.,  {Hernquist} L.,    {Sijacki}
  D.,  2008, \apj, 676, 33

\bibitem[\protect\citeauthoryear{{Efstathiou}, {Davis}, {White} \&
  {Frenk}}{{Efstathiou} et~al.}{1985}]{1985ApJS...57..241E}
{Efstathiou} G.,  {Davis} M.,  {White} S.~D.~M.,    {Frenk} C.~S.,  1985,
  \apjs, 57, 241

\bibitem[\protect\citeauthoryear{{Eisenstein} \& {Hu}}{{Eisenstein} \&
  {Hu}}{1998}]{1998ApJ...496..605E}
{Eisenstein} D.~J.,  {Hu} W.,  1998, \apj, 496, 605

\bibitem[\protect\citeauthoryear{Fedorenko}{Fedorenko}{1961}]{RPFedorenko_1961%
a}
Fedorenko R.~P.,  1961, Z. Vycisl. Mat. i. Mat. Fiz., 1, 922

\bibitem[\protect\citeauthoryear{{Hahn}, {Porciani}, {Carollo} \&
  {Dekel}}{{Hahn} et~al.}{2007}]{2007MNRAS.375..489H}
{Hahn} O.,  {Porciani} C.,  {Carollo} C.~M.,    {Dekel} A.,  2007, \mnras, 375,
  489

\bibitem[\protect\citeauthoryear{{Hahn}, {Teyssier} \& {Carollo}}{{Hahn}
  et~al.}{2010}]{2010MNRAS.405..274H}
{Hahn} O.,  {Teyssier} R.,    {Carollo} C.~M.,  2010, \mnras, 405, 274

\bibitem[\protect\citeauthoryear{{Hamilton}}{{Hamilton}}{2000}]{2000MNRAS.312.%
.257H}
{Hamilton} A.~J.~S.,  2000, \mnras, 312, 257

\bibitem[\protect\citeauthoryear{{Hockney} \& {Eastwood}}{{Hockney} \&
  {Eastwood}}{1981}]{1981csup.book.....H}
{Hockney} R.~W.,  {Eastwood} J.~W.,  1981, {Computer Simulation Using
  Particles}.
McGraw-Hill

\bibitem[\protect\citeauthoryear{{Hoffman} \& {Ribak}}{{Hoffman} \&
  {Ribak}}{1991}]{1991ApJ...380L...5H}
{Hoffman} Y.,  {Ribak} E.,  1991, \apjl, 380, L5

\bibitem[\protect\citeauthoryear{{Jenkins}}{{Jenkins}}{2010}]{2010MNRAS.403.18%
59J}
{Jenkins} A.,  2010, \mnras, 403, 1859

\bibitem[\protect\citeauthoryear{{Jing}}{{Jing}}{2005}]{2005ApJ...620..559J}
{Jing} Y.~P.,  2005, \apj, 620, 559

\bibitem[\protect\citeauthoryear{{Katz}, {Quinn}, {Bertschinger} \&
  {Gelb}}{{Katz} et~al.}{1994}]{1994MNRAS.270L..71K}
{Katz} N.,  {Quinn} T.,  {Bertschinger} E.,    {Gelb} J.~M.,  1994, \mnras,
  270, L71+

\bibitem[\protect\citeauthoryear{{Knollmann} \& {Knebe}}{{Knollmann} \&
  {Knebe}}{2009}]{2009ApJS..182..608K}
{Knollmann} S.~R.,  {Knebe} A.,  2009, \apjs, 182, 608

\bibitem[\protect\citeauthoryear{{Ma} \& {Bertschinger}}{{Ma} \&
  {Bertschinger}}{1995}]{1995ApJ...455....7M}
{Ma} C.,  {Bertschinger} E.,  1995, \apj, 455, 7

\bibitem[\protect\citeauthoryear{{Martin} \& {Cartwright}}{{Martin} \&
  {Cartwright}}{1996}]{martin1996}
{Martin} D.~F.,  {Cartwright} K.~L.,  1996, in Electronics Research Laboratory
  Memorandum UCB/ERL M96/66 Solving poisson's equation using adaptive mesh
  refinement.
University of California, Berkeley

\bibitem[\protect\citeauthoryear{Miniati \& Colella}{Miniati \&
  Colella}{2007}]{Miniati:2007:BSA:1297418.1297547}
Miniati F.,  Colella P.,  2007, J. Comput. Phys., 227, 400

\bibitem[\protect\citeauthoryear{{Munshi}, {Sahni} \& {Starobinsky}}{{Munshi}
  et~al.}{1994}]{1994ApJ...436..517M}
{Munshi} D.,  {Sahni} V.,    {Starobinsky} A.~A.,  1994, \apj, 436, 517

\bibitem[\protect\citeauthoryear{{Naoz}, {Yoshida} \& {Barkana}}{{Naoz}
  et~al.}{2010}]{2010arXiv1009.0945N}
{Naoz} S.,  {Yoshida} N.,    {Barkana} R.,  2010, arXiv:1009.0945

\bibitem[\protect\citeauthoryear{{Navarro} \& {White}}{{Navarro} \&
  {White}}{1994}]{1994MNRAS.267..401N}
{Navarro} J.~F.,  {White} S.~D.~M.,  1994, \mnras, 267, 401

\bibitem[\protect\citeauthoryear{{O'Shea}, {Bryan}, {Bordner}, {Norman},
  {Abel}, {Harkness} \& {Kritsuk}}{{O'Shea} et~al.}{2004}]{2004astro.ph..3044O}
{O'Shea} B.~W.,  {Bryan} G.,  {Bordner} J.,  {Norman} M.~L.,  {Abel} T.,
  {Harkness} R.,    {Kritsuk} A.,  2004, arXiv:astro-ph/0403044

\bibitem[\protect\citeauthoryear{{Padmanabhan} \& {Subramanian}}{{Padmanabhan}
  \& {Subramanian}}{1993}]{1993ApJ...410..482P}
{Padmanabhan} T.,  {Subramanian} K.,  1993, \apj, 410, 482

\bibitem[\protect\citeauthoryear{{Peebles}}{{Peebles}}{1982}]{1982ApJ...263L..%
.1P}
{Peebles} P.~J.~E.,  1982, \apjl, 263, L1

\bibitem[\protect\citeauthoryear{{Pen}}{{Pen}}{1997}]{1997ApJ...490L.127P}
{Pen} U.,  1997, \apjl, 490, L127+

\bibitem[\protect\citeauthoryear{{Pichon}, {Thi{\'e}baut}, {Prunet}, {Benabed},
  {Colombi}, {Sousbie} \& {Teyssier}}{{Pichon}
  et~al.}{2010}]{2010MNRAS.401..705P}
{Pichon} C.,  {Thi{\'e}baut} E.,  {Prunet} S.,  {Benabed} K.,  {Colombi} S.,
  {Sousbie} T.,    {Teyssier} R.,  2010, \mnras, 401, 705

\bibitem[\protect\citeauthoryear{{Press} \& {Davis}}{{Press} \&
  {Davis}}{1982}]{1982ApJ...259..449P}
{Press} W.~H.,  {Davis} M.,  1982, \apj, 259, 449

\bibitem[\protect\citeauthoryear{{Protogeros} \& {Scherrer}}{{Protogeros} \&
  {Scherrer}}{1997}]{1997MNRAS.284..425P}
{Protogeros} Z.~A.~M.,  {Scherrer} R.~J.,  1997, \mnras, 284, 425

\bibitem[\protect\citeauthoryear{{Prunet}, {Pichon}, {Aubert}, {Pogosyan},
  {Teyssier} \& {Gottloeber}}{{Prunet} et~al.}{2008}]{2008ApJS..178..179P}
{Prunet} S.,  {Pichon} C.,  {Aubert} D.,  {Pogosyan} D.,  {Teyssier} R.,
  {Gottloeber} S.,  2008, \apjs, 178, 179

\bibitem[\protect\citeauthoryear{{Salmon}}{{Salmon}}{1996}]{1996ApJ...460...59%
S}
{Salmon} J.,  1996, \apj, 460, 59

\bibitem[\protect\citeauthoryear{{Scoccimarro}}{{Scoccimarro}}{1998}]{1998MNRA%
S.299.1097S}
{Scoccimarro} R.,  1998, \mnras, 299, 1097

\bibitem[\protect\citeauthoryear{{Seljak} \& {Zaldarriaga}}{{Seljak} \&
  {Zaldarriaga}}{1996}]{1996ApJ...469..437S}
{Seljak} U.,  {Zaldarriaga} M.,  1996, \apj, 469, 437

\bibitem[\protect\citeauthoryear{{Silk}}{{Silk}}{1968}]{1968ApJ...151..459S}
{Silk} J.,  1968, \apj, 151, 459

\bibitem[\protect\citeauthoryear{{Sirko}}{{Sirko}}{2005}]{2005ApJ...634..728S}
{Sirko} E.,  2005, \apj, 634, 728

\bibitem[\protect\citeauthoryear{{Somogyi} \& {Smith}}{{Somogyi} \&
 {Smith}}{2010}]{2010PhRvD..81b3524S}
 {Somogyi} G., {Smith}, R.~E., 2010, \prd, 81, 023524

\bibitem[\protect\citeauthoryear{{Springel}}{{Springel}}{2005}]{2005MNRAS.364.%
1105S}
{Springel} V.,  2005, \mnras, 364, 1105

\bibitem[\protect\citeauthoryear{{Springel}, {White}, {Jenkins}, {Frenk},
  {Yoshida}, {Gao}, {Navarro}, {Thacker}, {Croton}, {Helly}, {Peacock}, {Cole},
  {Thomas}, {Couchman}, {Evrard}, {Colberg} \& {Pearce}}{{Springel}
  et~al.}{2005}]{2005Natur.435..629S}
{Springel} V.,  {White} S.~D.~M.,  {Jenkins} A.,  {Frenk} C.~S.,  {Yoshida} N.,
   {Gao} L.,  {Navarro} J.,  {Thacker} R.,  {Croton} D.,  {Helly} J.,
  {Peacock} J.~A.,  {Cole} S.,  {Thomas} P.,  {Couchman} H.,  {Evrard} A.,
  {Colberg} J.,    {Pearce} F.,  2005, \nat, 435, 629

\bibitem[\protect\citeauthoryear{{Stadel}, {Potter}, {Moore}, {Diemand},
  {Madau}, {Zemp}, {Kuhlen} \& {Quilis}}{{Stadel}
  et~al.}{2009}]{2009MNRAS.398L..21S}
{Stadel} J.,  {Potter} D.,  {Moore} B.,  {Diemand} J.,  {Madau} P.,  {Zemp} M.,
   {Kuhlen} M.,    {Quilis} V.,  2009, \mnras, 398, L21

\bibitem[\protect\citeauthoryear{{Talman}}{{Talman}}{1978}]{1978JCoPh..29...35%
T}
{Talman} J.~D.,  1978, Journal of Computational Physics, 29, 35

\bibitem[\protect\citeauthoryear{{Tatekawa} \& {Mizuno}}{{Tatekawa} \&
  {Mizuno}}{2007}]{2007JCAP...12..014T}
{Tatekawa} T.,  {Mizuno} S.,  2007, JCAP, 12, 14

\bibitem[\protect\citeauthoryear{{Teyssier}}{{Teyssier}}{2002}]{2002A&A...385.%
.337T}
{Teyssier} R.,  2002, \aap, 385, 337

\bibitem[\protect\citeauthoryear{{Trottenberg}, {Oosterlee} \&
  {Schuller}}{{Trottenberg} et~al.}{2001}]{trottenberg2001}
{Trottenberg} U.,  {Oosterlee} C.~W.,    {Schuller} A.,  2001, Multigrid.
Academic Press

\bibitem[\protect\citeauthoryear{{Turk}, {Abel} \& {O'Shea}}{{Turk}
  et~al.}{2009}]{2009Sci...325..601T}
{Turk} M.~J.,  {Abel} T.,    {O'Shea} B.,  2009, Science, 325, 601

\bibitem[\protect\citeauthoryear{{van Leer}}{{van
  Leer}}{1977}]{1977JCoPh..23..263V}
{van Leer} B.,  1977, Journal of Computational Physics, 23, 263

\bibitem[\protect\citeauthoryear{{Wise} \& {Abel}}{{Wise} \&
  {Abel}}{2008}]{2008ApJ...685...40W}
{Wise} J.~H.,  {Abel} T.,  2008, \apj, 685, 40

\bibitem[\protect\citeauthoryear{{Yamamoto}, {Sugiyama} \& {Sato}}{{Yamamoto}
  et~al.}{1998}]{1998ApJ...501..442Y}
{Yamamoto} K.,  {Sugiyama} N.,    {Sato} H.,  1998, \apj, 501, 442

\bibitem[\protect\citeauthoryear{{Yoshida}, {Omukai} \& {Hernquist}}{{Yoshida}
  et~al.}{2008}]{2008Sci...321..669Y}
{Yoshida} N.,  {Omukai} K.,    {Hernquist} L.,  2008, Science, 321, 669

\bibitem[\protect\citeauthoryear{{Yoshida}, {Sugiyama} \&
  {Hernquist}}{{Yoshida} et~al.}{2003}]{2003MNRAS.344..481Y}
{Yoshida} N.,  {Sugiyama} N.,    {Hernquist} L.,  2003, \mnras, 344, 481

\bibitem[\protect\citeauthoryear{{Zel'Dovich}}{{Zel'Dovich}}{1970}]{1970A&A...%
..5...84Z}
{Zel'Dovich} Y.~B.,  1970, \aap, 5, 84

\end{thebibliography}


\label{lastpage}
\end{document}